%% file: edge13co.tex
\documentclass[twocolumn]{aastex631}
\usepackage[caption=false]{subfig}
\shorttitle{Resolved \ttco \ in Nearby Galaxies}
\shortauthors{Cao et al.}
\graphicspath{{./}{figures/}}

\def\ttco{\mbox{$^{13}$CO}}
\def\twco{\mbox{$^{12}$CO}}
\def\twline{ $^{12}\mathrm{CO} (J = 1 \rightarrow 0)$}
\def\ttline{ $^{13}\mathrm{CO} (J = 1 \rightarrow 0)$}

\def\rdef{$\mathcal{R}_{12/13} \equiv I[^{12}\mathrm{CO} (J = 1 \rightarrow
  0)]/I[^{13}\mathrm{CO} (J = 1 \rightarrow 0)]$}
\def\fircolor{$F_{60}/F_{100}$}
\def\rqua{$r_{\rm 25}$}
\def\rtt{\mbox{$\mathcal{R}_{12/13}$}}
\def\grtt{\mbox{$\mathcal{R}_{12/13}^{\rm gal}$}}
\def\rstack{\mbox{$\mathcal{R}_{12/13}^{\rm stack}$}}

\def\un{\rm \ }
\def\itw{\mbox{$I_{12}$}}
\def\itt{$I_{13}$}

\def\itts{\mbox{$I_{13}^{\rm stack}$}}
\def\ftw{$F_{12}$}
\def\ftt{$F_{13}$}

\def\itw{$I_{12}$}
\def\itt{$I_{13}$}

\def\ha{\mathrm{H} \alpha}
\def\hb{\mathrm{H} \beta}
\def\logoh{\log{\mathrm{(O/H)}}}

\newcommand{\beq}{\begin{equation}}
\newcommand{\eeq}{\end{equation}}

\newcommand{\mc}[1]{\multicolumn{1}{c}{#1}}

\begin{document}

\title{The EDGE-CALIFA Survey: 
Spatially Resolved \ttco$(1-0)$ Observations and 
Variations in \twco$(1-0)$/\ttco$(1-0)$ in Nearby Galaxies on kpc Scales}

\input{affiliations.tex}

\input{authors.tex}



\begin{abstract}

We present \ttline\ observations for the EDGE-CALIFA survey, 
which is a mapping survey of 126 nearby galaxies at a typical spatial resolution of 1.5 kpc. 
Using detected \twco\ emission as a prior, we detect \ttco\ in {41} galaxies via integrated line flux {over the entire galaxy}, and in {30} galaxies via 
integrated line intensity {in resolved synthesized beams}.
Incorporating our CO observations and optical IFU spectroscopy, we perform a systematic comparison between the line ratio \rdef\ and the properties of the stars and ionized gas.
Higher \rtt\  values are found in interacting galaxies than in non-interacting galaxies. 
The global \rtt\ slightly increases with infrared color \fircolor,  
but appears insensitive to other host galaxy properties such as morphology, stellar mass, or galaxy size. 
We also present annulus-averaged \rtt\  profiles for our sample up to a galactocentric radius of  $0.4r_{25}$\ ($\sim6\rm\ kpc$), taking into account the \ttco \  non-detections by spectral stacking.
The radial profiles of \rtt\ are quite flat across our sample. 
Within galactocentric distances of  $0.2$\rqua, azimuthally-averaged \rtt\ increases with star formation rate. However, the Spearman rank correlation tests show the azimuthally-averaged \rtt\ does not strongly correlate with any other gas or stellar properties in general, especially beyond $0.2$\rqua\ from the galaxy centers. 
Our findings suggest that in the complex environments in galaxy disks,  \rtt\ is not a sensitive tracer for ISM properties. Dynamical disturbances, like galaxy interactions or the presence of a bar, also have an overall impact on \rtt, which further complicate the interpretations of \rtt\ variations. 

\end{abstract}



\section{Introduction}  \label{sec:intro}

The molecular interstellar medium (ISM) plays a critical role in galaxy evolution, serving as the fuel for star formation in galaxies. Large scale processes accumulate the gas and regulate the abundance and properties of molecular clouds. In turn, the gas properties on local cloud scales control the star formation process which drives the evolution of the entire galaxy. It is therefore important to understand how local and global properties of galaxies affect the molecular gas. 

In nearby galaxies, 
\twline \ is the most commonly used tracer of molecular gas. 
However, \twline\ is typically optically thick, and its optical depth can vary. Because the isotopologue \ttco\ is much less abundant in galaxies {\citep[e.g.][]{WilsonRood1994, Henkel2014}}, its lines tend to be optically thin and offer a powerful complement to \twline. 
The line intensity ratio  \rtt$ \equiv I[^{12}\mathrm{CO} (J = 1 \rightarrow
  0)]/I[^{13}\mathrm{CO} (J = 1 \rightarrow 0)]$ \ can be used to trace variations in optical depth of molecular gas, with the caveat that \rtt \  can also be influenced by chemical processes. 

Because the emission of \ttco \ is weaker and more difficult to observe, interferometric mapping studies of this line have been only made in a small number of galaxies, often focusing on a single target each time \citep[e.g.][]{Meier2004, M64, Aalto2010,Pety2013, Sliwa2017, Gallagher2018}. 
Resolved \rtt \ values ranging from $\sim$4 to $\sim$25 have been 
reported in these studies. 
Previous single dish studies found that the galaxy-integrated \rtt\ 
correlates with \textit{IRAS} IR color and the average star formation rate (SFR) surface density,  
indicating lower opacity in more active, hotter systems \citep{Y&S1986, Aalto1991, SageIsbell1991, Davis2014}.  
In contrast to these galaxy-integrated studies, studies comparing resolved \rtt\ estimates to the SFR surface density within galaxies have not found a strong correlation \citep[e.g.][]{Cao2017, Cormier2018}.  
Alternatively, high \rtt\ in a galactic region may indicate the presence of more diffuse molecular gas than in low \rtt\ regions \citep[e.g.][]{R&B1985, Goldsmith2008, Pety2013, Roman-Duval2016}.
These findings suggest that changes in \rtt\ may reflect variations in the density of molecular gas in different regions within a galaxy. 

On the other hand, variations in \rtt \ could reflect variations in \ttco \  abundance resulting from changes in the production or destruction of $^{13}$C or \ttco. 
While $^{12}\rm C$ is produced by all stars through various mass loss mechanisms, 
$^{13}\rm C$  is ejected into the ISM in the AGB phase of low mass stars mainly via CNO processing of $^{12}\rm C$. The isotope fractional abundance $[^{12}\rm C$/$^{13}\rm C]$ is therefore an important tracer of the chemical evolution of galaxies. 
\textcolor{black}{
$[^{12}\rm C$/$^{13}\rm C]$ in nearby galaxies could range from 20 to 150 \citep[][and references therein]{WilsonRood1994, Henkel2014, Tang2019}.  
Due to the high opacity of CO (typically $>1$ in extragalactic environments), \rtt \ sets a lower bound on  $[^{12} \rm C$/$^{13} \rm C]$ \citep{Henkel1993, Henkel2014}. }
Abnormally high \rtt \ values are found in 
(U)LIRGs and high redshift galaxies and have been suggested to result from recent star formation in these galaxies \citep{Danielson2013,Sliwa2017}. 
Meanwhile, studies using chemical evolution modeling suggests \rtt \ could be used as a diagnostic of the stellar initial mass function 
\citep{Romano2017, Zhang2018}. 
In addition, chemical modeling studies suggest that the carbon  fractionation in CO, [\twco/\ttco], is less affected by the ISM chemistry over long time {scales} and in typical physical conditions, thus making it a plausible tracer of primordial $[^{12} \rm C$/$^{13} \rm C]$ in galactic environment \citep{Viti2020, Colzi2020}. 
Nonetheless, our overall understanding of how \rtt\ relates to the chemical evolution of galaxies remains poorly constrained, and may benefit from inclusion of stellar and gas metallicity data.



In this paper, we present a systematic study of \ttline~ and \rtt~ variations down to kpc scales in the nearby universe, using the CARMA Extragalactic Database for Galaxy Evolution (EDGE) sample which targeted a total of 126 galaxies \citep{Bolatto2017}. 
Combining our \twco\ and \ttco\  observations with optical spectroscopic data, 
we investigate how \rtt~ could be affected by global galaxy properties and how its variations relate to the underlying stellar population and ionized gas on kpc scales. 
Our systematic study of \rtt~ over a wide variety of galaxies could provide useful insights about molecular gas properties in response to local star formation activity, chemical evolution, and to large scale dynamical processes.
We describe the \ttco \ observations and the ancillary data from the CALIFA IFU survey in Section \ref{sec:eobs}.
In Section \ref{sec:eresults}, we present measurements of \rtt \ and relate them to global and resolved properties. 
We discuss the possible causes of \rtt \ variations and the caveats of this study in Section \ref{sec:ediscuss}.

\section{Observations and Data Description}  
\label{sec:eobs}

\subsection{EDGE \ttco \ observations}
\ttco \  was simultaneously observed with \twco \ in the EDGE survey. 
For the purpose of this study focusing on \ttco, we summarize the essential characteristics of \ttco \ observations.  
Full details of the survey, observations, and data reduction 
for the EDGE sample are described in \citet{Bolatto2017}. 

The observations of CARMA EDGE were carried out from late 2014 to mid 2015 in the CARMA D and E arrays. 
The targets are 126 galaxies from the CALIFA sample selected for high 22 $\mu$m flux. 
The redshifted \ttco \ line, with a rest frequency of 110.201 GHz, is covered in the lower sideband (LSB). 
To cover 
a velocity range of $3800 \rm \ km/s$ with $14.3  \rm \ km/s$ resolution,
the LSB setup uses 3 spectral windows with bandwidth of 500 MHz each, 
whereas for \twco \ the bandwidth is 250 MHz and 5 spectral windows are used for  
a combined velocity range of $3000 \rm \ km/s$.
Therefore, 
\ttco \ has a lower spectral resolution of $14.3 \rm \ km/s$ than the \twco \ with $3.4 \rm \ km/s$ resolution. 

A 7-point hexagonal mosaic with centers separated by half {the} 
primary beam size of $27 \arcsec$ 
was used, to cover a field-of-view (FoV) with radius $\sim 50 \arcsec$. 
The integration time of each galaxy is about 40 minutes in E-array, 
and $\sim 3.5$ hours in D-array. 

The visibility data were calibrated by the 
automatic pipeline developed for STING 
using MIRIAD \citep{Rahman2011,Rahman2012,Wong2013}.
The uncertainty of the flux calibration is estimated at $\sim$10\% (see \citealt{Bolatto2017} for details). 
The calibrated data were deconvolved and imaged using the 
MIRIAD tasks MOSSDI2 and INVERT with Briggs's weighting 
robustness parameter of 0.5. 
The resulting typical beam size for \ttco \ is $\sim 5\arcsec$. 
Cubes were generated with a pixel size of $1 \arcsec$ and 
a velocity range is $860 \rm \  km \ s^{-1}$
($1600 \rm \  km \ s^{-1}$ for Arp 220) 
with a channel spacing of $20 \rm \  km \ s^{-1}$. 

We align the \ttco \ cube to match the pixel grid of the \twco \ cube 
for each galaxy. 
We choose a common resolution of $7 \arcsec$ 
and smooth all the \twco \ and \ttco  \ cubes to 
a FWHM beam size of $7 \arcsec$ for further analysis. 
{This resolution corresponds to physical scale of  0.68-4.4 kpc, given the range of distances of our sample is 20-130 Mpc. }
The typical channel noise for \ttco \  and \twco\ at this resolution are 8.5 mK and 16.1 mK respectively.

\subsection{Auxiliary Maps from EDGE-CALIFA} 
\subsubsection{EDGE \twco\ Moment Maps}

The \twco\ integrated intensities are derived from their respective cubes after applying the \twco\ dilated mask from \citet{Bolatto2017} in order to reject regions of the cube dominated by noise.  The mask was generated by identifying pixels with \twco\ emission above $3.5 \sigma$ in two adjacent channels of the cube, and expanding to contiguous regions with \twco\ brightness above $2\sigma$. The mask was expanded by an additional ``guard'' band of 1 pixel in all three dimensions in order to capture additional low-level emission.  The first three moments of the emission within the mask (moment-0, 1, and 2) were then calculated, representing the integrated intensity, intensity-weighted mean velocity, and intensity-weighted velocity dispersion respectively.  

\subsubsection{The CALIFA IFU Maps}

The galaxies in the EDGE sample are selected from the CALIFA survey DR3, 
an optical IFU survey consisting of $\sim 600$ nearby galaxies 
observed by the  $3.5 \un m$ telescope at the Calar Alto Observatory \citep{CalifaI,CalifaDR3}. 
The spatial resolution of CALIFA is typically $2.5 \arcsec$, 
corresponding to a scale of $\sim 700 \un pc$ at a distance of $64 \un Mpc$ \citep{Bolatto2017}. 
The spectra are observed in two settings, with 
a low resolution of $6$\  \AA \ across an unvignetted spectral range of 4240--7140 \AA\
and a medium resolution of $2.3$ \AA \ covering 3650--4620 \AA\ \citep{CalifaDR3}.  

The spatially resolved properties of the stellar populations 
and emission lines are derived from the CALIFA observations by 
Pipe3D \citep{Pipe3D}. 
The Pipe3D data products are stored as maps with a pixel size of $1\arcsec$ for each galaxy. 
Within Pipe3D, the resolved stellar mass, age, and metallicity are derived from the stellar populations modeling of the stellar emission. 
Fluxes, velocities, and line widths of the emission lines of each pixel are also derived by Pipe3D, and are used to derive the resolved ionized gas properties (Section \ref{sec:equations}). 

\subsection{Global Properties of Galaxies}

\subsubsection{\twco \ Velocity Dispersion}

\textcolor{black}{
Although the \twco\ moment-2 map provides a nominal estimate of the velocity dispersion at each location in the galaxy, it does not correct for the beam smearing of the galaxy's rotation, which can be especially important in the central regions of galaxies where the velocity gradient is largest.  We therefore obtained a separate estimate of the \twco\ velocity dispersion as a function of radius by fitting a circular rotation model directly to the masked \twco\ data cube, using the 3D-Barolo software package \citep{Teodoro2015}.  The fitting was performed with the center position fixed to the value provided by NED and with initial estimates of disk position angle (P.A.) and inclination provided by \citet{Levy2018}.  For roughly a dozen galaxies, the center was shifted to the \twco \  intensity peak if a clearly defined peak was present and this resulted in {smaller residuals and thus}
a substantially improved kinematic fit.  The program runs in two stages, in which the systemic velocity and P.A. are allowed to vary with radius in the first stage, and are then fixed to their mean fitted values in the second stage.  The rotation velocity and velocity dispersion are allowed to vary with radius in both stages.  Up to 8 rings of width 3\arcsec\ are fitted for each galaxy, although not all rings provide successful fits.  We characterize the face-on velocity dispersion of each galaxy as the median value of the dispersion, obtained at the second fitting stage, for all rings in the galaxy. 
}

\subsubsection{Other Host Galaxy Properties}
The Pipe3D data products are used to derive the global properties for each galaxy in CALIFA, either by integrating or taking the characteristic values at the effective radius \citep{Sanchez2018}. 
The global stellar mass of galaxy ($M_*$) is calculated by integrating all the pixels in the FoV with $\mathrm{S/N}>1$ in the continuum. 
The global SFR is also derived by integration. 
On the other hand, the global age, metallicity, and dust extinction are the corresponding mean values of the pixels at the effective radius $r_e$. 
In particular, the global age is the luminosity weighted age of the stellar population at $r_e$,
the gas metallicity is derived using the \citet{Marino2013} O2N3 calibration (Equation \ref{eq:o3n2}) at $r_e$, and 
the global $\mathrm{A_v}$ is the dust attenuation in the $V$ band derived from the $\ha/\hb$ line ratio at $r_e$. 
In addition, we obtain \textit{IRAS} IR colors $F_{60}/F_{100}$ from NED for each galaxy and list them in Table \ref{table:edgeg1}.
{Galaxy parameters such as distance,  $M_*$, SFR, and metallicity for the EDGE-CALIFA survey can be found in Table 1 and 3 in \citet{Bolatto2017}. 
The rest of the global parameters we use in this paper are presented in the Appendix. 
}

\citet{Barrera-Ballesteros2015} classified a sample of 103 CALIFA galaxies into different interacting stages 
using $r$-band Sloan Digital Sky Survey images. 
{We use the stages they identified;  for our sample that are not in these 103 galaxies, we classify the stage using the same method as in \citet{Barrera-Ballesteros2015}. }
For this study, we define a galaxy as ``interacting'' if it is in the merging stage (galaxy in a binary system with clear signatures of interaction) or post-merger stage (with the nuclei coalesced and evident tidal features) following \citet{Barrera-Ballesteros2015}.

\subsection{Resolved Data Points}
\subsubsection{Data alignment and extraction}
%
We align and convolve the CALIFA Pipe3D maps  
to match the EDGE CO maps and build a database 
that combines all the resolved properties of 
stars, ionized gas, and molecular gas. 
The Pipe3D maps are first regridded to the EDGE CO cubes
using the IDL routine \textit{hastrom} so that all 
the images of each galaxy are in the same celestial and velocity frame. 
Subsequently, we convolve the regridded Pipe3D maps  
to $7 \arcsec$ using a Gaussian kernel to match the typical 
resolution of the EDGE CO observations,   
assuming a Gaussian PSF of $2.5 \arcsec$ for the CALIFA cubes.  
We note this convolution is only correct for surface brightnesses, therefore properties such as SFR and metallicity are derived after smoothing following the methods described in Section \ref{sec:equations}.

After the CALIFA and EDGE images are aligned and smoothed, 
we extract values from nboth the regridded and smoothed 
images along a hexagonal sampling grid.
We generate the hexagonal 
grid with a nearest-neighbor spacing of $3.5 \arcsec$, 
that is, half of the FWHM beam size of the convolved images. 
Pixel values in radio maps usually oversample the beam and hence are not independent; 
this resampling reduces the number of pixels by a factor of 10 while still preserving a modest degree of oversampling (from \S\ref{sec:maps} there are 55 pixels per beam area compared to approximately 5 hexagonal grids per beam).
We adopt a hexagonal grid because it leads to equal spacing between all adjacent pixels, and we expect that the adopted sampling scheme has negligible impact on our results.
We take the values of the $1\arcsec \times 1\arcsec$ 
pixel nearest to each hexagonal grid point
and extract them from the images for the  
analysis in this paper.
Throughout the paper, we use the term ``grids" to refer to 
the hexagonally sampled pixels to distinguish them from 
the original pixels of the maps.

\subsubsection{Derived Resolved Properties}
\label{sec:equations}

%

We calculate the star formation rate (SFR) from the extinction corrected H$\alpha$ flux of each grid. 
The attenuation of $\ha$  is calculated from 
the flux ratio of ${\ha}/{\hb}$ using Eq. (1) 
from  \citet{Catalan2015}:
\beq 
\label{eq:Aa}
	A(\mathrm{H}_\alpha) = 
    \frac{K_{\ha}}
    	{-0.4 \times (K_{\ha}-K_{\hb})} 
    \times 
     \log{\frac{\ha/{\hb}}{2.86}}, 
\eeq 
where $K_{\ha} = 2.53 , K_{\hb} = 3.6$ are the extinction coefficients defined in \citet{Cardelli1989} for the Galactic extinction curve, 
and the constant $2.86$ is the intrinsic Balmer ratio for 
case B recombination. 
After the extinction correction, $\ha$ flux is  converted to SFR following Eq.\ (2) 
from \citet{RG2002} assuming a Salpeter IMF: 
\beq 
\label{eq:sfr}
SFR(\mathrm{M_{\odot}} \un{yr^{-1}}) 
	 = 7.9 \times 10^{-42} \ 
     	L_{\ha} (\mathrm{erg} \un {s^{-1}}) . 
\eeq
We use the Salpeter IMF for historical reasons related to CALIFA data processing \citep{Pipe3D}. 
\textcolor{black}{
For a more realistic SFR comparable to other recent values in the literature using IMF of \citet{Chabrier2003} or \citet{Kroupa2001}, we suggest to divide the quoted value by $\sim 1.6$ \citep[][and references therein]{Bolatto2017}}.

Both of these equations used for the SFR calculations are applicable 
to star forming regions only. 
We therefore identify the star formation regions by applying  the criterion suggested by \citet{SM2016}: 
we use the \citet{Kewley2001} demarcation line to exclude AGN-like regions, 
and require an $\ha$ equivalent width of $EW(\ha) > 6 $ \AA \ 
to exclude weak AGNs and regions ionized by 
low-mass evolved stars  \citep{CF2011}. 
Only for the grids classified as star forming regions is the star formation rate calculated using Equations (\ref{eq:Aa}) and (\ref{eq:sfr}). 
The surface densities are deprojected for disk inclination $i$ using a correction factor of $\cos{i}$ where $i$ is tabulated in \citet{Bolatto2017} Table 1.

We also calculate 
the oxygen abundance $12 + \logoh$  
for each star forming grid from 
the O3N2 index, $\mathrm{O3N2} = \log([\mathrm{O III}]\lambda5007/\hb) - 
							\log([\mathrm{NII}]\lambda6583/\ha)$.  
We use the empirical relation calibrated 
from $T_{e}$-based abundance measurements by \citet{Marino2013}:           
\beq 
12 + \logoh = 8.53 - 0.21 \times \mathrm{O3N2}. 
\label{eq:o3n2}
\eeq 

Besides SFR and oxygen abundance, the remaining IFU-based resolved properties (stellar surface density, stellar mass metallicity, and dust extinction) used in this study are derived by Pipe3D directly, and are hence at the native CALIFA resolution and gridding.

\subsection{Azimuthally averaged IFU properties}
\label{sec:stackprop}
\textcolor{black}{We derive azimuthally averaged stellar and ionised gas properties based on the corresponding resolved IFU maps. 
We use normalized radial bins $r_{\rm gal}/r_{25}$ with a step of $0.1$. 
These values are used in Section \ref{sec:erstackcorr} to investigate the correlations between azimuthally stacked line ratio \rstack \ and local properties. 
Since \rstack \ are measured from the stacked spectra which average the CO emissions for a given radial bin (see Section \ref{sec:specstack}), we use the mean value of each property to represent the bin, except for the sSFR, molecular gas fraction, and ionized metallicity because these are ratios. 
For the sSFR and molecular gas fraction, we use the mean values of stellar surface density, star formation surface density, and the molecular gas surface density to derive the corresponding ratios for the radial bin. 
For the ionized metallicity, we choose the median of resolved $12+\log(\rm O/H)$ to represent the annulus instead of the mean.  
}

\section{\ttco \ measurements}
\subsection{\ttco~ Moment-0 Maps}
\label{sec:maps}

\begin{figure*}[ht!]
\includegraphics[width=0.95\textwidth]{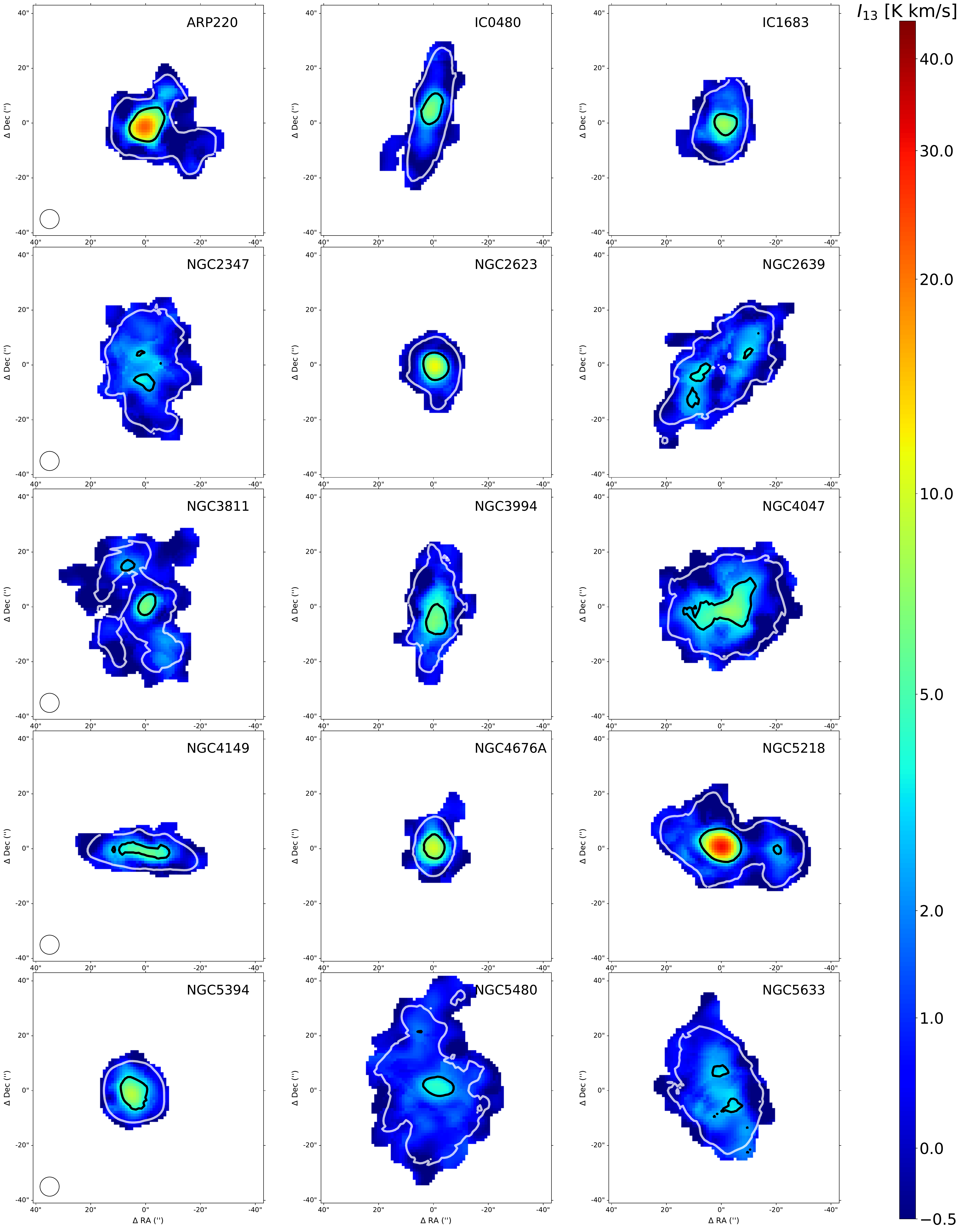}
\centering
\caption{\ttco\ intensity maps of the {30} galaxies with resolved \ttco\ detected from the EDGE survey. 
The black contours overlaid show the \ttco\ intensity observed with 
{S/N $>$4}, 
while the white contours levels show the \twco \ intensity with 
{S/N $>$3}.
{The black circles in the first panel of each row show the beam size.}
{We use a single color palette for all the galaxies, and the}
color scale is adjusted to emphasize the \ttco~ emission. 
\label{fig:maps}}
\end{figure*}

\begin{figure*} \ContinuedFloat
\epsscale{1}
\includegraphics[width=0.95\textwidth]{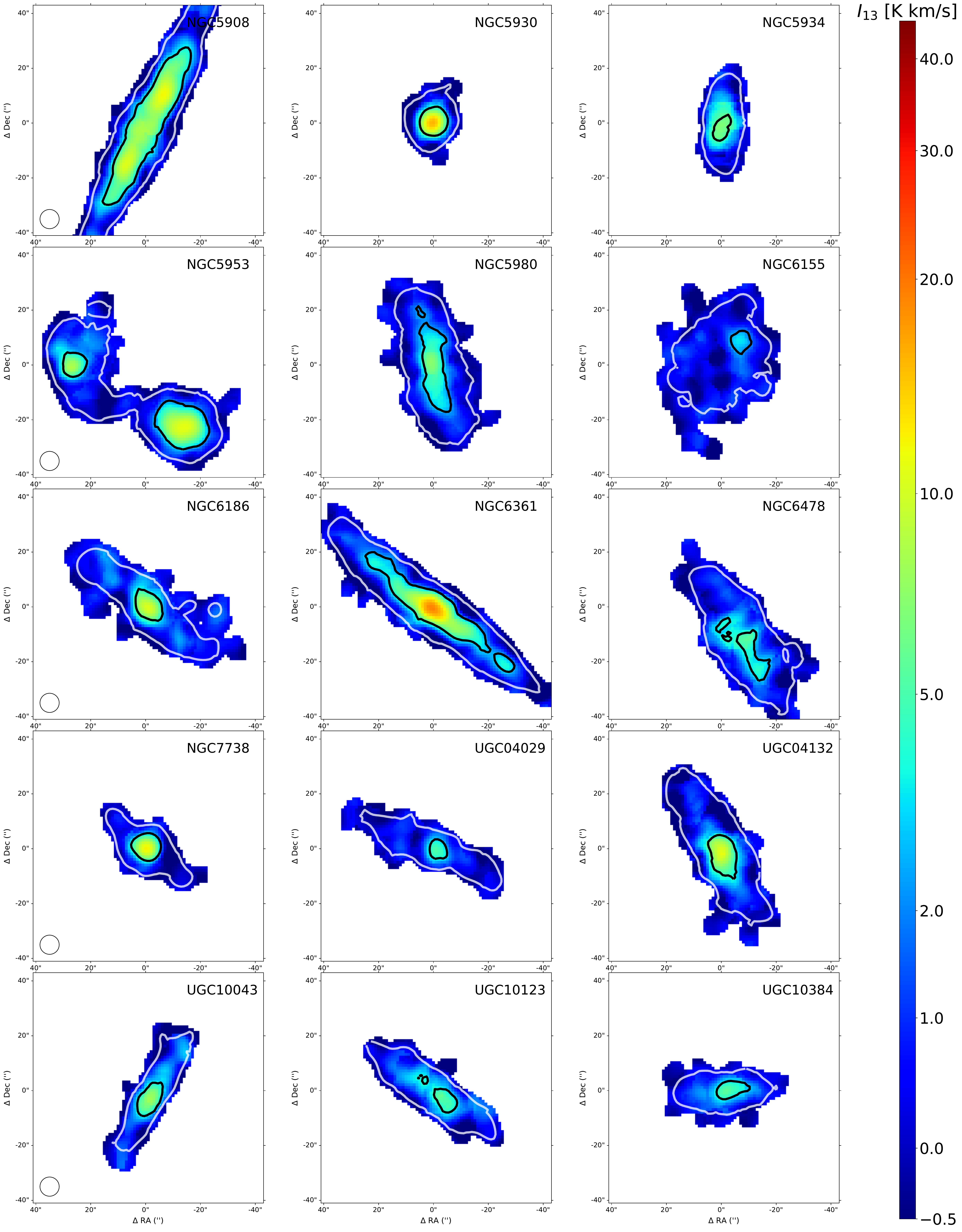}
\centering
\caption{(Continued).
\label{fig:maps2}
}
\end{figure*}

%
We use the same \twco \ dilated mask from \citet{Bolatto2017} 
to generate the  \ttco \ integrated intensity maps. 
We detect resolved {\ttco \ intensity }\itt~ in {30} galaxies (Figure \ref{fig:maps});  
in each of these galaxies, there are more
than {27} pixels (corresponding to {half} the size of 
a Gaussian beam of $7 \arcsec$) detected {in \ttco \ with S/N $>4$}. 
{Since \twco~ masks are used to derive the \ttco~ maps, negative values are occasionally present especially in the outer disks, mainly due to the random noise fluctuations of the faint \ttco \  emission.} 
All galaxies with \itt \ detected  
are also considered detections in integrated flux of \ttco \ ($F_{13}$, {see Section \ref{sec:fluxspec}}), except for {IC 0480 and UGC 04029.}  
In these two galaxies, \ttco \ is well detected {in galaxy centers.} 
However, the overall \ttco \ emission is weak;
negative \ttco \ fluxes from the noisy regions {on the edges of the maps}
diminish {$F_{13}$ integrated in both velocity and position} below the detection limit. 

The \ttco \ intensity maps of the {30} galaxies with resolved \itt \ detected  
are shown in Figure \ref{fig:maps}. 
Because \ttco \ is much weaker and has lower S/N 
than \twco \ in our observations, 
{\ttco \ intensity (\itt) detections with S/N $>4$}   
(shown in the black 
contours) {are mostly located within regions with brightest \twco}. 
 {Regions where we have reliable resolved \itt~ and \rtt~ are very limited and biased. Therefore, to capture the more robust averaged features, we stack the spectra over multiple regions/grids (see Section \ref{sec:specstack})}. 

\subsection{\ttco~  Flux Spectra}
\label{sec:fluxspec}

\begin{figure*}[htb!]
\includegraphics[width=0.95\textwidth]{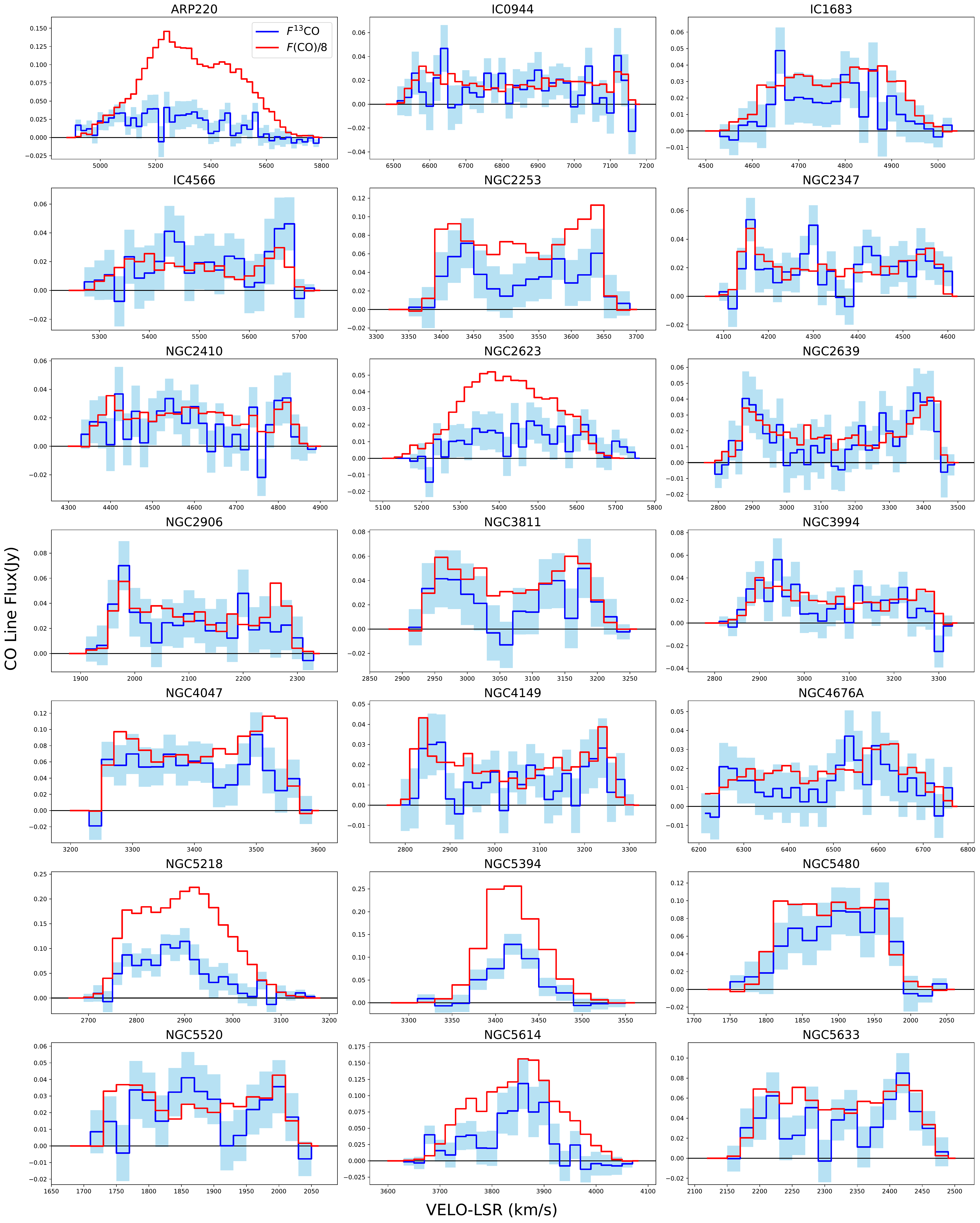} 
\centering
\caption{
\ttco \  flux spectra for the {41} galaxies 
with $F_{13}$ detected. 
The blue lines show the \ttco \ flux in each channel 
obtained using the dilated \twco\ masks, and red 
lines are the \twco \ flux resulting from the 
same masks scaled down by a factor of 8. 
The light blue shaded regions represent 
$1\sigma$ uncertainty of \ttco \ flux.  
}
\label{fig:efluxspec}
\end{figure*} 

\begin{figure*}[htb!]\ContinuedFloat
\includegraphics[width=0.95\textwidth]{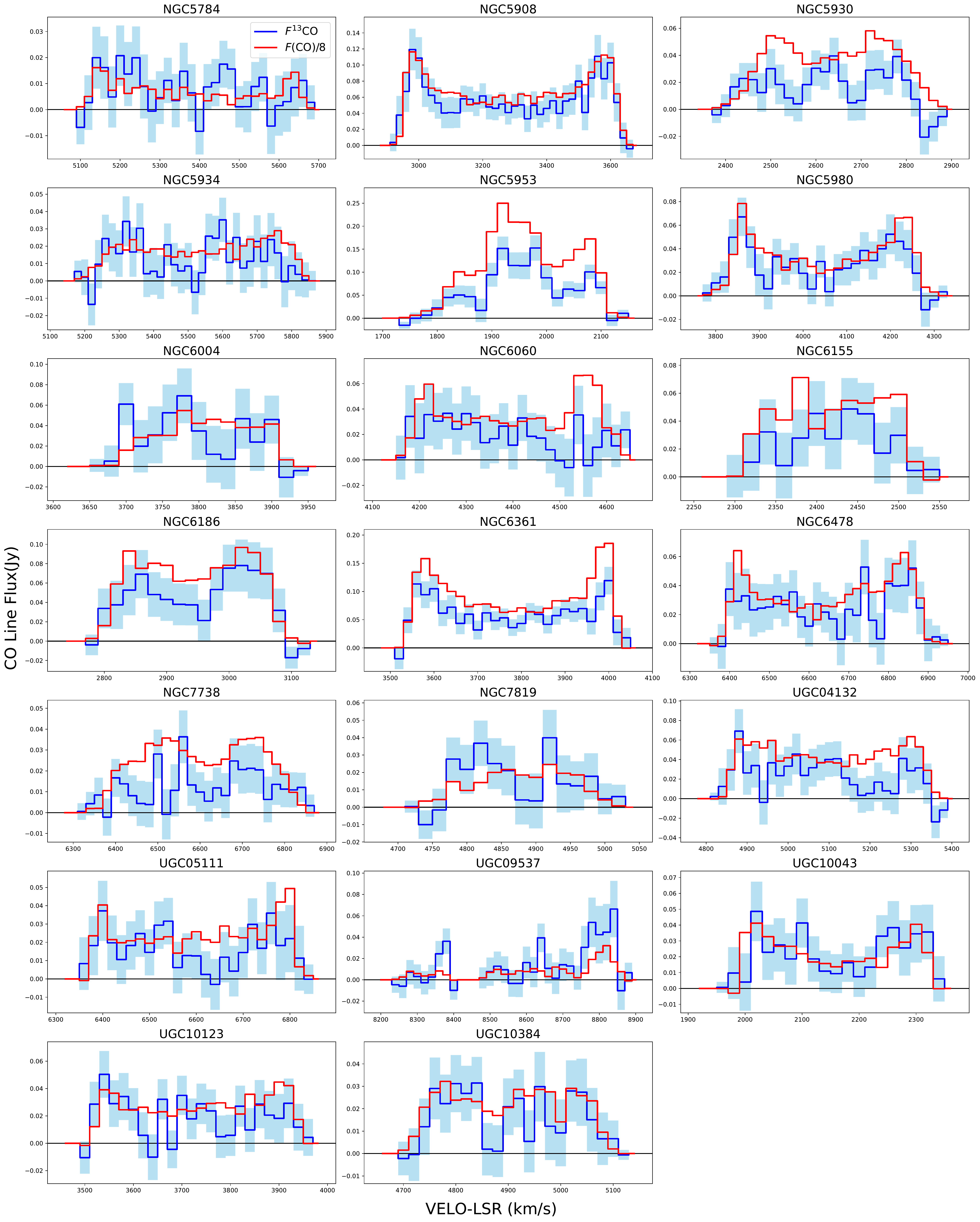} 
\centering
\caption{(Continued).
}
\end{figure*}

We sum all of the \twco\ and \ttco\ spectra within the \twco\ dilated mask to obtain integrated spectra for each galaxy. 
In other words, the flux in each channel is the summed intensity for all valid pixels in the mask. 
We also calculate velocity-integrated \twco\ and \ttco\ fluxes ($F_{12}$ and $F_{13}$) from these masked spectra.
The uncertainty in the spectra and fluxes are calculated by error propagation, taking into account the oversampling of the beam and using root-mean-square (RMS) noise values estimated from the signal-free velocity channels.

We list in Table \ref{table:edgeg1} the galaxy properties and \ttco \ fluxes of
{the 41 galaxies with S/N$(F_{13}) > 4$}. 
The other {64} galaxies detected in $F_{12}$ but not  $F_{13}$ are 
still used to provide lower limits on {the galaxy integrated line ratio \grtt \ } in the 
investigation of global correlations (Figure \ref{fig:f13vsglb}).
{We provide the upper limits of $F_{13}$ and the lower limit of \grtt \ {for} these galaxies in 
Table \ref{table:edgeg1}. }
The remaining 21 galaxies {in the EDGE-CALIFA sample} are not detected in \ftw~ at the $3\text{-}\sigma$ level, 
{and we omit these galaxies in the rest of the paper}.
The \ttco \ and \twco \ flux 
spectra of the 56 $F_{13}$ detected galaxies 
are shown in 
Figure \ref{fig:efluxspec}. 
We note that because the mask is a function of velocity, the specific flux and its uncertainty at a given channel depends strongly on the emission mask.  The $\pm$1$\sigma$ uncertainties in the \ttco\ spectra are indicated by the blue shading in Figure \ref{fig:efluxspec}.

\subsection{Radial stacked \ttco \ spectra}
\label{sec:specstack}
Since both \itw \ and \itt \ 
{generally} 
decrease exponentially with radius, 
few individual grids at large radius have detectable \itt, leaving {parts of the parameter space} unexplored in our resolved study of \rtt. 
To improve the S/N compared to what can be achieved in individual grids, {we first deproject the position of each grid on to the galactic plane using the galaxy's inclination and position angle, and calculate its distance to the galaxy's center $r_{\rm gal}$}.  
We then obtain the radial profiles of \rtt\ by stacking the spectra of \itw \ and \itt \ 
in normalized radial bins {in units of }$r_{\rm gal}/r_{25}$. 
{In a given radial bin}, the \ttco \ and \twco \ spectra 
for all the grids with \itw \ detected are selected. 
All of the masked spectra in {the} radial bin are shifted to a 
common central velocity (the weighted 
mean velocity of the \twco\ line provided by the moment-1 map), 
and are then averaged.
The channel noise of the stacked spectrum is calculated by taking the RMS of all the spectra that are stacked, and then scaling by a factor of $\sqrt{5.23}$ to account for the oversampling rate of 5.23 of the hexagonal grids.   
We fit the stacked \twco \ and \ttco \ spectra with a Gaussian function, and obtain the integrated fluxes from the Gaussian integral.  
The uncertainty of the integrated flux is calculated by 
\beq
\epsilon_{\rm stack} = \sigma_{\rm ch} \sqrt{{\rm FWHM} \cdot \Delta V_{\rm ch}}. 
\label{eq:stack_error}
\eeq
where $\sigma_{\rm ch}$ and $\Delta V_{\rm ch}$ are the channel noise and the channel width of the stacked spectrum respectively, and FWHM is obtained from the Gaussian fitting. 

{We start from the initial radial bins with a step of $0.1 (r_{\rm gal}/r_{25})$. If the stacked \ttco \ integrated flux \itts \ is below the detection threshold of S/N $=4$, we merge the bin with the next in the direction of increasing $r_{\rm gal}$, until \itts \ has S/N $>4$, the bin size reaches an upper limit of $0.4 (r_{\rm gal}/r_{25})$, or there are no more bins to merge beyond $0.8 (r_{\rm gal}/r_{25})$, and repeat the stacking.} 
We show the stacked spectra of two galaxies as examples in Appendix  \ref{sec:arradius}. 
{The Gaussian fitted fluxes resulting from the stacked \twco \ and \ttco \ spectra in these adjusted bins are used to derive the radial profiles of stacked line ratio \rstack. (see Section \ref{sec:eradial}). }

{In total, we detect \itts \ in 41 galaxies in our sample. All of the galaxies with resolved \itt \ detected are also detected with the spectral stacking method, except for the five galaxies for which we adopt inclinations of $90 ^{\circ}$ (IC0480, UGC04029, UGC10043, UGC10123, UGC10384). For these galaxies, the galactic distance $r_{\rm gal}$ cannot be calculated, so we omit them from the azimuthal stacking.}

\startlongtable
\begin{deluxetable*}{l r r c r r r r c r r c c}
\tabletypesize{\scriptsize}
\tablecaption{\label{table:edgeg1} \ttco~ and 
\rtt\ measurements in the EDGE-CALIFA Survey}
\tablehead{
\colhead{Galaxy} &
\colhead{$F_{13}$} &
\colhead{\grtt} &
\colhead{$\left<\sigma(I_{13})\right>$} &
\colhead{$\mathcal{N}_{13}$} & 
\colhead{$\mathcal{N}_{12}$} & 
\colhead{$\mathcal{N}_{12}^{\mathcal{R}}$} &
\colhead{\rtt} & 
\colhead{$\mathcal{N}_{13}^{\rm stack}$} & 
\colhead{\rstack} &
\colhead{$F_{60}/F_{100}$} & 
\colhead{Inter} & 
\colhead{Bar} \\
\colhead{} &
\colhead{(Jy km/s)} &
\colhead{} &
\colhead{(K km/s)} & 
\colhead{} & 
\colhead{} & 
\colhead{} & 
\colhead{} & 
\colhead{} & 
\colhead{} & 
\colhead{} & 
\colhead{} & 
\colhead{} \\
\colhead{(1)} &
\colhead{(2)} &
\colhead{(3)} &
\colhead{(4)} & 
\colhead{(5)} & 
\colhead{(6)} & 
\colhead{(7)} & 
\colhead{(8)} &
\colhead{(9)} & 
\colhead{(10)} &
\colhead{(11)} & 
\colhead{(12)} & 
\colhead{(13)}}
\startdata
\input{edge_gts.tex} 
\enddata
\tablecomments{ 
(1) Galaxy name; 
(2) \ttco~ integrated flux or its upper limit; 
(3) Integrated (global) \twco-to-\ttco~ line ratio or its lower limit; (4) The mean $1\sigma$ noise of resolved \itt~ using the dilated \twco \ mask. 
(5) Number of hexagon grids with resolved \itt~ detected with {S/N$>4$};
{(6) Number of hexagon grids with resolved \itw~ detected with S/N$>3$;
(7) Number of hexagon grids with resolved \itw~ that are \grtt~ times brighter than the \itt~ detection threshold;}
(8) Median of resolved line ratio \rtt; 
(9) Number of bins with azimuthally stacked $\mathcal{N}(I_{13}^{\rm stack})$ detected with with {S/N$>4$}; 
(10) Median of azimuthally stacked line ratio \rstack; 
(11) Far-IR flux ratio at 60 and 100 from NED \citep{https://doi.org/10.26132/ned1}; 
(12) The interacting galaxies classified in merging or post-merger stage by \citet{Barrera-Ballesteros2015} are assigned to ``1'', and the others are shown with ``0''
(13) Galaxies with bar present from HyperLEDA are indicated with number ``1'', and the others are ``0''. 
}
\end{deluxetable*}

\section{\twco$(1-0)$/\ttco$(1-0)$ variations}
\label{sec:eresults}

\subsection{Fairly Constant \rtt~ Across the Sample}

\begin{figure*}[htb!]
\includegraphics[width=0.32\textwidth]{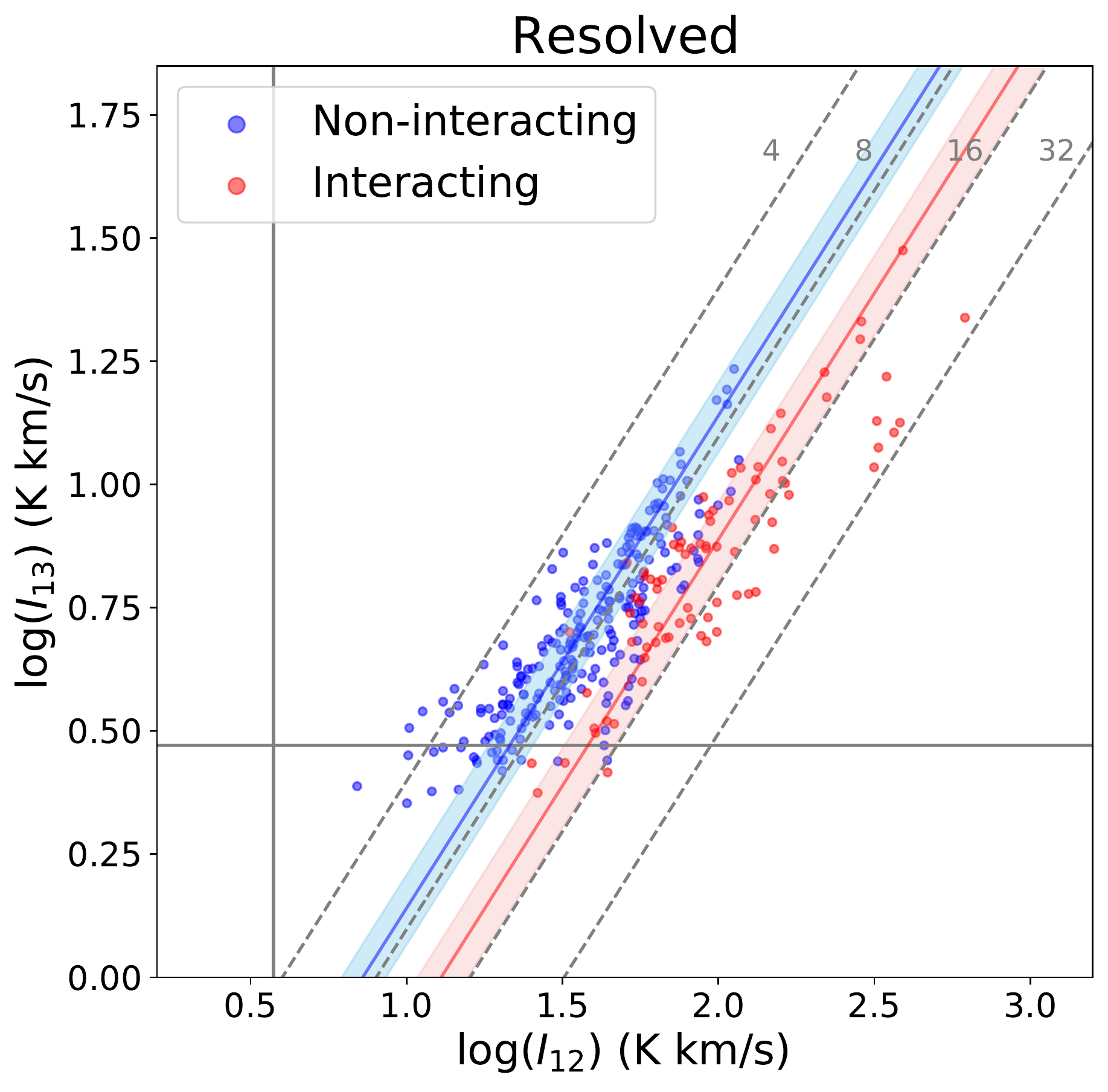} 
\includegraphics[width=0.33\textwidth]{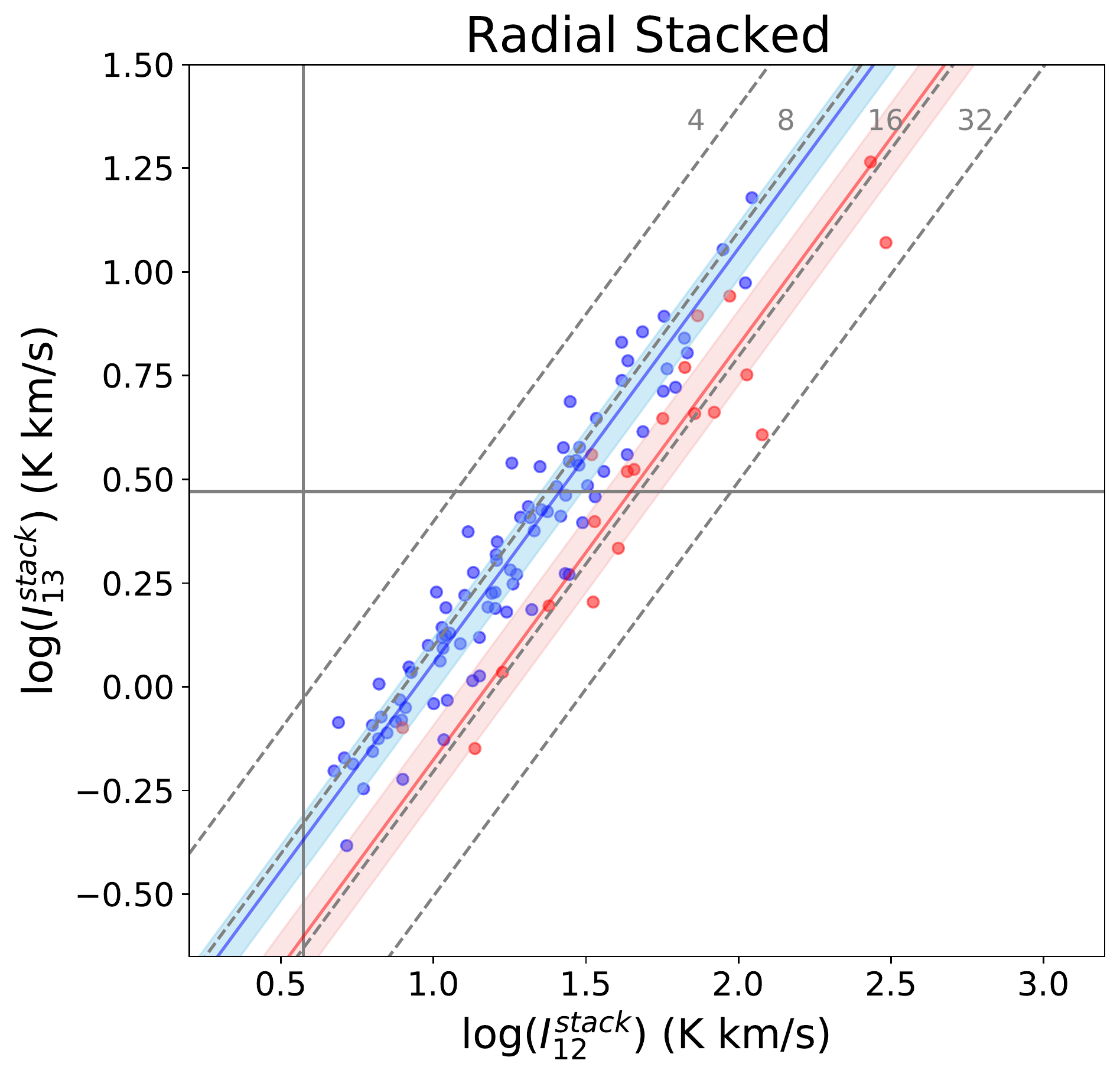} \includegraphics[width=0.32\textwidth]{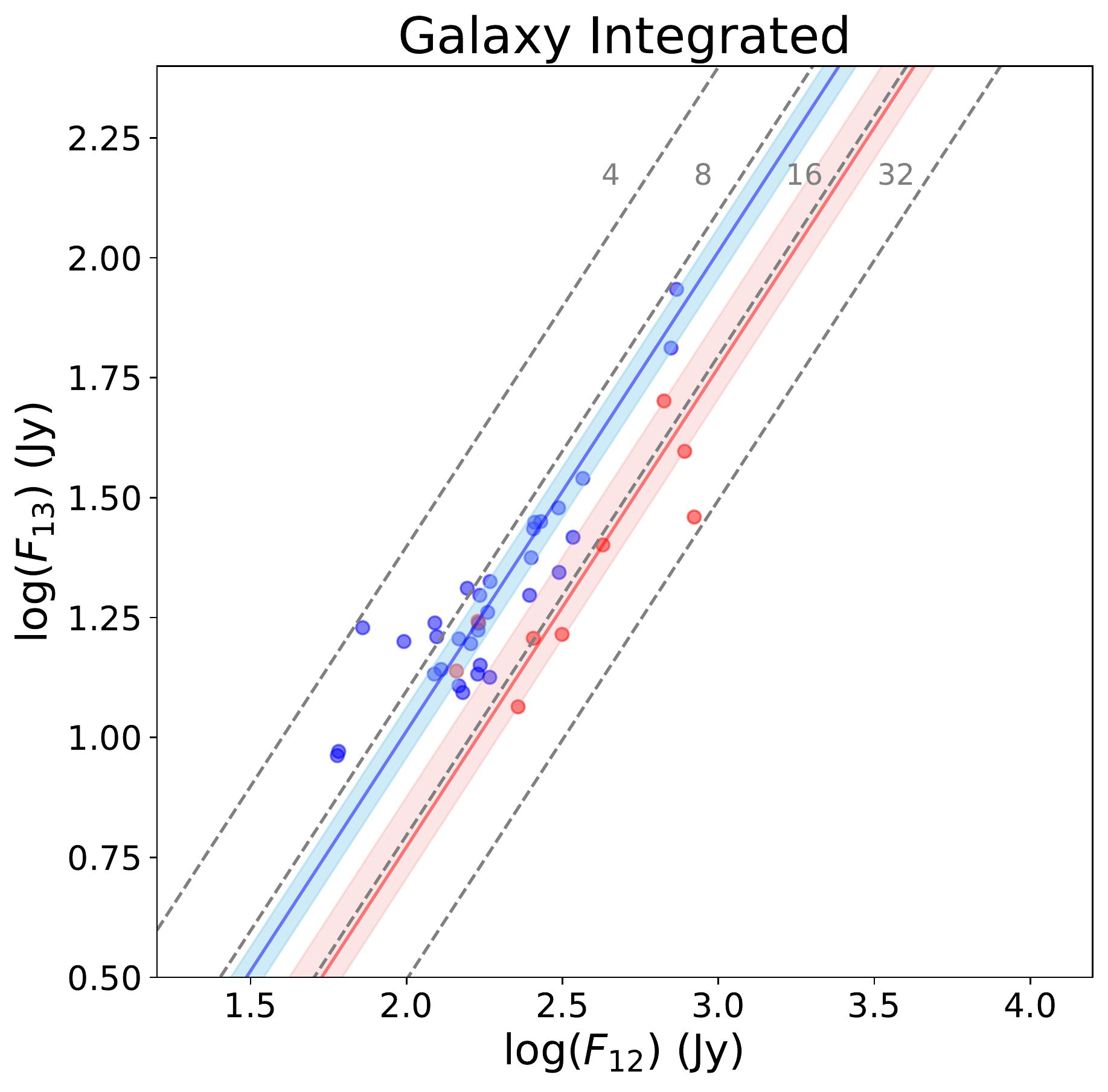} 
\caption{
\ttco \ emissions, measured on individual grid (left panel), radial annulus,  
(middle) and the entire galaxy (right panel),  
as functions of their \twco \ counterparts. 
The blue dots show the \ttco  \ detections with {S/N $>4$} in the non-interacting galaxies, while the red ones are those in the interacting galaxies.
The blue and red solid lines show the median of the \rtt \ of the non-interacting and interacting galaxies, with shaded regions show the first and third quartiles of the corresponding measurements.  
Constant \rtt \ values of [4, 8, 16, 32] are shown as gray dotted  
lines in each panel. 
The typical detection limits of resolved \itt \ and \itw \  are shown 
as horizontal and vertical  solid lines respectively, in the left and middle panels. Note in the middle panel, radial stacked $I_{13}^{\rm stack}$ approaches ranges well below this limit.   
All three types of measurements are closely distributed around  constant values of \rtt. 
}
\label{fig:ei1312all}
\end{figure*}

In Figure \ref{fig:ei1312all}, we present the integrated global $F_{13}$ flux, radially stacked $I_{13}^{\rm stack}$ and resolved 
\itt\ as functions of their \twco \ counterparts. 
$F_{13}$,  $I_{13}^{\rm stack}$,  and \itt\ 
 follow \twco~ emission quite well.  
In the low \itw \ regime, resolved \itt \ appears {0.2-0.7 dex} 
higher than the {stacked} measurements because of their limited sensitivity shown as the horizontal gray line in the left panels.  In the middle panel, we repeat this line to show the \itt \ sensitivity of the individual grid, and find that the $I_{13}^{\rm stack}$ measurements {extend} well below the line. 
By stacking the spectra over multiple grids, $I_{13}^{\rm stack}$ reveals the presence of \ttco \  that is not detectable in individual grid samples.

From our integrated, stacked, and resolved \ttco, we measure three types of \twco-to-\ttco~ line ratios, all expressed in brightness temperature units:
integrated line ratio $\grtt = 0.912 \times F_{12}/F_{13}$, 
radial stacked ratio \rstack $\equiv I_{12}^{\rm stack}/I_{13}^{\rm stack}$, 
and resolved line ratio $\mathcal{R}_{12/13} = I_{12}/I_{13}$. 
For the integrated line ratio, 
the factor $0.912$ accounts for the 
different rest frequencies of \ttco \ and \twco. 
\grtt\ indicates the overall line ratio of a galaxy, comparable with the unresolved \rtt\ obtained from single-dish measurements,  
assuming missing flux from the interferometric observation is negligible.
The values and $1\text{-}\sigma$ uncertainties of \grtt\ are tabulated in the fourth column 
of Table \ref{table:edgeg1}. {\twco~ missing flux of $\sim 10\text{-}50\%$ will propagate an additional  $\sim 0.5\text{-}10\%$ uncertainty to the line ratio we obtain. 
On the other hand, we do not expect \ttco \ flux is lost because it is mainly originated from clumpy structures. 
} 

\begin{deluxetable*}{l r r r r c r r r r c r r r r}[htb!]
\tablecaption{\label{tb:r13sta}
\rtt \ Measurement Statistics}
\tablehead{
\colhead{} &  
 \multicolumn{4}{c}{All galaxies } & &
  \multicolumn{4}{c}{Non-interacting galaxies } & &
 \multicolumn{4}{c}{Interacting galaxies }  \\
 \cline{2-5} \cline{7-10} \cline{12-15}\\
 &
\mc{$\mathcal{N}_{\rm gal}$ }&
\mc{$\mathcal{N}_{\rm data}$ }  &
\mc{Median} &
\mc{SIQR} &
& 
\mc{$\mathcal{N}_{\rm gal}$} &
\mc{$\mathcal{N}_{\rm data}$} &
\mc{Median} &
\mc{SIQR} &
& 
\mc{$\mathcal{N}_{\rm gal}$} &
\mc{$\mathcal{N}_{\rm data}$} &
\mc{Median} &
\mc{SIQR} 
}
\startdata
 \rtt & 30 &  $323$ & $ 7.97$ & $2.13$ & & $22$ &$241$ &$7.24$ & $1.20$ & 
&$8$ & $82$ & $12.92$ & $2.44$  \\
 \rstack & $41$& $109$ & $9.25$ & $2.00$ & & $32$& $89$& $8.78$& $1.38$& 
& $9$ & $20$& $15.00$& $3.18$  \\
 \grtt & $41$ & $41$ & $10.21$ & $1.74$ & & $32$ & $32$ & $9.70$& $1.19$& 
& $9$ & $9$ &$16.92$& $3.18$ \\ 
\enddata
\tablecomments{Number of galaxies, number of measurements, median, and semi-interquartile range (SIQR) of 
resolved \rtt, azimuthally averaged \rstack, and galaxy integrated \grtt \ in our sample. 
}
\end{deluxetable*}

We provide statistics of the line ratios in Table \ref{tb:r13sta}. 
Individual line {flux or intensity} measurements are plotted in Figure \ref{fig:ei1312all}, with the 
solid colored lines showing the median values for interacting and non-interacting galaxies and the shaded regions delineating the first and third quartiles. 
Remarkably, the line ratios are fairly constant among different galaxies in our sample. 
For the resolved \rtt \ with {323} measurements from {30} galaxies, 
the semi-interquartile range (SIQR) is only {2.13}. 

For all three types of measurements, the line ratios in the interacting galaxies are typically higher and exhibit larger scatter than those in the non-interacting galaxies. 
This implies the gas conditions could be systematically different due to the interaction process on large scales, therefore in the following investigations of \rtt, we always split our sample into interacting and non-interacting galaxies. 
We discuss possible mechanisms for this difference in Section \ref{sec:dint}. 

The median line ratios increase slightly when going from resolved \rtt \ on kpc scales to global \grtt \  on {the entire galaxies}. 
In Figure \ref{fig:ei1312all}, resolved \rtt \ show largest scatter around a constant value indicated by the diagonal dashed lines. 
Stacked values $\mathcal{R}_{12/13}^{\rm stack}$ shift to higher values, 
while the integrated \grtt \ {exhibit} highest values.  
\textcolor{black}{
Because the \rstack\ measurements are mostly within $0.4 r/r_{25}$ (see Section \ref{sec:eradial} and Figure \ref{fig:rstack_all}),  
the difference between $\mathcal{R}_{12/13}^{\rm stack}$ and \rtt \ is likely due to the bias of the non-detections of \itt \ within $0.4 r/r_{25}$, 
while the difference between \rstack \  and \grtt \  implies the non-detections beyond $0.4 r/r_{25}$ on average should have higher line ratios.}

\subsection{Integrated \grtt\ and galaxy global properties}
\label{sec:eglobal}

\begin{figure*}[tb!]
\epsscale{1.05}
\plotone{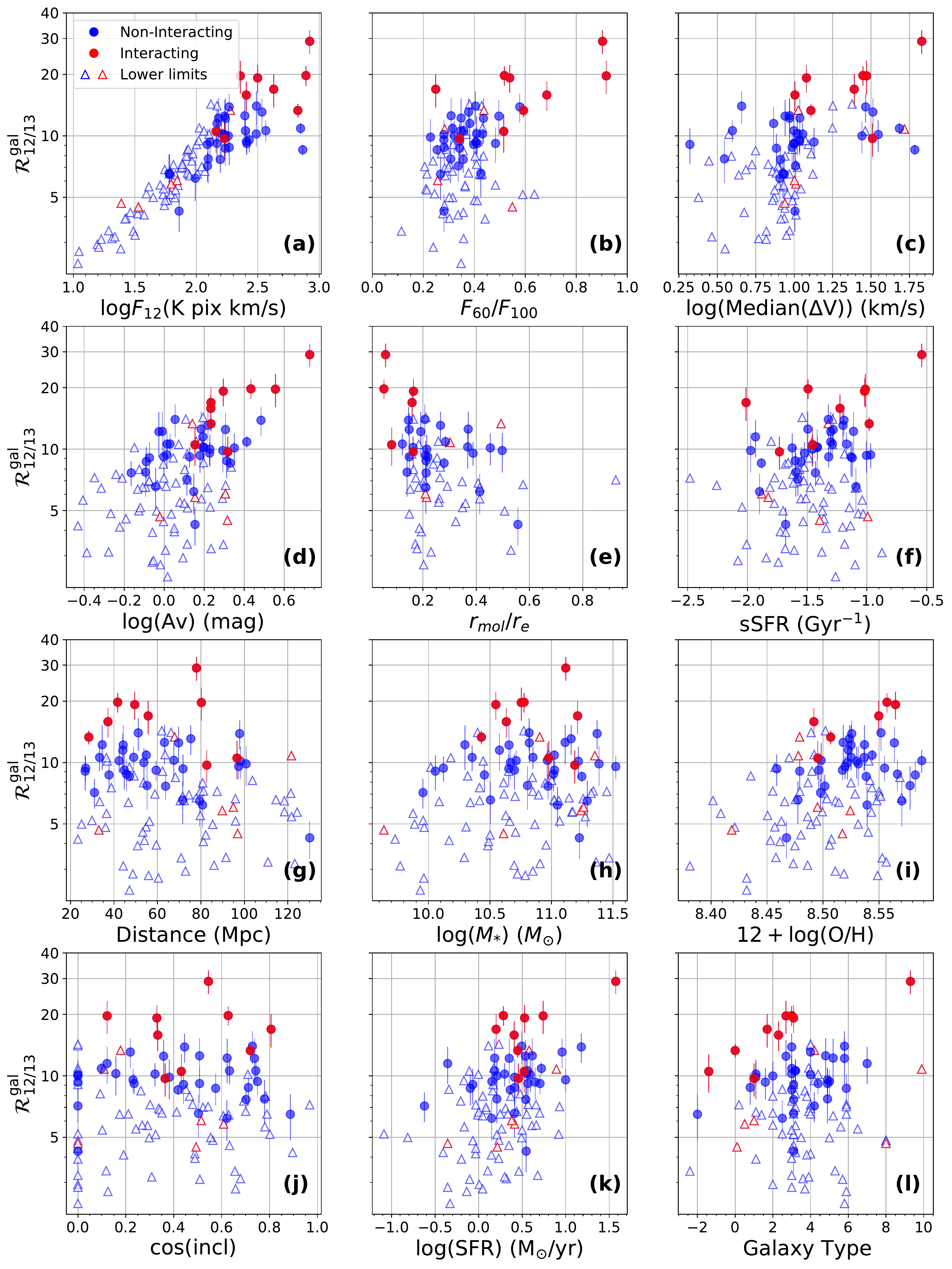}
\vspace{-0.3cm}
\caption{
Integrated line ratio  \grtt\  as functions of global galaxy parameters {for the 105 galaxies 
in which $F_{12}$ is detected with S/N $>3$}. 
Blue symbols show the isolated galaxies and the red symbols show the interacting galaxies. 
The circles with error bars show the \grtt~ and its uncertainty  for the {41 galaxies with $F_{13}$ detected with S/N $>$4}. 
{}.
The lower limits of \grtt~ of the other {64} 
galaxies with $F_{12}$ detected but $F_{13}$ below the detection limits are shown as triangles. 
The Spearman rank correlation test results are shown in Table \ref{tb:rsglb}. 
%
}
\label{fig:f13vsglb}
\end{figure*}

\begin{deluxetable*}{l c c r r c r r c r r}[htp!]
\tablecaption{\label{tb:rsglb} Spearman's rank correlation coefficients between 
global \grtt~ and host galaxy properties}
\tablehead{
\colhead{Global property} & 
Note & Ref & 
 \multicolumn{2}{c}{All galaxies } & &
 \multicolumn{2}{c}{Non-interacting galaxies }& & 
 \multicolumn{2}{c}{Interacting galaxies } \\
 \cline{4-5} \cline{7-8}  \cline{10-11} 
 & & &
\mc{$r_{\rm s} $} &
\mc{$P_{0}$ } &
& 
\mc{$r_{\rm s} $ } &
\mc{$P_{0}$ } & 
&
\mc{$r_{\rm s} $ } &
\mc{$P_{0}$ } 
}
\startdata
\input{f1213_corr_tb.tex}
\enddata
\tablenotetext{a}{Radius enclosing 50\% of the \twco~ flux ($r_{\rm mol}$) divided by equivalent radius ($r_{\rm e}$)}
\tablecomments{ References: 1. \citet{Bolatto2017}; 2. IRAS IR fluxes are from NED.  3. \citet{Sanchez2018}; 
4. Wen et al. in prep.;5. HyperLEDA.
}
\end{deluxetable*}

We carry out Spearman's rank correlation tests between \grtt~ and the other galaxy global properties. 
We first calculate Spearman's rank correlation coefficients $r_s$ and p-values $P_{0}$ for all the {41} galaxies with $F_{13}$ detected. 
{Since there are likely systematic differences of the line ratio due to interaction status, we then split the sample into non-interacting and interacting galaxies, and calculate $r_s$ and $P_0$ respectively.}
The results are listed in Table \ref{tb:rsglb}, ordered by absolute values of $r_s$ of all galaxies. 
Figure \ref{fig:f13vsglb} shows  \grtt~ as functions of the first 12 global galaxy parameters in Table \ref{tb:rsglb}. 
Besides the {41} galaxies with $F_{13}$ detected, for which we perform the correlation tests, we also show the lower limits on \grtt~ with triangle symbols for the other {62} galaxies in which $F_{12}$ is detected with {S/N $>4$} but $F_{13}$ is not detected; the lower limits are calculated as
{$0.912 F_{12}/(4\sigma(F_{13}))$}. 
We highlight the interacting galaxies in red in the figure; the isolated galaxies are shown in blue.

In Table \ref{tb:rsglb}, the absolute values $r_s$ of non-interacting  galaxies are generally smaller than that of all galaxies, and the $P_0$ values of non-interacting galaxies are also often larger than that of all galaxies: excluding the interacting galaxies, the correlations between \grtt~ and global properties are weaker and less significant.
We consider significant correlations with $P_{0} < 0.05$, i.e. the significant levels of such correlations are above $2 \sigma$.  
\textcolor{black}{
Given the sample size and uncertainty in our measurements, correlations with $|r_s| < 0.4$ are mostly indistinguishable from random fluctuations in our data. Because the {sample size of} interacting galaxies are much smaller comparing to the non-interacting galaxies, their rank correlation coefficients are less reliable than the ``all galaxies'' and ``non-interacting galaxies'' samples.  
We proceed only discussing global correlations with $|r_s| > 0.4$ and $P_{0} < 0.05$ for the samples of ``all galaxies'' and ``non-interacting galaxies'', with the caution that the interacting galaxies may play important roles in some correlations found in the ``all galaxies'' sample. }

For all the {41 galaxies with $F_{13}$ detected, there is no strong correlation ($|r_s| \geq 0.7$)} between \grtt~ and any of the the global properties we investigate in this study.  
\grtt~ moderately correlates with $F_{12}$, infrared color $F_{60}/F_{100}$, median velocity dispersion, dust attenuation, and molecular gas concentration. 
In the non-interacting galaxies, 
\grtt~ only moderately correlates with {$F_{12}$ and $F_{60}/F_{100}$}. 
As shown in Figure \ref{fig:f13vsglb}, interacting galaxies tend to have higher velocity dispersion, dust attenuation and molecular gas concentration, which might account for the significantly higher \grtt~ than the non-interacting galaxies, and lead \grtt~ to correlate with these three parameters when they are included in the correlation tests. 
 
The {increasing \rtt \ with higher \twco~ flux} is likely due to the requirement of high \ftw\ in order to detect \ftt, which limits the \ftt-detected galaxies to low values of \grtt~ when \ftw\ is low.  Galaxies not detected in \ftt\ (shown as triangles) could potentially have higher values of \grtt~. \textcolor{black}{The detection bias will also affect quantities that strongly correlate with \ftt, such as $\rm A_{v}$ and possibly $\Delta \rm V$ (i.e. cases where the non-detections are on one side of the graph)}.

As shown in panel (b) of Figure \ref{fig:f13vsglb}, \grtt~ increases  
with $F_{60}/F_{100}$; higher \grtt~ are  associated with higher dust temperature implied by larger  $F_{60}/F_{100}$ values.
This correlation is primarily driven by the starbursting galaxies which are also interacting. Such correlation still appears significant ($P_0<0.01$) when the interacting galaxies are excluded.
However, the large number of galaxies with $F_{13}$ not detected  when $F_{60}/F_{100}$ is lower might also make the correlation ambiguous without the interacting galaxies.  

\subsection{Radial variations in \rtt}
\label{sec:eradial}

\begin{figure*}[tbh!]
\includegraphics[width=0.33\textwidth]{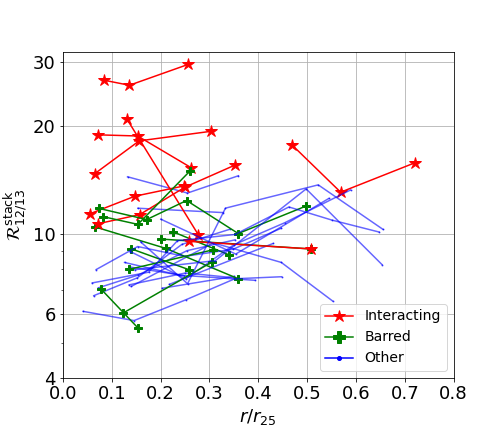} 
\includegraphics[width=0.33\textwidth]{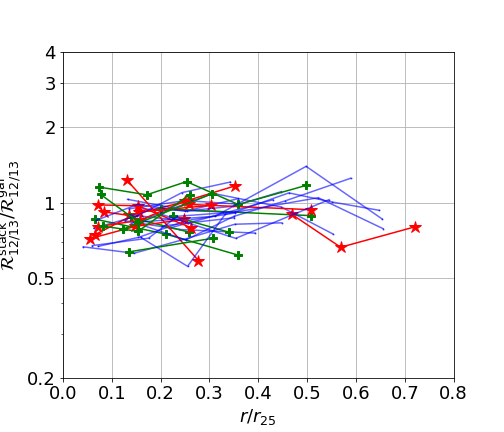} \includegraphics[width=0.33\textwidth]{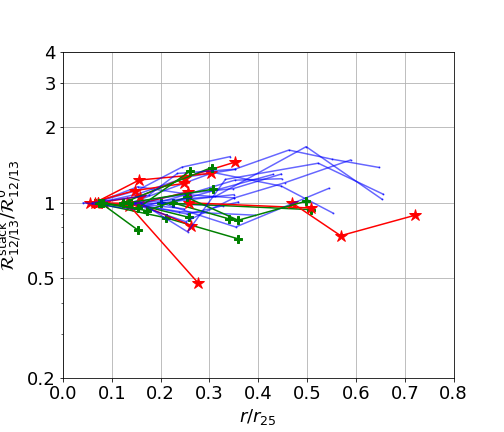}
\caption{
Summary of the {34} line ratio \rstack \ radial profiles in our sample.
{
Absolute values of \rstack~ are shown in the left panel, 
the middle panel shows the \rstack~  normalized 
by the each galaxy's global line ratio \grtt, and the right panel shows  \rstack~ devided by \rstack \ at the central radial bin ($\mathcal{R}_{12/13}^{\mathrm{0}}$). }
Interacting galaxies are shown in red star symbols, and barred, non-interacting galaxies are shown in green crosses. 
The other unbarred, non-interacting galaxies are shown in blue lines.
Within  an non-interacting, unbarred galaxy, 
\rstack \ are fairly constant as a function of radius, 
with possible tendency of slight increase beyond $0.2\text{-}0.3 r_{25}$;  
outliers from the general trend are exclusively found in barred or interacting galaxies. }
\label{fig:rstack_all}
\end{figure*}

Figure \ref{fig:rstack_all} summarizes the \rstack \
radial profiles from our sample. 
{
For all the 41 galaxies with \rstack \ detected, 7 galaxies are removed in this figure because within each of galaxies, only one bin is above the detection threshold}. 
The radial profile of each {of the 41 galaxies} are shown in Appendix \ref{sec:arradius}). 
In general, interacting galaxies show systematically higher \rstack \ at all radii than the non-interacting galaxies. 
{
For 2/3 of the non-interacting galaxies, we are able to 
detect the radial profiles of \rstack \ up to
0.4 $r_{25}$ (corresponding to a galactocentric radius of typically $\sim 6 \un kpc$).  
The remaining 1/3 are mostly barred galaxies, within which  \rstack \ are still only detected in the centers; \rstack \ in disks of these barred galaxies below the detection limit should be higher than the detected \rstack \ in the other non-interacting galaxies. These} suggest that dynamic disturbance like interaction and bar presence could leave imprints on the radial profiles of \rstack, which implies radial changes of gas conditions and/or chemical abundance.    

We find a wide variety of \rstack~ radial trends across our sample. 
In the galaxies without bars or interacting signatures (blue lines in Figure \ref{fig:rstack_all}), \rstack \ present flat or slightly increasing trends away from the galactic centers.  
This general trend we find is qualitatively consistent with our previous work of 11 galaxies from the CARMA STING survey \citep{Cao2017} and a recent study of \ttco \ of 9 nearby galaxies on kpc scales by \citet{Cormier2018}. 
Increasing \rtt \ with galactocentric radius are also 
reported in the Milky Way \citep{Roman-Duval2016} 
and in some nearby galaxies {\citep{Meier2004, M64, M101}}. 
Our results are in contrast to an earlier single dish study by \citet{Paglione2001} that found \rtt \ decreases away from the galaxy center for about half of their sample galaxies. 
We notice that higher \rstack \ in the central regions or decreasing \rstack \ radial trends are {often} found in barred or interacting galaxies, which may be due to large scale dynamical processes like gas flows caused by bar or interactions.

{In the middle and right panels of Figure \ref{fig:rstack_all}}, we divide \rstack \ by the each galaxy's global line ratio \grtt \ (listed in column (6) in Table \ref{table:edgeg1}) and by the \rstack \ measured in the first radial bin.  
In the {left} panel, \rstack/\grtt \ show less scatter ($\sim 0.2$ dex) comparing to \rstack \ shown in Figure \ref{fig:rstack_all} ($\sim 0.7$ dex)  in galaxy centers within $0.2 r_{25}$, while beyond $0.2 r_{25}$, the scatters of \rstack/\grtt \ are similar to those of \rstack. 
{This suggests that the central variations contribute the most to galaxy-to-galaxy variations of the integrated flux ratios we measured.}
On the other hand, the internal variations of \rstack\ highlighted in the {right} panel shows larger scatter with increasing radii; the contribution from the local variations within a galaxy become more important further away from the center.  
In galactic disks (with galactocentric distance  $> 0.2 r_{25}$), the variations of \rstack~ mixes both differences among galaxies and specific radial trends within a galaxy.

\subsection{\rstack~ and azimuthally averaged local properties} \label{sec:eresolve}
\label{sec:erstackcorr}
\begin{figure*}[htb!]
\epsscale{1.02}
\plotone{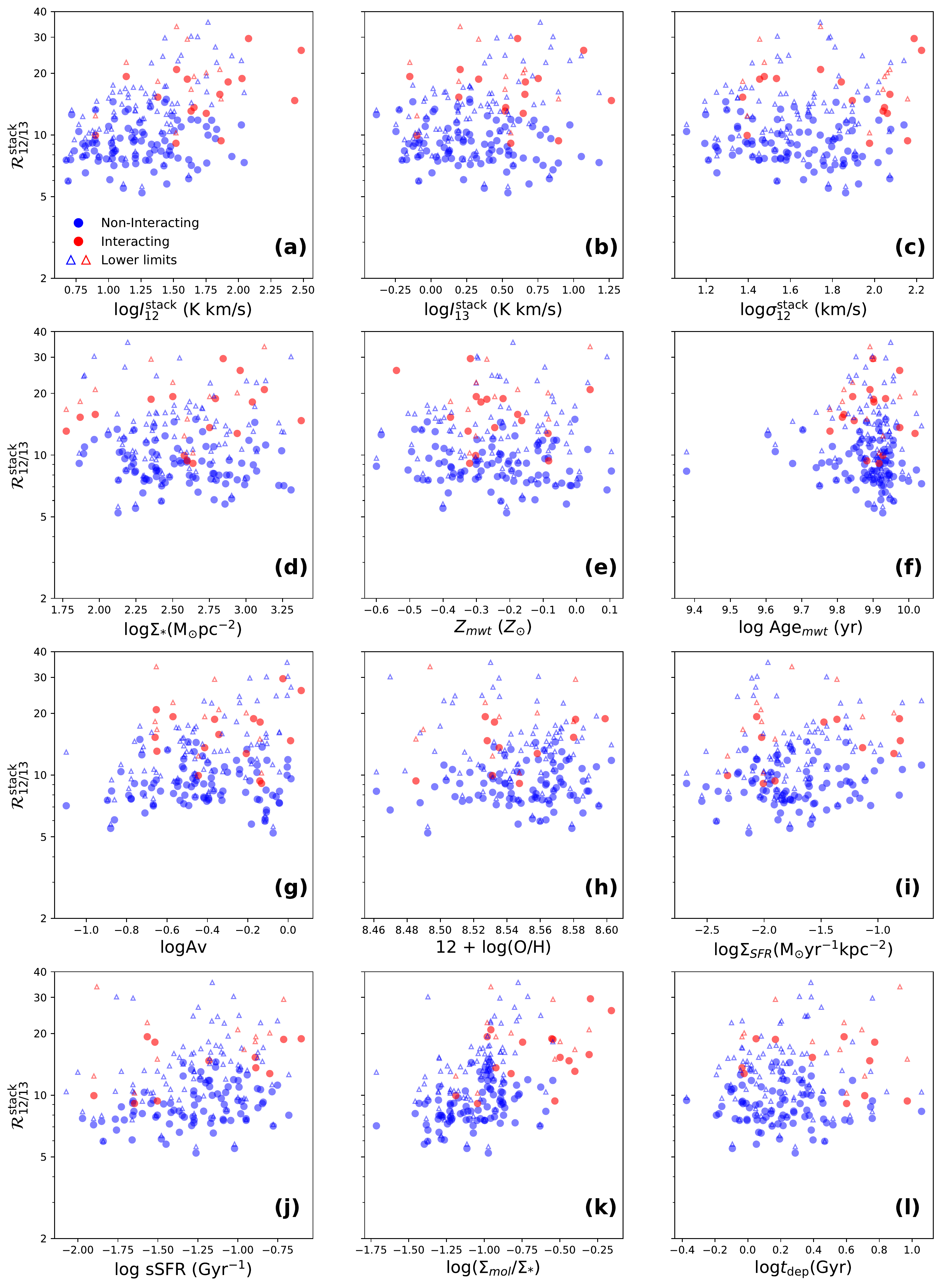}
\caption{
Radial-stacked line ratio \rstack\  as functions of azimuthally averaged properties {in the 41 galaxies with radial stacked \ttco \ detected with S/N $>$4} from our sample.
The explanations of x-axis labels and the Spearman rank correlation test results are shown in Table \ref{tb:rsstack}.
In panels (h), (i), (j), (l), we exclude the radius bin if 
it contains less than $50\%$ grids identified as star forming regions with the BPT diagnostics.  
The filled circles show the individual azimuthally averaged \rstack \ measurements for {annuli with stacked \ttco \ detected with S/N $>$4 in the} non-interacting (blue) and interacting (red) galaxies. 
{The lower limits of \rstack \ for annuli with stacked \ttco \ undetected are shown in triangles. }
 }
\label{fig:r13stk}
\end{figure*}

\begin{deluxetable*}{l c c r r c r r c r r}[htb!]
\tablecaption{\label{tb:rsstack} Spearman's rank correlation coefficients between \rstack \  and azimuthally averaged local parameters}
\tablehead{
\colhead{Parameter} & 
Note & Ref & 
 \multicolumn{2}{c}{All galaxies } & &
 \multicolumn{2}{c}{Non-interacting galaxies }& & 
 \multicolumn{2}{c}{Interacting galaxies } \\
 \cline{4-5} \cline{7-8}  \cline{10-11} 
 & & &
\mc{$r_{\rm s} $} &
\mc{$P_{0}$ } &
& 
\mc{$r_{\rm s} $ } &
\mc{$P_{0}$ } & 
&
\mc{$r_{\rm s} $ } &
\mc{$P_{0}$ } 
}
\startdata
\input{all_stk_rs.tex}

\enddata
\tablecomments{ All these local properties are averaged azimuthmally.
References: 1. \citet{Bolatto2017}; 2. This work;  
3. \citet{Pipe3D}. 
}
\end{deluxetable*}

In this section, we investigate how \rstack \ correlates with 
azimuthally averaged local 
properties (Section \ref{sec:stackprop}). 
Figure \ref{fig:r13stk} summarizes the bivariate distribution of 
\rstack \ and the local properties averaged azimuthally.
In our sample, there are {109}
measurements of \rstack \  from {41} galaxies available.
For the investigations of gas phase metallicity, star formation rate, specific star formation, and the depletion time, we exclude the annuli where less than half of the grids are identified as star forming regions. 
We show the non-interacting and interacting galaxies in different colors. 

We calculate the Spearman's rank correlation coefficients and list the results in Table \ref{tb:rsstack}. 
Combining all \rstack \ from all the {41} galaxies together,  \rstack \ moderately correlates with $I_{12}$, specific star formation rate, and the molecular-gas-to-stellar mass fraction ($\Sigma_{\mathrm{mol}}/\Sigma_{*}$) with 
$ 0.2 < |r_s| <  0.6 $ and  {$P_{0} < 0.003$ (corresponding to $3\text{-}\sigma$ significance)}. 
{The first and to some extent the third correlations 
tend} to be driven by the systematically higher \twco \ in the interacting galaxies (show in the red filled circles in Figure \ref{fig:r13stk}. 
Excluding these interacting galaxies,  
\rstack \ in non-interacting galaxies does not significantly correlate with  $I_{12}$. 
However, the positive correlations between \rstack \ and sSFR, as well as $\Sigma_{\rm mol}/\Sigma_{*}$, remain significant regardless whether or not we exclude the interacting galaxies. 
In interacting galaxies, {there are no significant correlations above $2\sigma$ significance.}

Except for these four parameters, $I_{12}$, sSFR, and $\Sigma_{\rm mol}/\Sigma_{*}$, 
we do not find any other significant trends of  \rstack\ with the local properties averaged on each annulus. 
A positive trend between \rtt \ and SFR has been reported before 
by \citet{Davis2014} using galaxy-integrated measurements, {while 
our \rstack \ measurements suggest no correlation with SFR.}
Neither the interacting nor non-interacting sample has a $P_0$ supporting a significant correlation with SFR.   

For each of the correlations we investigate, we perturb the \rstack \ values by its measurement uncertainties 1000 times to obtain the $r_s$ and $P_0$ values distributions. We characterise the uncertainties of $r_s$ and $P_0$ by their SIQR from these distributions. Our conclusions on the significance of each correlation remain the same when taking the $r_{s}$ and $P_0$ uncertainty into consideration. 

\textcolor{black}{
Note that with the spectral stacking technique, we are able to measure \rstack\ values from $I_{13}^{\rm stack}$ that are well below the limit that the individual \itt \ detection threshold imposes (the horizontal gray lines in Figure \ref{fig:ei1312all}). 
Such detection threshold on \itt~ could introduce a strong positive trend of \rtt~ with increasing \itw, which is due to the biased \itt \ detections towards lower \itw \ end (as the horizontal lines in Figure \ref{fig:ei1312all} intersect with lower values of \rtt\ shown in dotted lines; see Appendix \ref{sec:eresolve} for more details about this bias effect on resolved measurements). Our \rstack~ show {no correlation with \itw~} for non-interacting galaxies; the detection bias is eliminated at some extent by stacking multiple spectra together. 
However, we should be cautious that these \rstack~ detections are still biased: in the EDGE-CALIFA survey, there are {remaining 63 galaxies} that we only detect $I_{12}^{\rm stack}$ but not  $I_{13}^{\rm stack}$ (or \rstack).
}

\section{Discussion}
\label{sec:ediscuss}
\textcolor{black}{
Using kpc scale CO interferometric observations in 105 nearby galaxies from the EDGE-CALIFA survey, we measure the \twco \ to \ttco \ line ratio \rtt \ from kpc scales to {entire galaxies}, and investigate how they correlate with global galaxy properties (Section \ref{sec:eglobal}), galactocentric distance (Section \ref{sec:eradial}), and other resolved properties (Section \ref{sec:eresolve}). 
Here we briefly describe the mechanisms that cause \rtt \ variations first. We then summarize our results to identify 
the parameters that are related the observed \rtt \ on kpc scales and the possible mechanisms of these relations.} 

\subsection{Possible causes of \rtt \ variations}
Causes of variations in \rtt\ can be grouped into the following three broad categories.

\paragraph{Abundance variations}
Changes in the isotopic abundance ratio ([$^{12}$C/$^{13}$C]) can directly impact \rtt\ if variations in optical depth are of secondary importance.  In the Milky Way, a positive gradient with Galactocentric distance has been observed in the isotopic abundance ratio \citep{Milam2005}, which may contribute to a positive observed gradient in \rtt\ \citep{Roman-Duval2016}.  The radial [$^{12}$C/$^{13}$C] gradient may reflect stellar population differences since $^{12}$C production skews toward massive stars while $^{13}$C production skews toward intermediate-mass stars.  In general, however, [$^{12}$C/$^{13}$C] is difficult to measure because of the much higher opacity of $^{12}$C-bearing species.  Furthermore, the {\it isotopologue} abundance ratio $[\rm ^{12}CO/^{13}CO]$ may depart from the isotopic ratio in the presence of chemical fractionation (which favors \ttco\ production in cold regions, \citealt{Watson1976}) or selective photodissociation (which tends to destroy \ttco\ in unshielded regions, \citealt{Bally1982}).  The abundance ratio of optically thinner isotopologues, e.g.\ [$^{13}$CO/C$^{18}$O], can provide a more unambiguous diagnostic \citep[e.g.][]{Jimenez-Donaire2017ApJ}, although it also involves the [$^{16}$O/$^{18}$O] ratio.  Abnormally low $^{13}$CO/C$^{18}$O intensity ratios in ULIRGs, for instance \citep[e.g.][]{Sliwa2017, Brown&Wilson2019}, appear inconsistent with chemical fractionation or selective photodissociation, and support the hypothesis that the high \rtt\ in these galaxies is due to recent ISM enrichment by massive stars \citep[see also][for theoretical evidence]{Viti2020}.

\paragraph{Opacity changes in LTE}
A simple prediction based on the assumptions of LTE and fixed \twco\ and \ttco\ abundances is that 
\begin{equation}
\label{eq:rttv}
\rtt \propto \frac{1}{\tau(^{13}{\mathrm{CO}})} = \left[\frac{^{12}{\rm CO}}{^{13}{\rm CO}}\right]\frac{1}{\tau(^{12}{\mathrm{CO}})} \propto \frac{T_{k}^2 \Delta v}{N_{\rm H_2}}\;,
\end{equation}
for optically thick \twco\ and optically thin \ttco\ \citep[e.g.,][]{Paglione2001}. 
Thus, opacity variations that stem from changes in molecular gas column density, temperature, and/or line widths could lead to variations in \rtt.  Specifically, \rtt\ should increase with increasing temperature or line width and decrease with increasing column density.
Previous studies have suggested that line broadening due to the stellar feedback may contribute to the increasing trend of \rtt \ with SFR \citep{Crocker2012, Davis2014}.   
A recent study using a sample of 80 galaxies observed by 45 m Nobeyama telescope find that \rtt~ increases with sSFR \citep{COMING_X}.
Even without significant feedback, line broadening may result from the higher velocity dispersion needed to support the gas disk in denser regions of galaxies \citep{Aalto1995}.
However, no significant correlation between \rtt \ and line width or SFR surface density has been found in several kpc-scale studies within individual galaxies \citep{Meier2004, Cao2017,  Cormier2018}.
These discrepant results may arise from higher gas column density  $N_{\rm H_2}$ compensating for the higher $T_k$ and $\Delta v$ in regions of high SFR. 

\paragraph{Non-LTE Effects}
An LTE interpretation of \rtt\ implicitly assumes that the density is high enough to thermalize both lines, and that both lines originate from the same volume and thus experience the same degree of beam dilution.
However, if the volume density is low, \rtt\ will increase since \ttco\ is mostly sub-thermally excited while \twco, with a lower effective critical density, remains bright. 
The presence of such low-density gas can strongly influence the observed line ratios if it fills a larger volume than the denser \ttco-emitting gas.
The model based on non-LTE radiative transfer and a log-normal volume density distribution within a beam developed by \citet{Leroy2017} demonstrates that for fixed opacity, temperature, and abundance, the differential excitation roughly accounts for 0.2 dex variations of \rtt \ below a mean density of 10$^{3}$\rm \ cm$^{-3}$. 
Above 10$^{3}$\rm \ cm$^{-3}$, this effect is quite negligible. 

In the non-LTE regime, the positive correlation between temperature and \rtt\ may be substantially reduced or even reversed compared to the LTE case \citep[e.g.,][]{Hirota2010}.
Roughly speaking, this is because the higher temperature causes the peak emissivity of both lines to shift to lower densities, allowing \ttco\ to emit more efficiently and reducing \rtt.
{Assuming 
molecular emission lines emerging from a {log-normal distribution of densities} 
at a mean density of 400 \rm \ cm$^{-3}$ (i.e.\ 10$^{2.6}$ cm$^{-3}$) and an initial temperature of 15~K, the non-LTE radiative transfer modeling shows that an increase 
of the temperature by a factor of 2 leads to a decrease of \rtt\ by roughly the same factor  \citep{Puschnig2020}.}

\subsection{Elevated \rtt~ in interacting galaxies}

\label{sec:dint}
\textcolor{black}{
In our sample, we find the most striking factor that affects the \rtt \ values in a galaxy is whether or not it is experiencing a strong interaction. \rtt \ in interacting galaxies tends to be systematically higher than non-interacting galaxies from kpc scale to the {entire galaxy} (Figure \ref{fig:ei1312all}). 
The \rstack \ radial profiles of interacting galaxies show a similar enhancement (Figure \ref{fig:rstack_all}). 
Moreover, we can infer from Figures \ref{fig:r13stk} and \ref{fig:r13res} that these differences between interacting and non-interacting galaxies are unlikely due to their differences in other resolved properties except for \itw \ and $\rm A_v$, since both of interacting and non-interacting samples share the similar parameter spaces. 
This similarity suggests that independently from other 
local parameters,  
the gas conditions are systematically different due to the interaction process on large scales.}

Higher global \grtt~ values in merging galaxies, especially (U)LIRGs have been reported by single dish surveys reported 
than in normal spiral galaxies 
\citep[e.g.][]{Aalto1991, Casoli1991, Casoli1992}. 
The observed resolved \rtt \ of merging galaxies
on smaller scales from different individual studies also tend to be higher than the typical \rtt \ observed in normal galaxies \citep[e.g.][]{Young1984, Taniguchi1998, Henkel2014, Aalto2010}.
In our previous systematic study of 
resolved \rtt \ from the STING survey, the highest \rtt \ is found in the interacting galaxy NGC 5713 \citep{Cao2017}. 
\textcolor{black}{Our results from the CARMA EDGE survey confirm that this.} 

During the interaction process, the gas is driven toward the inner few kpc of the galaxy and triggers active star formation. Both the stars and interstellar medium are strongly perturbed during the process.  
For example, the molecular clouds could be disrupted to diffuse gas due to the dynamical disturbance.
The physical conditions in molecular gas during the merging process are expected to be different from the gas in normal, non-interacting galaxies.

We notice that the systematically higher \rtt \ is due to enhanced \itw \ emission rather than reduced \itt \ in the interacting galaxies (Figures \ref{fig:ei1312all} and \ref{fig:r13stk} panels (a) and (b)), confirming the same finding by \citet{Casoli1992} but down to the kpc scales. 
With the similar \itt, we would expect that the \ttco \ column density in the interacting galaxies is also similar to that in the non-interacting galaxies. 
Assuming $\rm ^{13}C$ or \ttco \ abundance in the interacting galaxies remain the same, 
the elevated \rtt \ and enhanced \twco \ is mainly due to the reduced opacity of \twco~ emission. Under the LTE assumption, higher temperature and/or broader line width would reduce the \twco~ opacity. 

Unfortunately, we do not have the direct temperature or the velocity dispersion measured for the molecular gas. We can only infer these effects from indirect parameters. 
On the galaxy scale, we show that higher \grtt~ are associated with highest IR colors in the interacting galaxies in panel (b) of Figure \ref{fig:f13vsglb}, suggesting that temperature could be part of the reason. Meanwhile, two of the interacting galaxies with very low IR colors (NGC5614 and UGC08107) still show higher \grtt, implying that the temperature cannot be the only mechanism that account for the higher \grtt~ in interacting galaxies. On the other hand,  in panel (c) of Figure \ref{fig:f13vsglb} \grtt~ we find that the typical (median) velocity dispersion in the interacting galaxies
tend to be higher than the interacting galaxies; elevated \rtt \ in
interacting galaxies might be explained by the broadened line width in them. 

If it is the higher temperature and/or broader line width that reduce the \twco \ opacity and thus elevate the \rtt \ in the interacting galaxies, we should also see similar effects on kpc scales. 
However, such effects are not prominent on kpc scales: except for Arp 220, the line widths of the azimuthally stacked \twco \  in interacting galaxies are not significantly higher than that in the non-interacting galaxies (panel (c) of Figure \ref{fig:r13stk}). 
In addition, both the line width and gas temperature are expected to increase with star formation, but we do find that SFR surface density in interacting galaxies is higher than the others (panel (i) of Figure \ref{fig:r13stk}). The apparent discrepancy on galaxy and kpc sales implies that the non-LTE effects and/or abundance anomaly are more likely the reasons for the systematic difference in \rtt~ in the interacting galaxies. 

One explanation to our observed higher \rtt \ in interacting galaxy on all scales is that the large-scale gas inflows induced by the interactions make the non-LTE effects or abundance changes prominent in these systems, especially in the centers.  
In non-interacting galaxies, we find that radial profiles of \rtt \ generally show  increasing trends (Figure \ref{fig:rstack_all}), which are likely caused by the less dense or less processed gas at larger radii.  
During the interacting process, \rtt~ could be elevated by the inflow of gas from outer regions with higher \rtt. 
The inflow might last for a period of time, funnelling the gas to centers and  enhancing the molecular gas concentration gradually. 
Indeed, we see a wide spread of gas concentrations in the interacting galaxies (Figure \ref{fig:f13vsglb} panel (e)), with three of them show the highest gas concentrations in our sample.  
As the interaction proceeds, the gas inflowed should not stay in low density but form denser gas or more molecular clouds, and thus triggers more star formation in the center. The increased fraction of denser gas in later times is plausible, as the strongest \ttco \ emission which likely traces denser gas are found in the advanced mergers like Arp 220 , NGC 2623, and NGC 5218. 
It is also in the galaxies that we find the enhanced SFR; the triggered starburst can then in turn increase \rtt \ via opacity effects. 
In fact, the two late stage mergers (Arp 220 and NGC 2623) also have the highest IR color and median velocity dispersion, which may further enhance their \rtt~ values.    

The gas inflow could also induce lower abundance ratio of $[\rm ^{12}C/^{13}C]$. The metal-poor gas from the outer regions dilute the metallicity in interacting galaxies \citep[e.g.][]{Kewley2000}. In particular, this dilution is observed on large scales \citep{Barrera-Ballesteros2015b}, as the star formation triggered in centers could enrich the ionized gas abundance there. 
However, in our case of $[\rm ^{12}C/^{13}C]$, both the inflow of less processed gas and recent massive star formation could drive it to higher values,  which may raise \rtt \ both locally and on large scale.  


Building a matched control sample to compare the interacting and non-interacting galaxies in details is beyond the scope of this paper. Nevertheless, our data suggests that the \rtt~ in interacting systems tends to be higher than non-interacting {galaxies}, implying the physical conditions in the molecular gas or its chemical abundance are changed by the interacting process. 
Due to the limited sample, we also do not study impact of detailed stages of interactions.  Since interacting processes are complex and there are many different types of interacting galaxies and mergers, future multi-line observations including \ttco \ for a representative sample of interacting galaxies would be helpful to unravel more details about interactions.

\subsection{A complex picture of \rtt~ on kpc scales}
\label{sec:ddriver}
\rtt \ measured on {kpc scales and across the entire galaxy} are fairly constant 
among our sample. Although bearing some detection bias, this is still remarkable considering our sample's diversity. 
However,  this does not mean that the gas conditions remain constant. The lack of strong variations in \rtt \ reveals a complex picture of observed \rtt \ in galaxies. The variations in \rtt \ cannot be attributed to a single factor;  
instead, several different mechanisms determines \rtt \ from kpc to {the entire galaxies}. 

\begin{figure*}[hbt!]
\epsscale{0.5}
\plotone{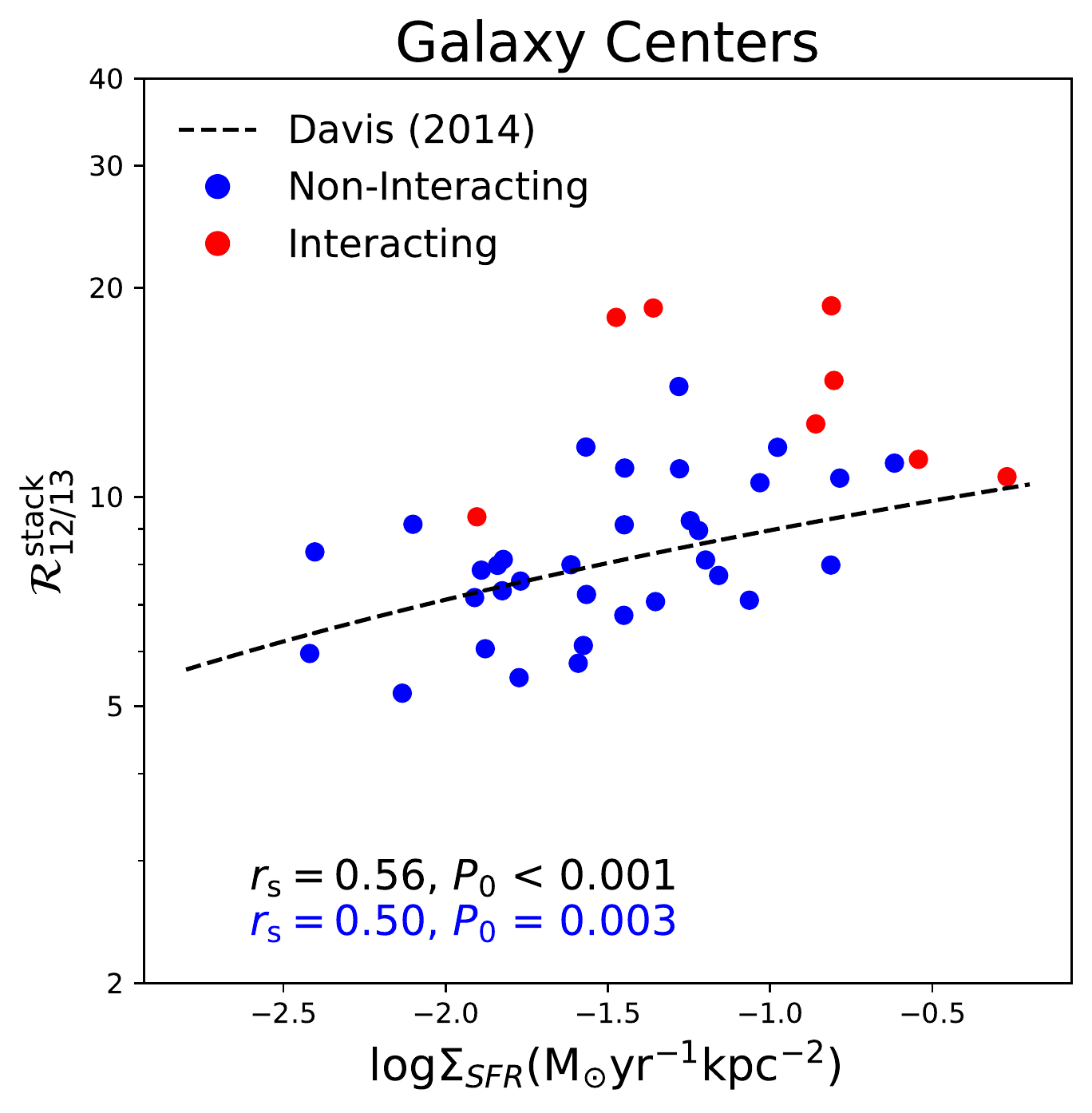}
\plotone{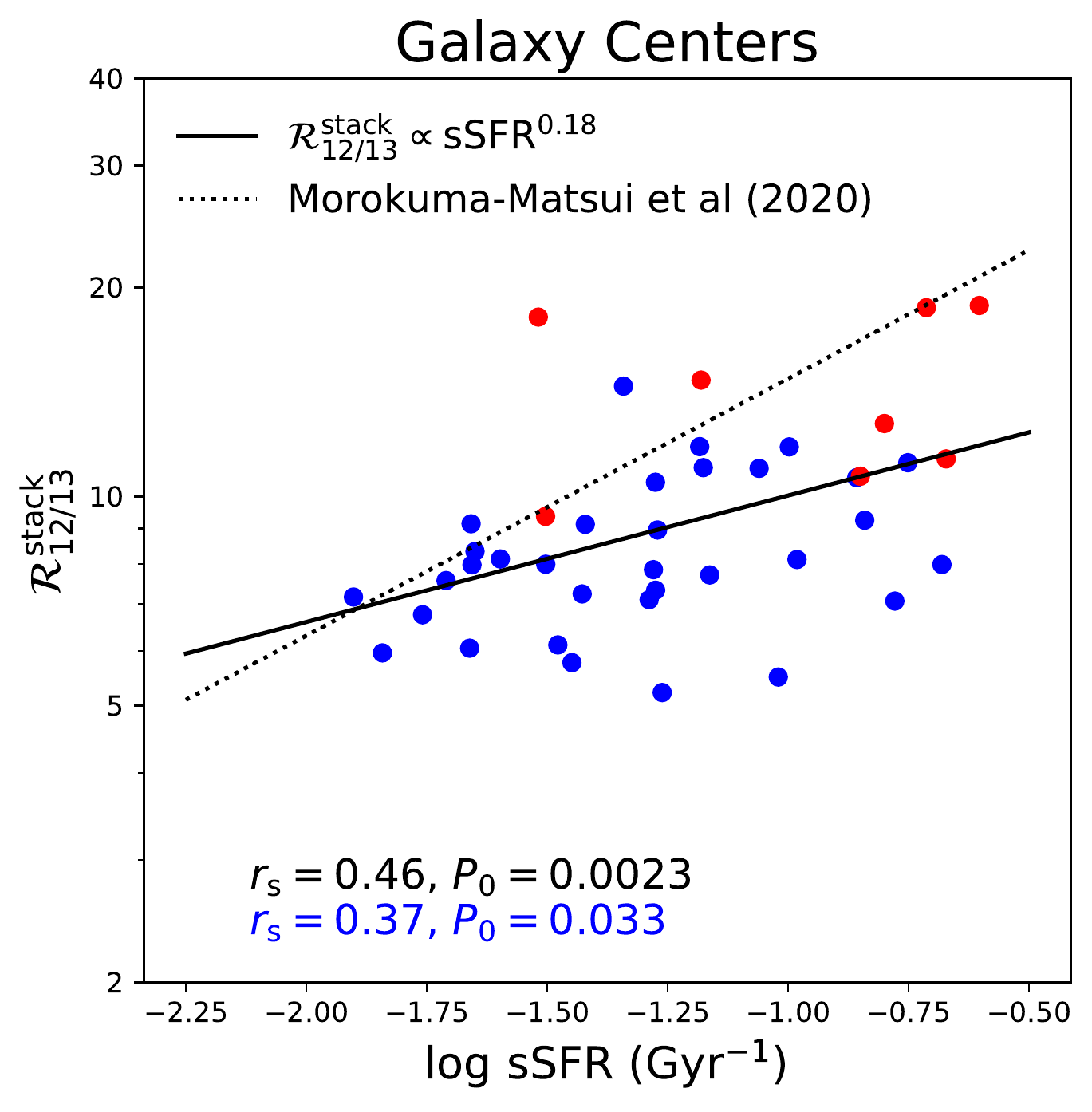}
\caption{Stacked line ratio \rstack~ vs. azimuthally averaged SFR (left) and sSFR (right) surface density in galaxy centers (within $0.2r_{25}$). The labels and colors are the same as in Figure \ref{fig:r13stk}. In the left panel, the dashed line shows empirical relation derived by \citet{Davis2014}. In the right panel, the solid line is the fitting based our sample, and the dotted line shows the fitting by \citet{COMING_X}. The Spearman rank coefficients are shown in black and blue color in the bottom for all the galaxies and for non-interacting galaxies respectively.}
\label{fig:rstack-sfr}
\end{figure*}

We discuss in Section \ref{sec:dint} that \rtt \ in interacting galaxies is likely elevated due to the merging process.  
For the remaining non-interacting galaxies, \rtt\ slightly increases with {the IR color $F_{60}/F_{100}$ (panel B in Figure \ref{fig:f13vsglb}). The higher dust temperature  is expected to be associated with active star formation}, so these results 
are consistent with previous galaxy surveys \citep{Davis2014, COMING_X}.  
However, we do not find that \grtt~ correlates with the characteristic SFR or sSFR measured at effective radius (panels (k) and (f) in Figure \ref{fig:f13vsglb}). 
We also do not find that \rtt \ correlates with SFR or sSFR surface densities measured on kpc scale annulus (Figure \ref{fig:r13stk}), which {generally}
agrees with the previous kpc scale studies \citep{Cao2017, Cormier2018}. 
The apparent discrepancy between our global and resolved results suggest that {the temperature} may cause the galaxy-to-galaxy variations in \rtt, but the variations within a galaxy due to other effects might wash it out when we combine all the resolved \rtt~ measurements.

The \rstack \ measured on annuli suggest that the \rstack \ variations in galaxy centers are likely {contribute most to the galaxy-to-galaxy differences (Figure \ref{fig:rstack_all})}. 
We perform the correlation tests for \rstack \ and other local properties following the same methods as in Section \ref{sec:eresolve}, restricting the measurements to be within $0.2 r_{25}$ instead of using all the detected values. 
We find that in the galaxy center, \rstack \ increase with the azimuthally averaged SFR and sSFR surface density (Figure \ref{fig:rstack-sfr}). 
For the SFR, the general trend roughly follows the empirical fitting from \citet{Davis2014}. 
The correlation between \rstack \ and sSFR in our sample is weaker and less significant than \citet{COMING_X}, shown as the dotted line in the right panel of Figure \ref{fig:rstack-sfr}. Note our \rstack \ over the central two annuli are typically lower than their \rtt~ stacked over the entire galaxy by $10\%$. 
These trends found in galaxy centers is generally consistent with the integrated  correlations in both our and other studies \citep[e.g.][]{YoungAndSanders1986, SageIsbell1991, Aalto1991, Aalto1995, Crocker2012, Davis2014, Vila-Vilaro2015, GoalsIRAM2019,  COMING_X}, where \grtt~, IR color, SFR, and sSFR measured integratedly all have large contributions from the galaxy centers. 
On the other hand, we do not find \rstack~ significantly correlate with SFR or sSFR in other radius bins beyond $0.2 r_{25}$. 
Therefore, it seems that that \rtt~ increases with stellar feedback, either through the reduced opacity due to higher temperature and/or velocity dispersion, but these effects are likely only prominent in galaxy centers.   
\textcolor{black}{Higher [\twco/\ttco] abundance  might also play a role there, as multi-line modeling suggests higher [\twco/\ttco] associated with higher star formation activities \citep[e.g.][]{Topal2020}. 
However,}
in galaxy disks, the variations of \rtt~ seem more complex that non-LTE effects must be taken into account.

Within a non-interacting galaxy, \rstack~ in general slightly increases with radius (Figure \ref{fig:rstack_all}). For massive main-sequence galaxies, the sSFR also generally increases with radius. 
 Indeed, we find that for the non-interacting galaxies without bars, within a galaxy, 
 both sSFR and \rstack \ show increasing trend, implying a positive correlation between 
 them.  
 This {correlation is} more likely due to the non-LTE effects, otherwise we would also see a correlation between \rstack \ and SFR or sSFR across different galaxies. As sSFR increases with radius as the volume density decreases, a substantial fraction of \ttco \ might be subthermally excited and thus \rtt~ increases. Indeed, our detections of resolved \ttco \ drastically drops with increasing galactocentric distances (Figure \ref{fig:maps}). 
Meanwhile, the increasing sSFR reflects the inside-out growth of a galaxy, from which we also expect positive radial $\rm [^{12}C/^{13}C]$ abundance gradient. Such abundance gradient could also explain the slightly increasing trend of \rtt \ with sSFR and galactocentric radii within a galaxy.

We also find \rstack~ increases with increasing molecular gas fraction (Figure \ref{fig:r13stk} panel(k)). 
Since we calculate the molecular gas using fixed $X$ factor based on \twco \ emission, this means what we find is essentially a positive correlation between \ttco \ emission and underlying stellar surface density. 
The increased \ttco~ with increasing stellar surface density seems to reflect the higher averaged density of gas due to the gravity fields.  
However, the lack of (anti-)correlation between \rstack~ and stellar surface density (Figure \ref{fig:r13stk} panel (d)) emphasizes that the gas density cannot be a dominating driver of \rtt~ variations. 
For example, stellar feedback from new stars forming with denser gas can reduce the \twco~ opacity and increase \rtt. 

In summary, our study reals a complex picture of observed \rtt~ on kpc scale  in galaxies. 
Variations in \rtt~ on kpc scales are not very sensitive to any particular local environments or the global galaxy properties in general; there seems to be no single driver of the \rtt~ variations. 
In our sample, the weak correlations we find suggest stellar feedback could increase \rtt \ in galaxy centers.   
In galaxy disks, non-LTE effects and/or abundance variations become important in determining \rtt. 
Moreover, dynamical disturbance like interacting process or bar presence could result in \rtt \  that deviate from the other galaxies.

\subsection{Caveats of this study}

One caveat of the this study is that we do not have single dish observations for the calculations of 
the line ratios. 
There could be \twco~ emissions that are missed in our interferometric observations: extended emission might be filtered out 
by the lack of short spacings, and 
weak \itw \ on smaller scales are missed due to the relatively low brightness sensitivity of interferometers compared to single dishes.  
The effect of spatial filtering on the intensity and 
\rtt \ measurement of each galaxy is unclear, but are more likely to affect \twco \ than \ttco. 
Although we do not see indications that EDGE misses a large fraction of the \twco \ flux \citep{Bolatto2017}, 
observing the sample galaxies in single dish telescopes such as IRAM and GBT will improve the absolute line ratio values measured in this study. 

When searching for correlations between \rtt \ and local properties, to reduce the detection bias due to the limited sensitivity of \ttco \ on resolved kpc scales, we use the azimuthal stacked spectra and the averaged local properties. 
This inevitably removes signatures of azimuthal variations of \rtt. Higher sensitivity \ttco \  observations on kpc-scales over a representative sample are required 
to fill the gap between cloud-scale studies and the galaxy evolution. 

In this study, we focus on \ttco \ and its comparison to \twco. We use their line ratio \rtt \ to infer the underlying variations of gas conditions. However, the mechanisms that determines \rtt \ are complex, which makes it impossible to infer the gas conditions under the simplest LTE assumptions. For non-LTE radiative transfer modeling to provide improved constraints on gas properties, additional high-J CO lines are strongly desired. 

Last but not least, we still do not have good understanding of the $\rm ^{13} C$ or $\rm ^{13} CO$ abundance in other galaxies. There are only a few direct measurements $\rm ^{13} C$ \citep[e.g.][]{Henkel2014, Tang2019}. 
Even with the multi-line diagnostics, this will still be a problem that could introduce large uncertainties on gas conditions constraints. 
Future observations on $\rm ^{13} C$ abundance will be desired to better interpret the \ttco \ observations and improve the multi-line modeling in extragalactic studies.  

\section{Conclusions}
We present \ttco \ observations for the EDGE-CALIFA survey, 
which is a mapping survey of 126 nearby galaxies at a typical spatial resolution of 1.5 kpc. 
We measure the \twco-to-\ttco~ line ratio \rtt~ resolved on kpc scales, averaged in annuli, and integrated over the entire galaxies. 
Combining our \ttco \ and \twco \ observations 
with optical spectroscopy IFU data from the CALIFA survey, 
we perform a systematic study of \rtt \ in relation to with local environments and their host galaxy properties for a wide variety of galaxies. 
Since \rtt~ variations could reflect the changes of physical conditions or chemical abundance of molecular gas, 
such a systematic study on \rtt~ provides useful implications for understanding the physics of molecular ISM and star formation in the context of galaxy evolution. 
Our main conclusions are as follows: 
\begin{enumerate}
\item 
    We detect resolved \ttco~ in 
    {30} galaxies from the 126 galaxies in the EDGE-CALIFA survey. 
    By stacking the spectra in annuli and integrating over regions detected in \twco, we also measure azimuthally averaged and integrated \ttco. 
    The \ttco \ emission measured on different scales closely associates the the strength of the \twco~ emission (Figure \ref{fig:ei1312all}). \rtt~ in our sample distribute in a narrow range, with median of {$7.97$} and semi-interquartile range (SIQR) of {$2.13$} (Table \ref{tb:r13sta}).   
\item 
    We find that the global, the azimuthally averaged, and the resolved line ratios \rtt~ in interacting galaxies are  systematically higher than in the non-interacting galaxies. 
    Thus the interaction process seems to enhance the \twco \ emission from {entire galaxies} down to kpc scales. 
    Inflow of less dense and less processed gas with higher $\rm [^{12}C/^{13}C]$ from large radii during the interacting process is a possible mechanism to drive the higher \twco \ and \rtt~ on all scales. 
\item 
    We present annulus-averaged \rtt \ radial profiles for our sample, taking into account the \ttco \  non-detections by spectral stacking (Figure \ref{fig:estack}). We find that roughly half of galaxies show increased \rtt\  beyond {$0.2R_{25}$}, suggesting more optically thin gas or lower $^{13}$C abundance 
    in the disks relative to the centers in general. 
    Decreasing radial trends of \rstack~ are mostly seen in interacting galaxies or barred galaxies, which may reflect the less dense or processed gas inflows from large radii due to large scale dynamic processes.   
\item 
    We study \rtt~ in relation to the stellar population and other ISM components within galaxies and to the global parameters of the galaxies. We do not find find strong correlation between \rtt~ and any local or global properties we investigate (Figure \ref{fig:f13vsglb}). In particular for non-interacting galaxies, while increased global \grtt~ with increasing IR color hints at reduced opacity due to higher temperature, we do not find the azimuthally averaged \rstack~ to be significantly correlated with local SFR  (Figure \ref{fig:r13stk}).  
    The lack of universal and strong correlations on different scales reveals a complex picture of molecular structure and the importance of multi-scale processes of the star formation in galaxies.  
    Our results therefore highlight the need of additional high-J CO lines for better constraints of molecular gas properties and for better understanding of the baryon cycle of the galaxy ecosystem. 
\end{enumerate}

\begin{acknowledgments}
Y.C. and T.W.  acknowledge support from the NSF through grants AST-1616199. R.C.L. acknowledges partial support by a NSF Astronomy and Astrophysics Postdoctoral Fellowship under award AST-2102625. J.B-B acknowledges support from the grant IA- 101522 (DGAPA-PAPIIT, UNAM) and funding from the CONACYT grant CF19-39578. DC acknowledges support by the Deutsche Forschungsgemein-schaft, DFG project number SFB956A. V. V. acknowledges support from the scholarship ANID-FULBRIGHT BIO 2016 - 56160020 and funding from NRAO Student Observing Support (SOS) - SOSPA7-014. V. V., acknowledge partial support from NSF-AST2108140.

This study makes use of data from the EDGE (\url{http://www.astro.umd.edu/EDGE/}) and CALIFA (\url{http://califa.caha.es/}) surveys and numerical values from the HyperLeda database (\url{http://leda.univ-lyon1.fr}). Support for CARMA construction was derived from the Gordon and Betty Moore Foundation, the Kenneth T. and Eileen L. Norris Foundation, the James S. McDonnell Foundation, the Associates of the California Institute of Technology, the University of Chicago, the states of California, Illinois, and Maryland, and the NSF. CARMA development and operations were supported by the NSF under a cooperative agreement and by the CARMA partner universities. This research is based on observations collected at the Centro Astron\'{o}mico Hispano-Alem\'{a}n (CAHA) at Calar Alto, operated jointly by the Max-Planck Institut f\"{u}r Astronomie (MPIA) and the Instituto de Astrofisica de Andalucia (CSIC).
This research has made use of the NASA/IPAC Extragalactic Database (NED) which is operated by the California Institute of Technology, under contract with the National Aeronautics and Space Administration. 
\end{acknowledgments}

\facilities{CARMA, CAO:3.5m}
\software{Miriad \citep{miriad}, Pipe3D \citep{Pipe3D}, 
IDL, matplotlib, astropy \citep{astropy}}


\bibliography{edge13co}
\bibliographystyle{aasjournal}



\appendix

\section{Additional global properties}
\label{sec:snrpeak}
The main galaxy parameters in the EDGE-CALIFA survey are summarized in the Tables 1-3 in \citet{Bolatto2017}. 
In this appendix, we provide 
the additional global properties used in this study in Table \ref{table:edgeg2}. 

\startlongtable
\begin{deluxetable*}{l r r c c c r r c}
\tabletypesize{\scriptsize}
\tablecaption{\label{table:edgeg2} Galaxy Parameters in the EDGE-CALIFA Survey}
\tablehead{
\colhead{Galaxy} &
\colhead{$F_{12}$} &
\colhead{$\rm A_v$} &
\colhead{$r_{\rm mol}/r_{\rm e}$ } &
\colhead{Median($\Delta \rm V$)} & 
\colhead{sSFR} & 
\colhead{Incl} & 
\colhead{PA} & 
\colhead{$\theta$} 
\\
\colhead{} &
\colhead{(Jy km/s)} &
\colhead{(mag)} &
\colhead{} & 
\colhead{(km/s)} & 
\colhead{(Gyr$^{-1}$)} &
\colhead{($^{\circ}$)} &
\colhead{($^{\circ}$)} &
\colhead{(kpc)} 
\\
\colhead{(1)} &
\colhead{(2)} &
\colhead{(3)} &
\colhead{(4)} & 
\colhead{(5)} & 
\colhead{(6)} & 
\colhead{(7)} & 
\colhead{(8)} &
\colhead{(9)} 
}
\startdata
\input{edge_gts_a.tex}
\enddata
\tablecomments{ 
(1) Galaxy name; 
(2) \twco~ integrated flux or its upper limit; 
(3) Nebular extinction from Pipe3D; 
(4) Median velocity dispersion;
(5) Specific star formation from Pipe3D; 
(6) Morphology inclination from HyperLEDA;
(7) Morphology position angle from CALIFA; 
(8) Linear resolution of the CO mapping. 
}
\end{deluxetable*}

\section{The peak signal-to-noise map of \ttco}
\label{sec:snrpeak}
Figure \ref{fig:snrmaps} show the peak signal-to-noise (SNR) of \ttco \ in all of the channels without any masking. We obtain it by finding the highest peak \ttco \ temperature in the spectrum along each spatial pixel, and then dividing by the corresponding sensitivity. The regions we detect integrated \ttco \ (shown as black contours in Figure \ref{fig:snrmaps}) enclose the high peak SNR pixels, which support the significance our detection. 

\begin{figure*}[ht!]
\includegraphics[width=0.931\textwidth]{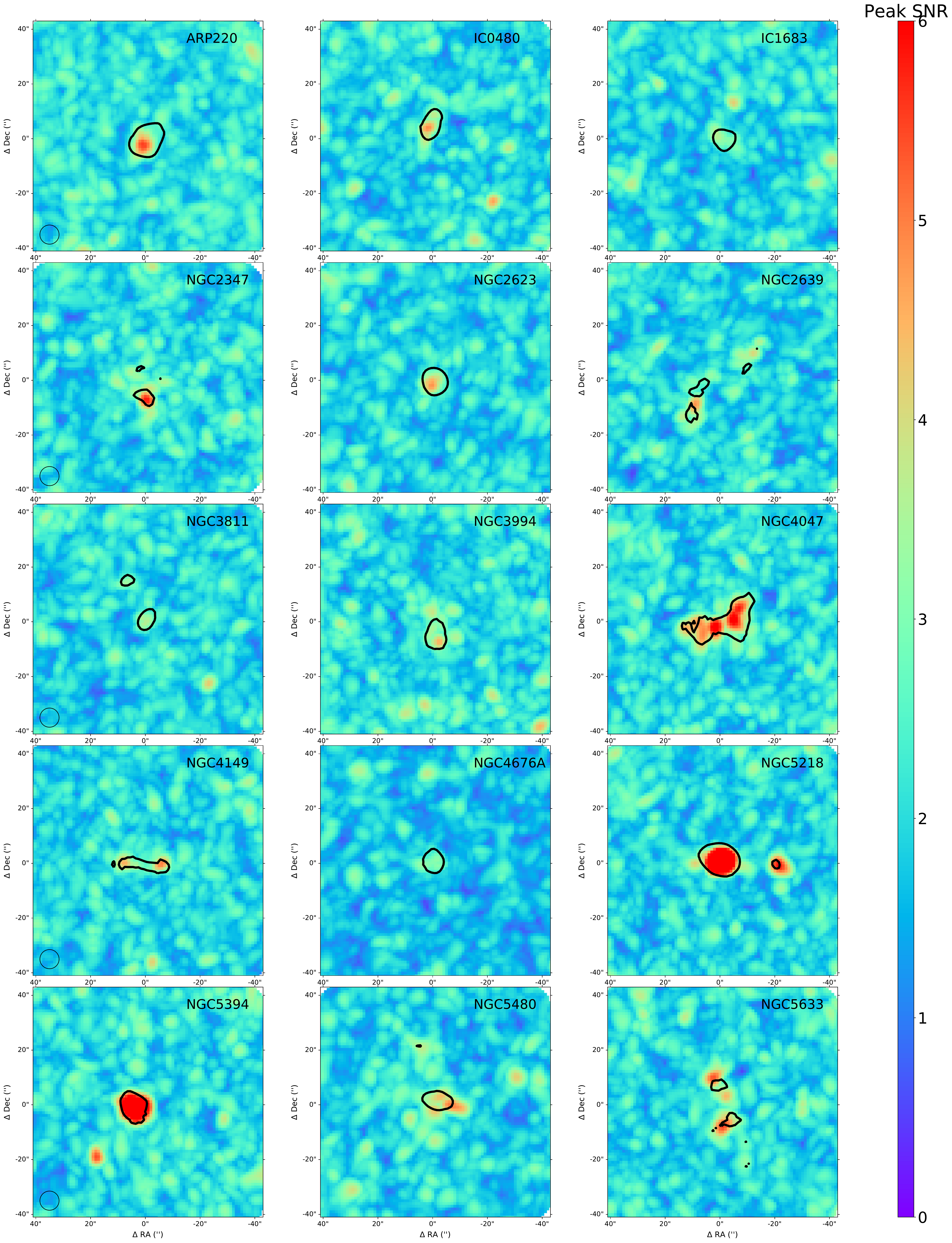} 
\centering
\caption{{The peak signal-to-noise map of \ttco \ for the {30} galaxies with resolved \ttco\ detected from the EDGE survey. 
The black contours overlaid show the \ttco\ intensity observed with 
{S/N $>$4}.  }
\label{fig:snrmaps}}
\end{figure*}

\begin{figure*} \ContinuedFloat
\epsscale{1}
\includegraphics[width=0.931\textwidth]{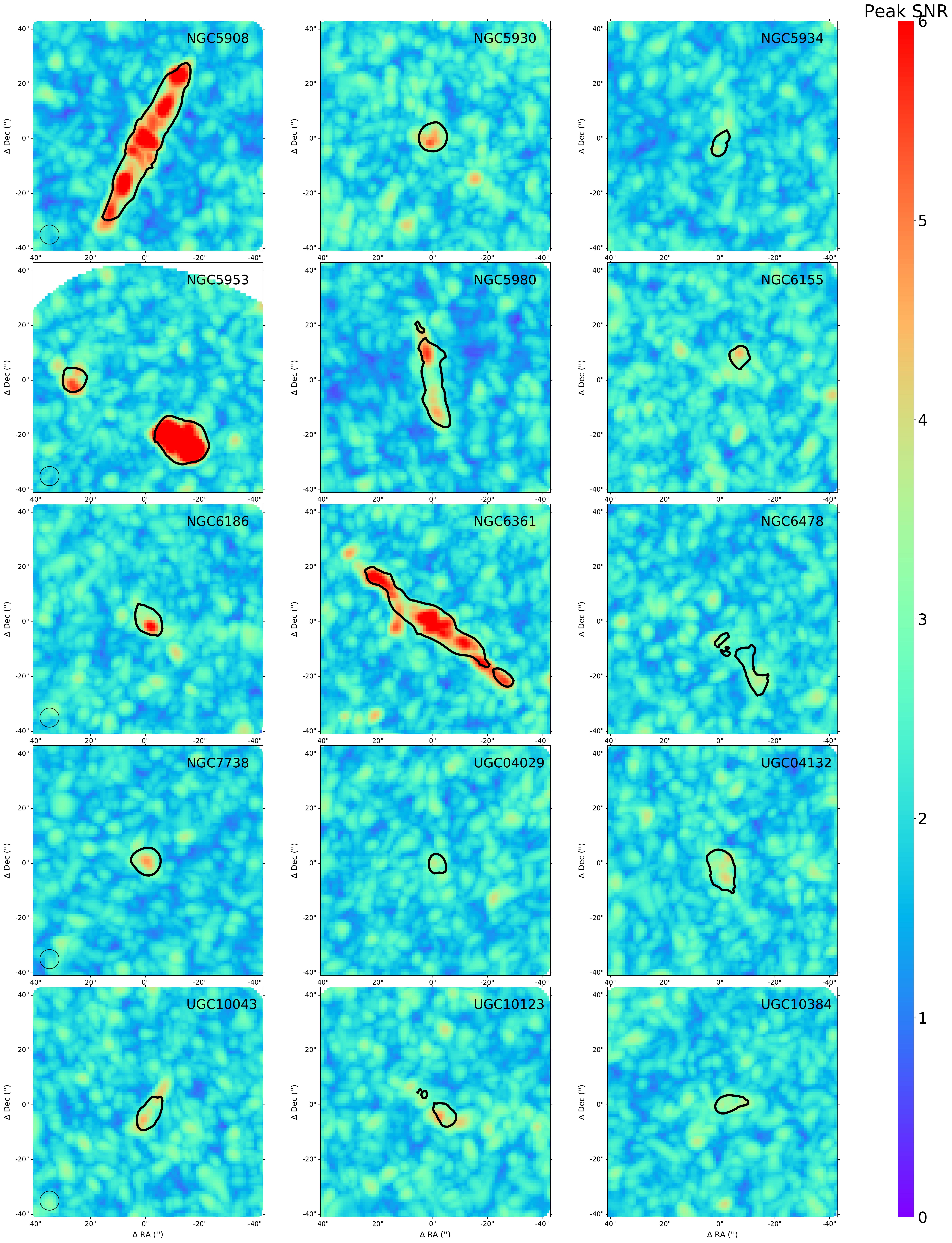} 
\centering
\caption{(Continued).
\label{fig:snrmaps2}
}
\end{figure*}

\section{Azimuthally stacked \rstack~ of each galaxy}
\label{sec:arradius}

\begin{figure*}[h!]
\includegraphics[width=1\textwidth]{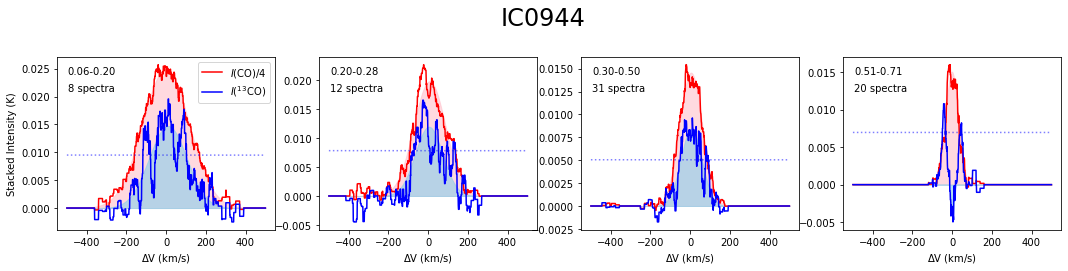}\\
\includegraphics[width=1\textwidth]{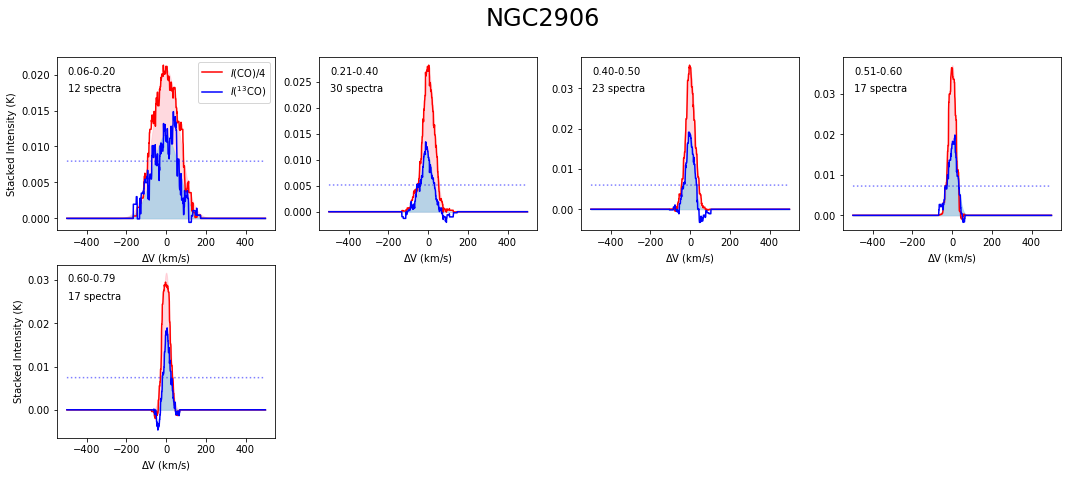}
\caption{
Azimuthally stacked spectra of IC 0944 (\textit{top}) and NGC 2906 (\textit{bottom}). 
{Each panel show the results of a radial bin adjusted following the method described in \ref{sec:specstack}.  
In the top left corner of each panel, we show the radial bin ranges 
in units of $r_{\rm gal}/r_{25}$ and the numbers of the spectra used in the stacking.}
The blue lines are the \ttco \ spectra and the red lines are the \twco \ spectra scaled {down} by a factor of 4. The blue vertical lines show the channel noise of the \ttco \ stacked spectrum derived following Equation \ref{eq:stack_error}. 
The red and blue shaded regions show the Gaussian fitted integrated fluxes of \twco\ and \ttco \  used for the stacked line ratios.
 }
\label{fig:estackspec}
\end{figure*}

Figure \ref{fig:estackspec} show two examples of the azimuthally stacked spectra from which we derive the \rstack.  
For IC 0944, we detect the integrated intensity of the stacked \ttco \ ($I_{13}^{\rm stack}$) in the 
{first three radii bins with S/N$>$4}. 
For NGC 2906, $I_{13}^{\rm stack}$ are detected {in the first four bins with S/N$>$4}. In these bins, the stacked \ttco \ spectra peak in the shifted central velocity, and are well fitted by Gaussian profiles (shown as the blue shaded regions).  

\begin{figure*}[h!]
\epsscale{1.165}
\plottwo{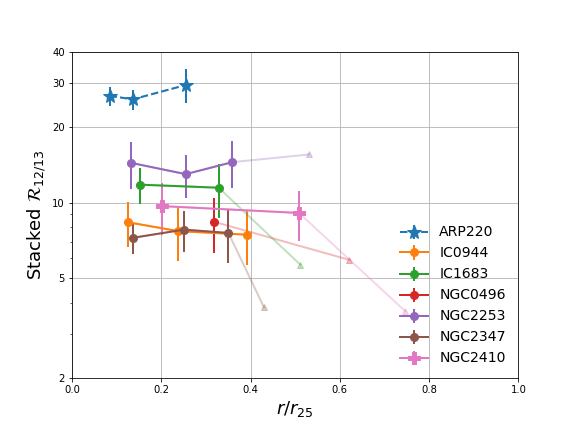}{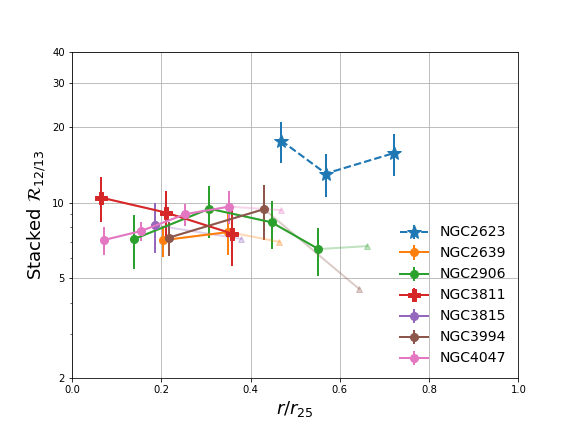}\\
\plottwo{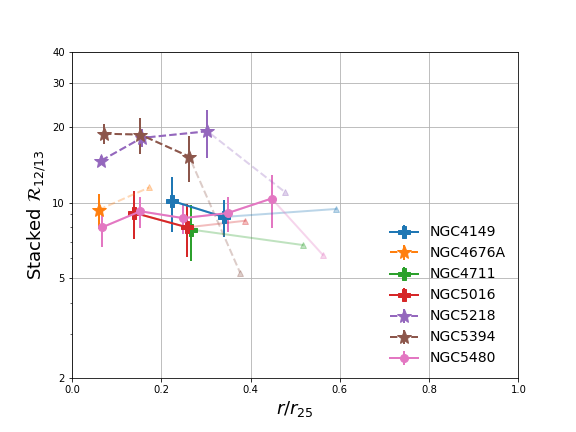}{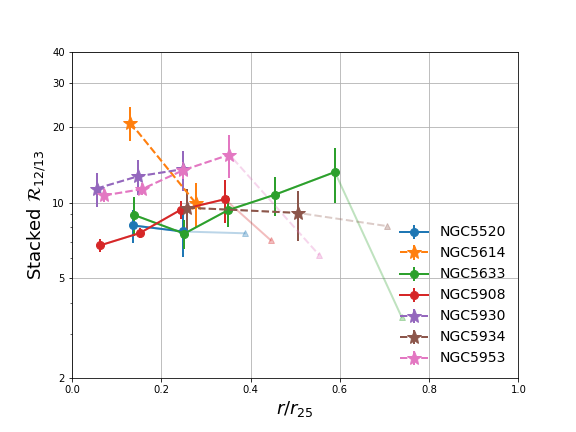}\\
\plottwo{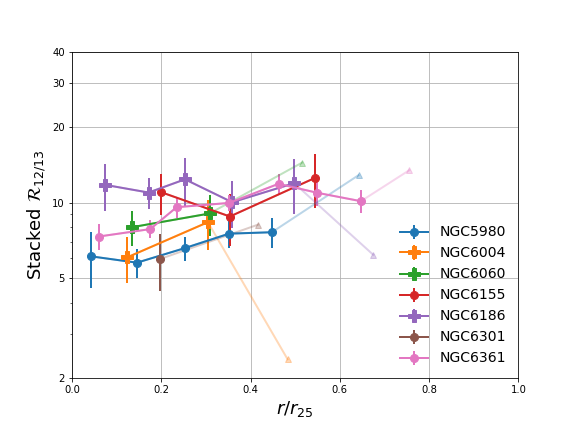}{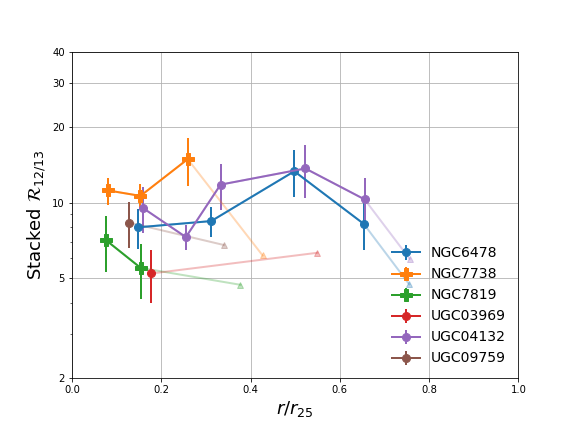}\\
\caption{
Radially averaged \rstack \  profiles of the {41} galaxies with $I_{13}^{\rm stack}$ detected {with S/N $>$4}
in our sample. 
The galactocentric radius for each galaxy
is normalized to its $r_{25}$. 
Spectra of \ttco \ and \twco \ are shifted and
stacked at each normalized radii bin to derive \rstack. 
Interacting galaxies are highlighted with star symbols. 
Cross symbols show the non-interacting galaxies with bars,  
and filled circles show the other non-interacting galaxies. 
{The lower limit of \rstack \ in the bins 
with $I_{13}^{\rm stack}$ below the detection threshold of 4 are shown in triangles}.
We are able to detect stacked \ttco \ out to a radius of
$\sim 0.4 r_{25}$ in most galaxies. }
\label{fig:estack}
\end{figure*}

Figure \ref{fig:estack} shows \rstack\ profiles of each individual galaxy of our sample.
We show the interacting galaxies with star symbols to distinguish them from the others. 
Across the sample, there are a wide variety of radial profiles. 
In each individual galaxy, the radial variations of \rstack \ are within $\sim 0.2$dex in most cases. 
\textcolor{black}{Any annulus with stacked \ttco \ below the detection limit of $3\sigma$ is omitted.} 
For the non-interacting galaxies, we show those with bars using cross symbols while the others are shown in filled circles.

\section{Correlations between kpc-scale \rtt~ and resolved properties}

\begin{figure*}
\epsscale{0.975}
\plotone{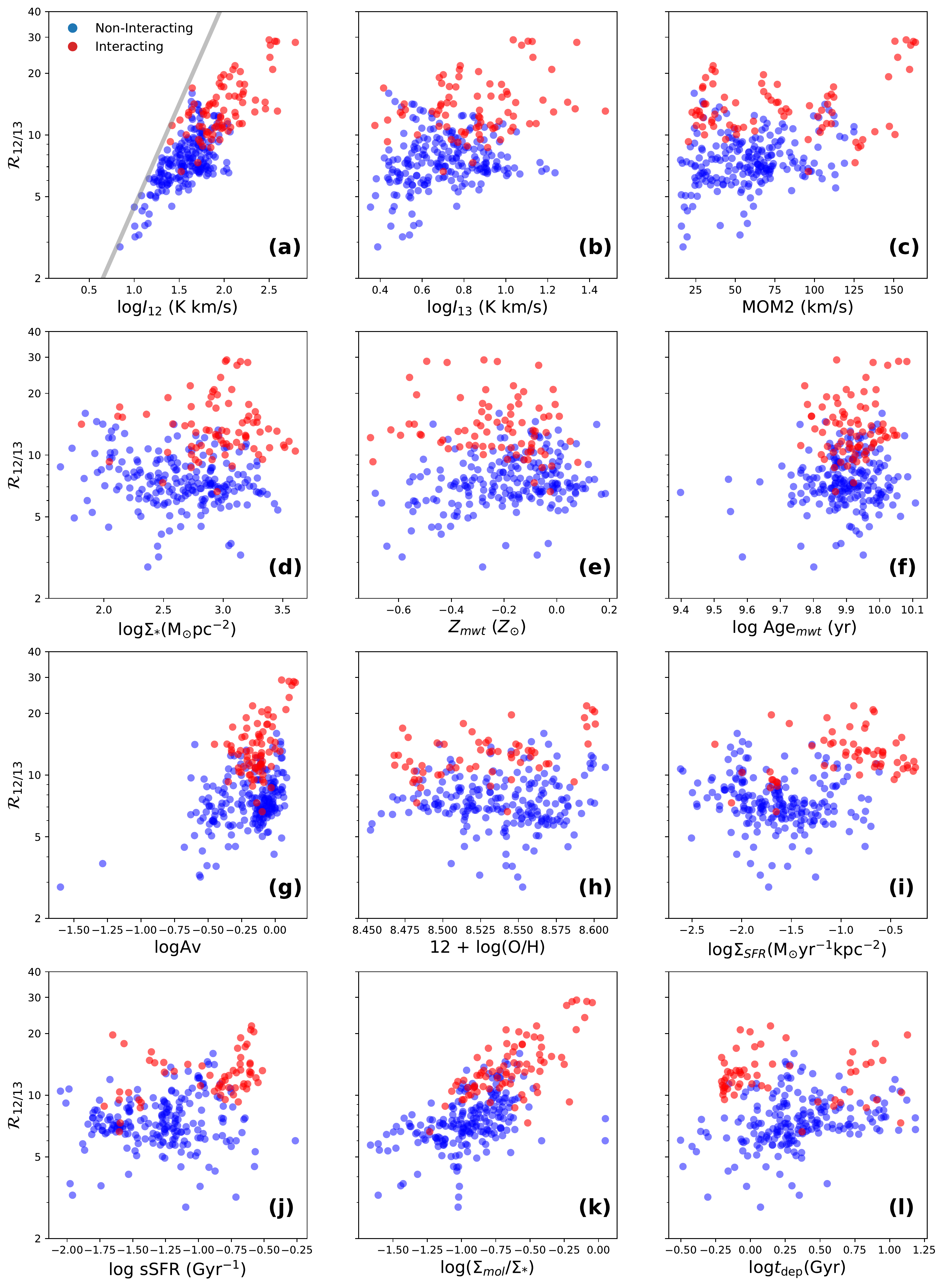}
\caption{
Resolved line ratio \rtt \ as functions local parameters {for regions where \itt \ is detected with S/N $>4$}. 
The Spearman rank correlation test results are shown in Table \ref{tb:rsres}.
The gray line in panel (a) corresponds to the upper boundary of our \rtt~ measurements imposed by a constant \itt~ detection of the sample (corresponding to the horizontal gray line in panel (a) of Figure \ref{fig:ei1312all}).    
In panels (h), (i), (j), (l), we exclude the grids which  are identified as star forming regions with the BPT diagnostics.  
The filled circles show the individual \rtt \  measurements for non-interacting (blue) and interacting (red) galaxies.
 }
\label{fig:r13res}
\end{figure*}

Figure \ref{fig:r13res} shows the correlations between resolved \rtt \ 
and other local properties for all the \itt \ detections in the non-interacting galaxies 
in the EDGE-CALIFA survey. 
The Spearman's rank correlation coefficients $r$ 
and the probabilities of no correlation $P_0$ 
are listed in Table \ref{tb:rsres}.

The strong apparent dependence on \itw \ is mainly due to the
sensitivity bias toward lower values in our  \rtt \ measurements 
when \itw~ is weak (panel (a)). 
In the lower \itw~ regime, only grids with strong \itt~ and lower \rtt~ are detected. 
Grids with \rtt$<5$ experience the most bias, and act to drive the strong positive correlation between \rtt~ and \itw. 
Since \twco \  moment 2 (panel (c)) and nebular extinction $\rm A_v$ (panel (g)) 
increase with \itw, 
the apparent increasing trends of \rtt \ with them 
could also result from bias due to 
limited sensitivity at lower \itw. 
Meanwhile, 
there is a tight correlation between \rtt \ and gas fraction shown in panel (k), 
which is not entirely due to detection bias, as the increasing trend is still obvious if we exclude grids with \rtt $<5$ from this panel. 
Similar to results based on \rstack,  this correlation may reflect the association of higher \itt~ with higher stellar surface density on kpc scales. 

\begin{deluxetable*}{l c c r r c r r c r r}[h!]
\tablecaption{\label{tb:rsres} Spearman's rank correlation coefficients between \rtt~ and local parameters}
\tablehead{
\colhead{Parameter} & 
Note & Ref & 
 \multicolumn{2}{c}{All galaxies } & &
 \multicolumn{2}{c}{Non-interacting galaxies }& & 
 \multicolumn{2}{c}{Interacting galaxies } \\
 \cline{4-5} \cline{7-8}  \cline{10-11} 
 & & &
\mc{$r_{\rm s} $} &
\mc{$P_{0}$ } &
& 
\mc{$r_{\rm s} $ } &
\mc{$P_{0}$ } & 
&
\mc{$r_{\rm s} $ } &
\mc{$P_{0}$ } 
}
\startdata
\input{res_tab.tex}

\enddata
\tablecomments{
References: 1. \citet{Bolatto2017}; 2. This work;  
3. \citet{Pipe3D}. 
}
\end{deluxetable*}

\end{document}

%% file: affiliations.tex
\newcommand{\UIUC}{\affil{Department of Astronomy, University of Illinois, Urbana, IL 61801, USA}}

\newcommand{\OSU}{\affil{Department of Astronomy, The Ohio State University, 140 West 18th Avenue, Columbus, Ohio 43210, USA}}

\newcommand{\Alberta}{\affil{Department of Physics, University of Alberta, Edmonton, AB T6G 2E1, Canada}}

\newcommand{\JHU}{\affil{Department of Physics and Astronomy, The Johns Hopkins University, Baltimore, MD 21218, USA}}

\newcommand{\Maryland}{\affil{Department of Astronomy, University of Maryland, College Park, MD 20742, USA}}

\newcommand{\MPE}{\affil{Max-Planck-Institut f\"{u}r extraterrestrische Physik, Giessenbachstra{\ss}e 1, D-85748 Garching, Germany}}

\newcommand{\NRAO}{\affil{National Radio Astronomy Observatory, 520 Edgemont Road, Charlottesville, VA 22903-2475, USA}}

\newcommand{\LAM}{\affil{
Aix Marseille Universit\'{e}, CNRS, CNES, LAM (Laboratoire d’Astrophysique de Marseille), F-13388 Marseille,
France}}

\newcommand{\UNAM}{\affil{Instituto de Astronom\'{\i}a, Universidad Nacional Aut\'{o}noma de M\'{e}xico, A.P. 70-264, 04510 M\'{e}xico, D.F.,  Mexico}}

\newcommand{\MPIfR}{\affil{Max-Planck-Institut f\"{u}r Radioastronomie, D-53121, Bonn, Germany}}

\newcommand{\UCB}{\affil{Department of Astronomy, University of California, Berkeley, CA 94720, USA}}

\newcommand{\Bonn}{\affil{Universit\"at Bonn, Argelander-Institut f\"ur Astronomie, Auf dem H\"ugel 71, D-53121 Bonn, Germany}}

\newcommand{\Chile}
{\affil{Departamento de Astronomía, Universidad de Chile,Casilla 36-D, Santiago, Chile}}

%% file: authors.tex
\correspondingauthor{Yixian Cao}
\email{ycao@mpe.mpg.de}

\author[0000-0001-5301-1326]{Yixian Cao}
\UIUC
\MPE

\author[0000-0002-7759-0585]{Tony Wong}
\UIUC

\author[0000-0002-5480-5686]{Alberto D. Bolatto}
\Maryland

\author[0000-0002-2545-1700]{Adam K. Leroy}
\OSU

\author[0000-0002-5204-2259]{Erik Rosolowsky}
\Alberta

\author[0000-0003-4161-2639]{Dyas Utomo}
\NRAO

\author[0000-0001-6444-9307]{Sebasti\'{a}n F. S\'{a}nchez}
\UNAM

\author[0000-0003-2405-7258]{Jorge K. Barrera-Ballesteros}
\UNAM

\author[0000-0003-2508-2586]{Rebecca C. Levy}
\altaffiliation{NSF Astronomy and Astrophysics Postdoctoral Fellow}
\affiliation{Steward Observatory, University of Arizona, Tucson, AZ 85721, USA}

\author[0000-0001-6498-2945]{Dario Colombo}
\MPIfR

\author{Leo Blitz}
\UCB

\author[0000-0002-8765-7915]{Stuart N. Vogel}
\Maryland

\author[0000-0003-1111-3951]{Johannes Puschnig}
\Bonn

\author[0000-0002-5877-379X]{Vicente Villanueva}
\Maryland

\author
{Monica Rubio}
\Chile

%% file: edge_gts.tex
ARP220 & $14.1 \pm 1.9 $  & $29.0 \pm 3.9 $  & $ 0.8 $  & $ 10$ & $ 11$ & $ 68$ &  $ 25.7 \pm 4.9 $  & $2$ &  $ 27.7 \pm 2.7 $  & $0.903 \pm 0.001 $    & $1 $  & $0 $  \\
IC0480 & $ \le 5.1 $  & $ \ge 14.0 $  & $ 0.7 $  & $ 6$ & $ 0$ & $ 47$ &  $ 7.8 \pm 1.2 $  & $ 0$ &  \nodata & $0.43 \pm 0.04 $  & $0 $  & $ 0 $  \\
IC0540 & $ \le 3.7 $  & $ \ge 5.2 $  & $ 0.7 $  & $ 0$ & $ 3$ & $ 12$ &  \nodata  & $ 0$ &  \nodata & $0.24 \pm 0.04 $  & $0 $  & $ 0 $  \\
IC0944 & $8.4 \pm 1.8 $  & $9.8 \pm 2.2 $  & $ 0.7 $  & $ 0$ & $ 2$ & $ 71$ &  \nodata  & $3$ &  $ 7.7 \pm 1.8 $  & $0.23 \pm 0.03 $   & $0 $  & $0 $  \\
IC1199 & $ \le 4.8 $  & $ \ge 8.5 $  & $ 0.6 $  & $ 0$ & $ 0$ & $ 59$ &  \nodata  & $ 0$ &  \nodata & $0.36 \pm 0.07 $  & $0 $  & $ 0 $  \\
IC1683 & $6.6 \pm 1.3 $  & $12.5 \pm 2.5 $  & $ 0.7 $  & $ 4$ & $ 5$ & $ 43$ &  $ 10.4 \pm 2.3 $  & $2$ &  $ 11.6 \pm 1.7 $  & $0.50 \pm 0.05 $   & $0 $  & $0 $  \\
IC2247 & $ \le 4.5 $  & $ \ge 14.3 $  & $ 0.5 $  & $ 0$ & $ 0$ & $ 61$ &  \nodata  & $ 0$ &  \nodata & $0.38 \pm 0.04 $  & $0 $  & $ 0 $  \\
IC2487 & $ \le 5.4 $  & $ \ge 9.1 $  & $ 0.6 $  & $ 0$ & $ 0$ & $ 55$ &  \nodata  & $ 0$ &  \nodata & $0.32 \pm 0.04 $  & $0 $  & $ 0 $  \\
IC4566 & $7.7 \pm 1.7 $  & $6.2 \pm 1.4 $  & $ 0.7 $  & $ 0$ & $ 0$ & $ 81$ &  \nodata  & $0$ &  \nodata & $1.0 \pm 1.4 $   & $0 $  & $1 $  \\
IC5376 & $ \le 2.6 $  & $ \ge 2.9 $  & $ 0.6 $  & $ 0$ & $ 0$ & $ 12$ &  \nodata  & $ 0$ &  \nodata & $1.0 \pm 1.4 $  & $0 $  & $ 0 $  \\
NGC0447 & $ \le 4.4 $  & $ \ge 7.2 $  & $ 0.6 $  & $ 0$ & $ 2$ & $ 30$ &  \nodata  & $ 0$ &  \nodata & $1.0 \pm 1.4 $  & $0 $  & $ 1 $  \\
NGC0477 & $ \le 6.6 $  & $ \ge 5.0 $  & $ 0.7 $  & $ 0$ & $ 0$ & $ 50$ &  \nodata  & $ 0$ &  \nodata & $0.32 \pm 0.05 $  & $0 $  & $ 1 $  \\
NGC0496 & $ \le 5.2 $  & $ \ge 7.2 $  & $ 0.6 $  & $ 0$ & $ 0$ & $ 52$ &  \nodata  & $ 1$ &  $ 8.4 \pm 2.1 $  & $0.30 \pm 0.05 $  & $0 $  & $ 0 $  \\
NGC0523 & $ \le 6.9 $  & $ \ge 13.3 $  & $ 0.7 $  & $ 0$ & $ 0$ & $ 71$ &  \nodata  & $ 0$ &  \nodata & $0.44 \pm 0.05 $  & $1 $  & $ 0 $  \\
NGC0551 & $ \le 5.3 $  & $ \ge 6.9 $  & $ 0.5 $  & $ 0$ & $ 0$ & $ 78$ &  \nodata  & $ 0$ &  \nodata & $0.28 \pm 0.04 $  & $0 $  & $ 1 $  \\
NGC1167 & $ \le 4.4 $  & $ \ge 3.4 $  & $ 0.6 $  & $ 0$ & $ 0$ & $ 16$ &  \nodata  & $ 0$ &  \nodata & $0.11 \pm 0.03 $  & $0 $  & $ 0 $  \\
NGC2253 & $10.8 \pm 2.0 $  & $14.0 \pm 2.6 $  & $ 0.6 $  & $ 0$ & $ 1$ & $ 144$ &  \nodata  & $3$ &  $ 14.4 \pm 3.1 $  & $0.40 \pm 0.03 $   & $0 $  & $0 $  \\
NGC2347 & $10.0 \pm 1.4 $  & $7.7 \pm 1.1 $  & $ 0.5 $  & $ 4$ & $ 0$ & $ 90$ &  $ 6.2 \pm 1.1 $  & $3$ &  $ 7.6 \pm 1.8 $  & $0.36 \pm 0.03 $   & $0 $  & $0 $  \\
NGC2410 & $7.7 \pm 1.6 $  & $10.2 \pm 2.2 $  & $ 0.6 $  & $ 0$ & $ 0$ & $ 86$ &  \nodata  & $2$ &  $ 9.4 \pm 1.5 $  & $0.41 \pm 0.03 $   & $0 $  & $1 $  \\
NGC2480 & $ \le 2.6 $  & $ \ge 4.7 $  & $ 0.5 $  & $ 0$ & $ 0$ & $ 16$ &  \nodata  & $ 0$ &  \nodata & $1.0 \pm 1.4 $  & $1 $  & $ 1 $  \\
NGC2487 & $ \le 7.6 $  & $ \ge 6.5 $  & $ 0.7 $  & $ 0$ & $ 1$ & $ 47$ &  \nodata  & $ 0$ &  \nodata & $0.21 \pm 0.03 $  & $0 $  & $ 1 $  \\
NGC2623 & $5.7 \pm 1.0 $  & $19.7 \pm 3.6 $  & $ 0.7 $  & $ 5$ & $ 4$ & $ 31$ &  $ 15.5 \pm 2.0 $  & $2$ &  $ 14.4 \pm 2.1 $  & $0.917 \pm 0.004 $   & $1 $  & $0 $  \\
NGC2639 & $10.3 \pm 1.8 $  & $8.8 \pm 1.6 $  & $ 0.7 $  & $ 5$ & $ 0$ & $ 93$ &  $ 3.7 \pm 1.2 $  & $2$ &  $ 7.4 \pm 0.9 $  & $0.28 \pm 0.02 $   & $0 $  & $0 $  \\
NGC2730 & $ \le 5.9 $  & $ \ge 4.8 $  & $ 0.5 $  & $ 0$ & $ 0$ & $ 27$ &  \nodata  & $ 0$ &  \nodata & $0.40 \pm 0.05 $  & $0 $  & $ 1 $  \\
NGC2906 & $9.7 \pm 1.7 $  & $8.7 \pm 1.5 $  & $ 0.6 $  & $ 0$ & $ 0$ & $ 99$ &  \nodata  & $4$ &  $ 7.8 \pm 1.5 $  & $0.34 \pm 0.03 $   & $0 $  & $0 $  \\
NGC2916 & $ \le 5.9 $  & $ \ge 4.6 $  & $ 0.7 $  & $ 0$ & $ 0$ & $ 34$ &  \nodata  & $ 0$ &  \nodata & $0.28 \pm 0.03 $  & $0 $  & $ 0 $  \\
NGC3303 & $ \le 5.2 $  & $ \ge 5.8 $  & $ 0.9 $  & $ 0$ & $ 5$ & $ 25$ &  \nodata  & $ 0$ &  \nodata & $1.0 \pm 1.4 $  & $1 $  & $ 0 $  \\
NGC3381 & $ \le 3.4 $  & $ \ge 4.2 $  & $ 0.5 $  & $ 0$ & $ 0$ & $ 15$ &  \nodata  & $ 0$ &  \nodata & $0.36 \pm 0.03 $  & $0 $  & $ 1 $  \\
NGC3811 & $6.9 \pm 1.7 $  & $12.2 \pm 3.1 $  & $ 0.6 $  & $ 3$ & $ 2$ & $ 80$ &  $ 8.9 \pm 2.3 $  & $3$ &  $ 9.1 \pm 2.2 $  & $0.44 \pm 0.44 $   & $0 $  & $1 $  \\
NGC3815 & $ \le 5.2 $  & $ \ge 8.0 $  & $ 0.7 $  & $ 0$ & $ 0$ & $ 48$ &  \nodata  & $ 1$ &  $ 8.1 \pm 1.8 $  & $0.44 \pm 0.44 $  & $0 $  & $ 0 $  \\
NGC3994 & $7.8 \pm 1.5 $  & $9.2 \pm 1.8 $  & $ 0.7 $  & $ 6$ & $ 3$ & $ 52$ &  $ 5.8 \pm 1.0 $  & $1$ &  $ 9.4 \pm 2.3 $  & $0.48 \pm 0.07 $   & $0 $  & $0 $  \\
NGC4047 & $16.9 \pm 2.0 $  & $10.6 \pm 1.3 $  & $ 0.7 $  & $ 21$ & $ 6$ & $ 118$ &  $ 7.0 \pm 1.5 $  & $4$ &  $ 8.4 \pm 1.0 $  & $0.36 \pm 0.18 $   & $0 $  & $0 $  \\
NGC4149 & $6.2 \pm 1.3 $  & $11.5 \pm 2.4 $  & $ 0.7 $  & $ 6$ & $ 3$ & $ 39$ &  $ 8.7 \pm 1.9 $  & $1$ &  $ 8.8 \pm 1.5 $  & $0.38 \pm 0.03 $   & $0 $  & $1 $  \\
NGC4185 & $ \le 4.2 $  & $ \ge 2.8 $  & $ 0.7 $  & $ 0$ & $ 0$ & $ 14$ &  \nodata  & $ 0$ &  \nodata & \nodata  & $0 $  & $ 1 $  \\
NGC4210 & $ \le 6.0 $  & $ \ge 7.5 $  & $ 0.5 $  & $ 0$ & $ 0$ & $ 83$ &  \nodata  & $ 0$ &  \nodata & $0.24 \pm 0.03 $  & $0 $  & $ 1 $  \\
NGC4211N & $ \le 3.7 $  & $ \ge 4.5 $  & $ 0.8 $  & $ 0$ & $ 3$ & $ 12$ &  \nodata  & $ 0$ &  \nodata & $0.55 \pm 0.10 $  & $1 $  & $ 0 $  \\
NGC4470 & $ \le 5.1 $  & $ \ge 6.4 $  & $ 0.6 $  & $ 0$ & $ 0$ & $ 45$ &  \nodata  & $ 0$ &  \nodata & $0.43 \pm 0.02 $  & $0 $  & $ 0 $  \\
NGC4644 & $ \le 4.0 $  & $ \ge 6.8 $  & $ 0.5 $  & $ 0$ & $ 0$ & $ 48$ &  \nodata  & $ 0$ &  \nodata & $1.0 \pm 1.4 $  & $0 $  & $ 1 $  \\
NGC4676A & $6.7 \pm 1.4 $  & $10.5 \pm 2.3 $  & $ 0.9 $  & $ 5$ & $ 7$ & $ 24$ &  $ 9.2 \pm 2.3 $  & $1$ &  $ 9.4 \pm 1.5 $  & $0.52 \pm 0.04 $   & $1 $  & $1 $  \\
NGC4711 & $ \le 5.3 $  & $ \ge 7.0 $  & $ 0.6 $  & $ 0$ & $ 0$ & $ 58$ &  \nodata  & $ 1$ &  $ 7.8 \pm 2.0 $  & $0.30 \pm 0.04 $  & $0 $  & $ 1 $  \\
NGC4961 & $ \le 3.6 $  & $ \ge 5.6 $  & $ 0.5 $  & $ 0$ & $ 0$ & $ 34$ &  \nodata  & $ 0$ &  \nodata & $0.36 \pm 0.07 $  & $0 $  & $ 1 $  \\
NGC5000 & $ \le 5.1 $  & $ \ge 8.0 $  & $ 0.6 $  & $ 0$ & $ 3$ & $ 39$ &  \nodata  & $ 0$ &  \nodata & $0.40 \pm 0.05 $  & $0 $  & $ 1 $  \\
NGC5016 & $ \le 6.1 $  & $ \ge 10.2 $  & $ 0.6 $  & $ 0$ & $ 0$ & $ 81$ &  \nodata  & $ 2$ &  $ 8.6 \pm 1.4 $  & $0.37 \pm 0.01 $  & $0 $  & $ 1 $  \\
NGC5056 & $ \le 5.2 $  & $ \ge 7.3 $  & $ 0.6 $  & $ 0$ & $ 0$ & $ 60$ &  \nodata  & $ 0$ &  \nodata & $0.41 \pm 0.05 $  & $0 $  & $ 0 $  \\
NGC5205 & $ \le 4.4 $  & $ \ge 5.0 $  & $ 0.6 $  & $ 0$ & $ 0$ & $ 28$ &  \nodata  & $ 0$ &  \nodata & $0.30 \pm 0.05 $  & $0 $  & $ 0 $  \\
NGC5218 & $19.3 \pm 2.1 $  & $19.7 \pm 2.2 $  & $ 0.7 $  & $ 13$ & $ 12$ & $ 101$ &  $ 14.4 \pm 1.8 $  & $3$ &  $ 18.1 \pm 1.9 $  & $0.518 \pm 0.004 $    & $1 $  & $1 $  \\
NGC5394 & $8.0 \pm 1.2 $  & $19.2 \pm 3.0 $  & $ 0.8 $  & $ 7$ & $ 8$ & $ 39$ &  $ 19.1 \pm 4.3 $  & $3$ &  $ 18.7 \pm 3.1 $  & $0.54 \pm 0.21 $   & $1 $  & $1 $  \\
NGC5406 & $ \le 8.5 $  & $ \ge 6.7 $  & $ 0.7 $  & $ 0$ & $ 0$ & $ 73$ &  \nodata  & $ 0$ &  \nodata & $0.31 \pm 0.04 $  & $0 $  & $ 1 $  \\
NGC5480 & $13.3 \pm 1.8 $  & $9.4 \pm 1.3 $  & $ 0.6 $  & $ 6$ & $ 0$ & $ 127$ &  $ 7.1 \pm 1.3 $  & $5$ &  $ 9.1 \pm 1.5 $  & $0.34 \pm 0.02 $   & $0 $  & $0 $  \\
NGC5520 & $6.6 \pm 1.3 $  & $9.1 \pm 1.8 $  & $ 0.6 $  & $ 0$ & $ 0$ & $ 56$ &  \nodata  & $2$ &  $ 7.9 \pm 1.0 $  & $0.40 \pm 0.03 $   & $0 $  & $0 $  \\
NGC5614 & $12.3 \pm 2.3 $  & $16.9 \pm 3.2 $  & $ 0.8 $  & $ 0$ & $ 4$ & $ 109$ &  \nodata  & $2$ &  $ 15.4 \pm 3.3 $  & $0.25 \pm 0.02 $   & $1 $  & $0 $  \\
NGC5633 & $11.6 \pm 1.6 $  & $10.6 \pm 1.5 $  & $ 0.6 $  & $ 4$ & $ 0$ & $ 97$ &  $ 6.8 \pm 1.2 $  & $5$ &  $ 9.3 \pm 1.6 $  & $0.35 \pm 0.01 $   & $0 $  & $0 $  \\
NGC5657 & $ \le 5.1 $  & $ \ge 5.2 $  & $ 0.9 $  & $ 0$ & $ 4$ & $ 20$ &  \nodata  & $ 0$ &  \nodata & $0.59 \pm 0.05 $  & $0 $  & $ 1 $  \\
NGC5732 & $ \le 3.9 $  & $ \ge 4.2 $  & $ 0.6 $  & $ 0$ & $ 0$ & $ 22$ &  \nodata  & $ 0$ &  \nodata & $0.28 \pm 0.03 $  & $0 $  & $ 0 $  \\
NGC5784 & $4.6 \pm 1.1 $  & $6.5 \pm 1.7 $  & $ 0.6 $  & $ 0$ & $ 0$ & $ 35$ &  \nodata  & $0$ &  \nodata & $0.27 \pm 0.01 $   & $0 $  & $0 $  \\
NGC5908 & $41.9 \pm 2.4 $  & $8.5 \pm 0.5 $  & $ 0.8 $  & $ 52$ & $ 53$ & $ 132$ &  $ 7.4 \pm 0.9 $  & $4$ &  $ 8.5 \pm 1.2 $  & $0.25 \pm 0.25 $   & $0 $  & $0 $  \\
NGC5930 & $7.8 \pm 1.3 $  & $15.8 \pm 2.6 $  & $ 0.8 $  & $ 7$ & $ 7$ & $ 30$ &  $ 12.3 \pm 2.1 $  & $2$ &  $ 13.2 \pm 1.6 $  & $0.68 \pm 0.01 $   & $1 $  & $1 $  \\
NGC5934 & $8.5 \pm 1.6 $  & $9.7 \pm 1.8 $  & $ 0.9 $  & $ 4$ & $ 9$ & $ 41$ &  $ 9.2 \pm 1.6 $  & $1$ &  $ 9.1 \pm 2.0 $  & $0.34 \pm 0.02 $   & $1 $  & $0 $  \\
NGC5947 & $ \le 3.2 $  & $ \ge 3.1 $  & $ 0.5 $  & $ 0$ & $ 0$ & $ 21$ &  \nodata  & $ 0$ &  \nodata & $0.36 \pm 0.05 $  & $0 $  & $ 1 $  \\
NGC5953 & $24.5 \pm 1.9 $  & $13.3 \pm 1.0 $  & $ 0.6 $  & $ 31$ & $ 25$ & $ 128$ &  $ 11.5 \pm 1.6 $  & $4$ &  $ 12.4 \pm 1.5 $  & $0.59 \pm 0.01 $   & $1 $  & $0 $  \\
NGC5980 & $13.7 \pm 1.5 $  & $9.2 \pm 1.0 $  & $ 0.6 $  & $ 22$ & $ 3$ & $ 101$ &  $ 6.2 \pm 1.1 $  & $5$ &  $ 6.6 \pm 1.0 $  & $0.41 \pm 0.03 $   & $0 $  & $0 $  \\
NGC6004 & $7.9 \pm 1.9 $  & $7.7 \pm 1.9 $  & $ 0.6 $  & $ 0$ & $ 0$ & $ 74$ &  \nodata  & $2$ &  $ 7.2 \pm 1.3 $  & $0.31 \pm 0.02 $   & $0 $  & $1 $  \\
NGC6060 & $9.7 \pm 2.1 $  & $12.5 \pm 2.7 $  & $ 0.6 $  & $ 0$ & $ 0$ & $ 158$ &  \nodata  & $2$ &  $ 8.5 \pm 1.1 $  & $0.31 \pm 0.04 $   & $0 $  & $1 $  \\
NGC6155 & $6.0 \pm 1.5 $  & $12.2 \pm 3.0 $  & $ 0.5 $  & $ 4$ & $ 0$ & $ 104$ &  $ 3.9 \pm 0.8 $  & $3$ &  $ 11.0 \pm 2.2 $  & $0.35 \pm 0.02 $   & $0 $  & $0 $  \\
NGC6168 & $ \le 5.6 $  & $ \ge 4.8 $  & $ 0.7 $  & $ 0$ & $ 0$ & $ 38$ &  \nodata  & $ 0$ &  \nodata & $0.41 \pm 0.04 $  & $0 $  & $ 0 $  \\
NGC6186 & $14.7 \pm 1.9 $  & $10.2 \pm 1.3 $  & $ 0.7 $  & $ 7$ & $ 11$ & $ 76$ &  $ 10.9 \pm 1.7 $  & $5$ &  $ 11.8 \pm 2.6 $  & $0.43 \pm 0.03 $   & $0 $  & $1 $  \\
NGC6301 & $ \le 7.6 $  & $ \ge 7.1 $  & $ 0.7 $  & $ 0$ & $ 0$ & $ 110$ &  \nodata  & $ 1$ &  $ 6.0 \pm 1.5 $  & $0.21 \pm 0.03 $  & $0 $  & $ 0 $  \\
NGC6310 & $ \le 2.8 $  & $ \ge 2.8 $  & $ 0.6 $  & $ 0$ & $ 0$ & $ 13$ &  \nodata  & $ 0$ &  \nodata & $0.24 \pm 0.04 $  & $0 $  & $ 0 $  \\
NGC6314 & $ \le 4.6 $  & $ \ge 6.6 $  & $ 0.6 $  & $ 0$ & $ 2$ & $ 36$ &  \nodata  & $ 0$ &  \nodata & \nodata  & $0 $  & $ 0 $  \\
NGC6361 & $31.6 \pm 2.0 $  & $10.9 \pm 0.7 $  & $ 0.7 $  & $ 39$ & $ 36$ & $ 124$ &  $ 8.8 \pm 1.9 $  & $7$ &  $ 10.0 \pm 1.2 $  & $0.31 \pm 0.01 $   & $0 $  & $0 $  \\
NGC6394 & $ \le 6.1 $  & $ \ge 5.7 $  & $ 0.8 $  & $ 0$ & $ 0$ & $ 42$ &  \nodata  & $ 0$ &  \nodata & $0.36 \pm 0.03 $  & $0 $  & $ 1 $  \\
NGC6478 & $13.8 \pm 1.8 $  & $9.6 \pm 1.3 $  & $ 0.7 $  & $ 12$ & $ 3$ & $ 101$ &  $ 6.4 \pm 1.2 $  & $4$ &  $ 8.3 \pm 1.3 $  & $0.28 \pm 0.02 $   & $0 $  & $0 $  \\
NGC7738 & $6.5 \pm 1.1 $  & $13.9 \pm 2.2 $  & $ 0.6 $  & $ 8$ & $ 6$ & $ 40$ &  $ 11.2 \pm 1.5 $  & $3$ &  $ 11.2 \pm 1.7 $  & $0.58 \pm 0.07 $   & $0 $  & $1 $  \\
NGC7819 & $4.5 \pm 1.0 $  & $6.6 \pm 1.6 $  & $ 0.6 $  & $ 0$ & $ 2$ & $ 27$ &  \nodata  & $2$ &  $ 6.3 \pm 1.2 $  & $0.43 \pm 0.05 $   & $0 $  & $1 $  \\
UGC00809 & $ \le 2.0 $  & $ \ge 2.7 $  & $ 0.5 $  & $ 0$ & $ 0$ & $ 9$ &  \nodata  & $ 0$ &  \nodata & \nodata & $0 $  & $ 0 $  \\
UGC03253 & $ \le 3.3 $  & $ \ge 3.9 $  & $ 0.6 $  & $ 0$ & $ 0$ & $ 16$ &  \nodata  & $ 0$ &  \nodata & $0.30 \pm 0.05 $  & $0 $  & $ 1 $  \\
UGC03539 & $ \le 5.4 $  & $ \ge 9.5 $  & $ 0.7 $  & $ 0$ & $ 0$ & $ 41$ &  \nodata  & $ 0$ &  \nodata & $0.40 \pm 0.07 $  & $0 $  & $ 1 $  \\
UGC03969 & $ \le 6.5 $  & $ \ge 6.5 $  & $ 0.8 $  & $ 0$ & $ 0$ & $ 43$ &  \nodata  & $ 1$ &  $ 5.2 \pm 1.3 $  & $0.42 \pm 0.06 $  & $0 $  & $ 0 $  \\
UGC03973 & $ \le 5.8 $  & $ \ge 5.2 $  & $ 0.7 $  & $ 0$ & $ 0$ & $ 48$ &  \nodata  & $ 0$ &  \nodata & $0.64 \pm 0.07 $  & $0 $  & $ 1 $  \\
UGC04029 & $ \le 4.7 $  & $ \ge 10.9 $  & $ 0.6 $  & $ 3$ & $ 0$ & $ 55$ &  $ 7.1 \pm 1.4 $  & $ 0$ &  \nodata & $0.43 \pm 0.04 $  & $0 $  & $ 1 $  \\
UGC04132 & $12.8 \pm 2.1 $  & $13.1 \pm 2.2 $  & $ 0.8 $  & $ 10$ & $ 7$ & $ 90$ &  $ 6.7 \pm 0.9 $  & $4$ &  $ 11.1 \pm 2.2 $  & $0.38 \pm 0.04 $   & $0 $  & $0 $  \\
UGC04280 & $ \le 3.3 $  & $ \ge 3.9 $  & $ 0.7 $  & $ 0$ & $ 1$ & $ 13$ &  \nodata  & $ 0$ &  \nodata & $0.44 \pm 0.07 $  & $0 $  & $ 0 $  \\
UGC04461 & $ \le 4.5 $  & $ \ge 6.5 $  & $ 0.7 $  & $ 0$ & $ 2$ & $ 31$ &  \nodata  & $ 0$ &  \nodata & $0.43 \pm 0.05 $  & $0 $  & $ 0 $  \\
UGC05108 & $ \le 4.3 $  & $ \ge 5.7 $  & $ 0.7 $  & $ 0$ & $ 2$ & $ 20$ &  \nodata  & $ 0$ &  \nodata & $0.48 \pm 0.07 $  & $0 $  & $ 1 $  \\
UGC05111 & $8.2 \pm 1.4 $  & $10.1 \pm 1.8 $  & $ 0.6 $  & $ 0$ & $ 0$ & $ 69$ &  \nodata  & $0$ &  \nodata & $0.28 \pm 0.08 $   & $0 $  & $0 $  \\
UGC05359 & $ \le 3.3 $  & $ \ge 3.2 $  & $ 0.5 $  & $ 0$ & $ 0$ & $ 24$ &  \nodata  & $ 0$ &  \nodata & $1.0 \pm 1.4 $  & $0 $  & $ 1 $  \\
UGC05598 & $ \le 4.5 $  & $ \ge 4.1 $  & $ 0.8 $  & $ 0$ & $ 0$ & $ 19$ &  \nodata  & $ 0$ &  \nodata & $0.34 \pm 0.05 $  & $0 $  & $ 0 $  \\
UGC07012 & $ \le 2.6 $  & $ \ge 3.1 $  & $ 0.5 $  & $ 0$ & $ 0$ & $ 11$ &  \nodata  & $ 0$ &  \nodata & $1.0 \pm 1.4 $  & $0 $  & $ 1 $  \\
UGC08107 & $ \le 6.8 $  & $ \ge 10.8 $  & $ 0.8 $  & $ 0$ & $ 2$ & $ 56$ &  \nodata  & $ 0$ &  \nodata & $0.28 \pm 0.05 $  & $1 $  & $ 1 $  \\
UGC08267 & $ \le 5.6 $  & $ \ge 8.2 $  & $ 0.8 $  & $ 0$ & $ 2$ & $ 35$ &  \nodata  & $ 0$ &  \nodata & $0.27 \pm 0.05 $  & $0 $  & $ 0 $  \\
UGC09067 & $ \le 6.3 $  & $ \ge 6.6 $  & $ 0.9 $  & $ 0$ & $ 0$ & $ 41$ &  \nodata  & $ 0$ &  \nodata & $0.38 \pm 0.06 $  & $0 $  & $ 0 $  \\
UGC09476 & $ \le 6.8 $  & $ \ge 6.4 $  & $ 0.6 $  & $ 0$ & $ 0$ & $ 58$ &  \nodata  & $ 0$ &  \nodata & $0.29 \pm 0.03 $  & $0 $  & $ 1 $  \\
UGC09537 & $8.3 \pm 1.7 $  & $4.3 \pm 0.9 $  & $ 0.8 $  & $ 0$ & $ 0$ & $ 52$ &  \nodata  & $0$ &  \nodata & $0.28 \pm 0.04 $   & $0 $  & $0 $  \\
UGC09542 & $ \le 4.8 $  & $ \ge 5.5 $  & $ 0.6 $  & $ 0$ & $ 0$ & $ 41$ &  \nodata  & $ 0$ &  \nodata & $0.21 \pm 0.03 $  & $0 $  & $ 0 $  \\
UGC09665 & $ \le 5.3 $  & $ \ge 10.6 $  & $ 0.7 $  & $ 0$ & $ 0$ & $ 53$ &  \nodata  & $ 0$ &  \nodata & $0.36 \pm 0.03 $  & $0 $  & $ 0 $  \\
UGC09759 & $ \le 4.9 $  & $ \ge 8.6 $  & $ 0.7 $  & $ 0$ & $ 3$ & $ 30$ &  \nodata  & $ 1$ &  $ 8.3 \pm 1.7 $  & $0.29 \pm 0.03 $  & $0 $  & $ 0 $  \\
UGC09873 & $ \le 2.8 $  & $ \ge 3.4 $  & $ 0.6 $  & $ 0$ & $ 0$ & $ 11$ &  \nodata  & $ 0$ &  \nodata & $1.0 \pm 1.4 $  & $0 $  & $ 0 $  \\
UGC09892 & $ \le 3.7 $  & $ \ge 4.4 $  & $ 0.6 $  & $ 0$ & $ 0$ & $ 31$ &  \nodata  & $ 0$ &  \nodata & $1.0 \pm 1.4 $  & $0 $  & $ 0 $  \\
UGC09919 & $ \le 2.2 $  & $ \ge 2.4 $  & $ 0.6 $  & $ 0$ & $ 0$ & $ 6$ &  \nodata  & $ 0$ &  \nodata & $0.35 \pm 0.08 $  & $0 $  & $ 0 $  \\
UGC10043 & $8.4 \pm 1.4 $  & $7.1 \pm 1.2 $  & $ 0.7 $  & $ 7$ & $ 5$ & $ 45$ &  $ 4.4 \pm 0.9 $  & $0$ &  \nodata & $0.34 \pm 0.03 $   & $0 $  & $0 $  \\
UGC10123 & $8.9 \pm 1.6 $  & $10.0 \pm 1.8 $  & $ 0.8 $  & $ 7$ & $ 0$ & $ 62$ &  $ 5.8 \pm 1.6 $  & $0$ &  \nodata & $0.32 \pm 0.02 $   & $0 $  & $0 $  \\
UGC10205 & $ \le 5.6 $  & $ \ge 6.0 $  & $ 0.8 $  & $ 0$ & $ 0$ & $ 39$ &  \nodata  & $ 0$ &  \nodata & $0.26 \pm 0.05 $  & $1 $  & $ 0 $  \\
UGC10380 & $ \le 4.8 $  & $ \ge 3.3 $  & $ 0.8 $  & $ 0$ & $ 4$ & $ 13$ &  \nodata  & $ 0$ &  \nodata & \nodata  & $0 $  & $ 0 $  \\
UGC10384 & $6.8 \pm 1.2 $  & $9.3 \pm 1.7 $  & $ 0.6 $  & $ 5$ & $ 2$ & $ 41$ &  $ 8.0 \pm 1.8 $  & $0$ &  \nodata & $0.41 \pm 0.04 $   & $0 $  & $0 $  \\
UGC10710 & $ \le 6.1 $  & $ \ge 5.4 $  & $ 0.8 $  & $ 0$ & $ 1$ & $ 33$ &  \nodata  & $ 0$ &  \nodata & $0.21 \pm 0.04 $  & $0 $  & $ 0 $  \\

%% file: f1213_corr_tb.tex
$F_{12} $ &\twco \ integrated flux & 1 & $0.65$ & $ < 0.001$ & &$0.56$ & $ < 0.001$ & &$0.70$ & $0.036$ \\
$F_{60}/F_{100}$ &  Far-IR flux ratio at 60 and 100 $\rm \mu m$ 
& 2 & $0.53$ & $ < 0.001$ & &$0.38$ & $0.034$ & &$0.52$ & $0.15$ \\
$\rm A_{v}$  & Nebular extinction from Pipe3D & 3 &
$0.52$ & $ < 0.001$ & &$0.34$ & $0.061$ & &$0.68$ & $0.042$ \\
$r_{\rm mol}/r_{\rm e}$  & Molecular gas concentration \tablenotemark{\footnotesize a} & 1, 3 &
$-0.45$ & $0.012$ & &$-0.26$ & $0.22$ & &$-0.54$ & $0.27$ \\
Median($\Delta$V)  &  Median velocity dispersion & 4
& $0.41$ & $0.01$ & &$0.21$ & $0.25$ & &$0.36$ & $0.39$ \\
sSFR & Specific star formation rate  &  3
& $0.35$ & $0.025$ & &$0.30$ & $0.1$ & &$0.40$ & $0.29$ \\
SFR   & Integrated star formation rate & 3 
& $0.28$ & $0.078$ & &$0.33$ & $0.074$ & &$0.27$ & $0.49$ \\
$12 + \mathrm{\log(O/H)}$  & O3N2 calibated gas metallicity at $r_e$ & 3
& $0.15$ & $0.36$ & &$0.14$ & $0.44$ & &$0.77$ & $0.072$ \\
Distance & Luminosity distance from gas line redshift   & 3  
& $-0.10$ & $0.55$ & &$-0.08$ & $0.66$ & &$-0.20$ & $0.61$ \\
Galaxy Type  & Morphological type & 5 
& $0.06$ & $0.71$ & &$0.25$ & $0.16$ & &$0.88$ & $0.0016$ \\
$M_{*}$ & Integrated stellar mass  & 3
& $-0.05$ & $0.77$ & &$0.01$ & $0.95$ & &$-0.03$ & $0.93$ \\
$\cos (\rm incl)$ & Morph inclination & 5 
 & $0.03$ & $0.84$ & &$-0.04$ & $0.81$ & &$-0.02$ & $0.97$ \\

%% file: all_stk_rs.tex
$I_{12}^{\mathrm{stack}}$ & Stacked \twco \ intensity & 1,2 &
 $0.32$ & $ < 0.001$ & & $0.11$ & $0.32$ & &$0.23$ & $0.33$ \\
$I_{13}^{\mathrm{stack}}$  & Stacked \ttco \ intensity & 2 &
 $0.00$ & $0.97$ & & $-0.18$ & $0.095$ & &$-0.11$ & $0.64$ \\
$\sigma_{12}^{\mathrm{stack}}$  & Stacked \twco~ line width & 1,2 & 
 $-0.00$ & $0.98$ & & $-0.15$ & $0.16$ & &$0.06$ & $0.82$ \\
$\Sigma_{*}$ & Mean stellar mass surface density & 3 & 
 $-0.04$ & $0.65$ & & $-0.20$ & $0.06$ & &$0.01$ & $0.96$ \\
$Z_{\rm mwt}$ & Mean mass weighted stellar metallicity&  3 & 
 $-0.17$ & $0.074$ & & $-0.17$ & $0.11$ & &$-0.08$ & $0.74$ \\
Age$_{\rm mwt}$  & Mean mass weighted stellar age & 3 & 
 $-0.11$ & $0.23$ & & $-0.16$ & $0.13$ & &$-0.13$ & $0.57$ \\
Av & Mean nebular extinction from Pipe3D  & 3 & 
 $0.12$ & $0.22$ & & $0.05$ & $0.61$ & &$-0.04$ & $0.87$ \\
 12 + $\log$(O/H) & Meadian gas phase metallicity from O3N2 & 3 & 
 $0.01$ & $0.9$ & & $0.09$ & $0.41$ & &$0.37$ & $0.17$ \\
$\Sigma_{\mathrm{SFR}}$ & Mean star formation rate surface density&  3 & 
 $0.17$ & $0.086$ & & $0.04$ & $0.69$ & &$-0.02$ & $0.94$ \\
 $sSFR$ & Mean specific star formation rate & 3 &
 $0.36$ & $ < 0.001$ & & $0.32$ & $0.0026$ & &$0.25$ & $0.37$ \\
$\Sigma_{\mathrm{mol}}/\Sigma_{*}$ & Mean molecular gas to stellar mass fraction & 2 &
 $0.56$ & $ < 0.001$ & & $0.43$ & $ < 0.001$ & &$0.43$ & $0.057$ \\
$t_{\mathrm{dep}}$ & Mean depletion time assuming constant $X_{\rm ^{12}CO}$ & 2 & 
 $-0.04$ & $0.7$ & & $-0.07$ & $0.55$ & &$0.01$ & $0.96$ \\

%% file: edge_gts_a.tex
ARP220 & $447.4 \pm 3.1 $  & $ 5.3 $  & $ 0.1 $  & $ 67.6 $  & $ -0.5 $  & $57 $  & $13 $  & $2.6 $  \\
IC0480 & $78.8 \pm 2.5 $  & $ 1.3 $  & $ 0.2 $  & $ 17.8 $  & $ -1.3 $  & $90 $  & $-12 $  & $2.2 $  \\
IC0540 & $21.0 \pm 2.1 $  & $ 1.7 $  & \nodata  & \nodata  & $ -2.0 $  & $90 $  & $-11 $  & $1.0 $  \\
IC0944 & $91.1 \pm 3.3 $  & $ 2.0 $  & $ 0.5 $  & \nodata  & $ -2.0 $  & $68 $  & $21 $  & $3.4 $  \\
IC1199 & $44.9 \pm 2.5 $  & $ 1.0 $  & $ 0.2 $  & $ 5.6 $  & $ -1.7 $  & $75 $  & $-17 $  & $2.3 $  \\
IC1683 & $90.5 \pm 2.5 $  & $ 2.0 $  & $ 0.1 $  & $ 8.7 $  & $ -1.3 $  & $69 $  & $-13 $  & $2.4 $  \\
IC2247 & $70.3 \pm 2.4 $  & $ 1.6 $  & $ 0.2 $  & $ 23.8 $  & $ -1.5 $  & $90 $  & $39 $  & $2.1 $  \\
IC2487 & $53.7 \pm 2.7 $  & $ 1.1 $  & $ 0.3 $  & $ 12.3 $  & $ -1.5 $  & $90 $  & $-17 $  & $2.1 $  \\
IC4566 & $52.5 \pm 2.7 $  & $ 1.4 $  & $ 0.4 $  & $ 8.1 $  & $ -1.9 $  & $52 $  & $-14 $  & $2.7 $  \\
IC5376 & $8.4 \pm 1.2 $  & $ 1.2 $  & \nodata  & \nodata  & $ -1.8 $  & $90 $  & $4 $  & $2.5 $  \\
NGC0447 & $34.8 \pm 1.8 $  & $ 0.6 $  & \nodata  & $ 10.3 $  & $ -2.5 $  & $15 $  & $31 $  & $2.7 $  \\
NGC0477 & $35.8 \pm 2.6 $  & $ 0.9 $  & $ 0.5 $  & $ 7.4 $  & $ -1.4 $  & $73 $  & $-16 $  & $2.9 $  \\
NGC0496 & $41.5 \pm 1.9 $  & $ 1.0 $  & $ 0.2 $  & $ 5.3 $  & $ -1.2 $  & $72 $  & $40 $  & $3.0 $  \\
NGC0523 & $101.2 \pm 3.4 $  & $ 1.4 $  & $ 0.5 $  & $ 10.7 $  & $ -1.3 $  & $80 $  & $-4 $  & $2.3 $  \\
NGC0551 & $40.2 \pm 2.3 $  & $ 0.8 $  & $ 0.3 $  & $ 8.0 $  & $ -1.6 $  & $70 $  & $-35 $  & $2.5 $  \\
NGC1167 & $16.4 \pm 1.9 $  & $ 2.0 $  & \nodata  & $ 6.6 $  & $ -2.0 $  & $49 $  & $-16 $  & $2.4 $  \\
NGC2253 & $165.0 \pm 3.9 $  & $ 1.1 $  & $ 0.2 $  & $ 4.6 $  & $ -1.3 $  & $43 $  & $16 $  & $1.7 $  \\
NGC2347 & $83.8 \pm 2.7 $  & $ 0.7 $  & $ 0.2 $  & $ 8.1 $  & $ -1.4 $  & $45 $  & $5 $  & $2.2 $  \\
NGC2410 & $86.0 \pm 3.3 $  & $ 1.6 $  & $ 0.4 $  & $ 10.5 $  & $ -1.4 $  & $81 $  & $26 $  & $2.3 $  \\
NGC2480 & $13.1 \pm 1.4 $  & $ 1.0 $  & \nodata  & $ 8.7 $  & $ -1.0 $  & $90 $  & $-22 $  & $1.1 $  \\
NGC2487 & $54.5 \pm 3.4 $  & $ 0.8 $  & \nodata  & $ 8.5 $  & $ -1.8 $  & $38 $  & $12 $  & $2.4 $  \\
NGC2623 & $121.9 \pm 1.7 $  & $ 3.6 $  & \nodata  & $ 29.6 $  & $ -1.0 $  & $83 $  & $-13 $  & $2.7 $  \\
NGC2639 & $99.1 \pm 3.7 $  & $ 2.1 $  & $ 0.2 $  & $ 9.7 $  & $ -1.6 $  & $45 $  & $-42 $  & $1.5 $  \\
NGC2730 & $31.2 \pm 2.6 $  & $ 0.5 $  & \nodata  & $ 7.2 $  & $ -1.1 $  & $56 $  & $-38 $  & $1.9 $  \\
NGC2906 & $91.9 \pm 3.8 $  & $ 0.8 $  & \nodata  & $ 7.7 $  & $ -1.6 $  & $55 $  & $4 $  & $1.3 $  \\
NGC2916 & $29.4 \pm 2.5 $  & $ 0.6 $  & \nodata  & $ 8.3 $  & $ -1.4 $  & $51 $  & $6 $  & $1.8 $  \\
NGC3303 & $33.4 \pm 2.1 $  & $ 1.4 $  & $ 0.2 $  & $ 10.3 $  & $ -1.8 $  & $53 $  & $42 $  & $3.0 $  \\
NGC3381 & $15.5 \pm 2.0 $  & $ 0.4 $  & \nodata  & $ 4.7 $  & $ -1.1 $  & $26 $  & $-11 $  & $0.8 $  \\
NGC3811 & $92.2 \pm 3.5 $  & $ 0.9 $  & $ 0.2 $  & $ 9.3 $  & $ -1.3 $  & $43 $  & $34 $  & $1.5 $  \\
NGC3815 & $45.7 \pm 2.4 $  & $ 0.8 $  & $ 0.2 $  & $ 8.7 $  & $ -1.6 $  & $62 $  & $-19 $  & $1.8 $  \\
NGC3994 & $78.8 \pm 2.9 $  & $ 1.0 $  & $ 0.1 $  & $ 9.9 $  & $ -1.1 $  & $59 $  & $13 $  & $1.5 $  \\
NGC4047 & $196.6 \pm 3.9 $  & $ 1.1 $  & $ 0.1 $  & $ 4.0 $  & $ -1.3 $  & $42 $  & $-1 $  & $1.7 $  \\
NGC4149 & $78.8 \pm 2.5 $  & $ 1.6 $  & \nodata  & $ 7.4 $  & $ -1.9 $  & $83 $  & $-4 $  & $1.5 $  \\
NGC4185 & $13.0 \pm 1.9 $  & $ 0.9 $  & \nodata  & $ 3.6 $  & $ -1.7 $  & $49 $  & $-15 $  & $1.9 $  \\
NGC4210 & $49.2 \pm 2.8 $  & $ 0.6 $  & \nodata  & $ 2.8 $  & $ -1.6 $  & $46 $  & $-33 $  & $1.3 $  \\
NGC4211N & $18.1 \pm 1.6 $  & $ 2.1 $  & \nodata  & \nodata  & $ -1.4 $  & $60 $  & $-35 $  & $3.3 $  \\
NGC4470 & $36.1 \pm 2.7 $  & $ 0.5 $  & \nodata  & $ 4.4 $  & $ -1.1 $  & $51 $  & $-2 $  & $1.1 $  \\
NGC4644 & $29.6 \pm 1.8 $  & $ 1.1 $  & $ 0.4 $  & $ 4.9 $  & $ -1.6 $  & $70 $  & $-39 $  & $2.4 $  \\
NGC4676A & $77.3 \pm 2.6 $  & $ 1.4 $  & $ 0.1 $  & \nodata  & $ -1.5 $  & $64 $  & $-24 $  & $3.3 $  \\
NGC4711 & $40.9 \pm 2.5 $  & $ 0.8 $  & $ 0.2 $  & $ 10.5 $  & $ -1.5 $  & $59 $  & $43 $  & $2.0 $  \\
NGC4961 & $22.0 \pm 2.0 $  & $ 0.4 $  & \nodata  & $ 8.5 $  & $ -1.0 $  & $39 $  & $25 $  & $1.2 $  \\
NGC5000 & $44.3 \pm 2.1 $  & $ 1.3 $  & \nodata  & $ 3.9 $  & $ -1.6 $  & $38 $  & $-3 $  & $2.7 $  \\
NGC5016 & $67.8 \pm 3.0 $  & $ 0.7 $  & \nodata  & $ 3.8 $  & $ -1.4 $  & $45 $  & $-21 $  & $1.3 $  \\
NGC5056 & $41.0 \pm 2.2 $  & $ 0.4 $  & $ 0.3 $  & $ 6.8 $  & $ -1.8 $  & $61 $  & $-12 $  & $2.7 $  \\
NGC5205 & $23.9 \pm 2.5 $  & $ 0.7 $  & \nodata  & $ 2.4 $  & $ -1.7 $  & $56 $  & $20 $  & $0.9 $  \\
NGC5218 & $416.9 \pm 4.6 $  & $ 2.7 $  & $ 0.1 $  & $ 28.0 $  & $ -1.5 $  & $51 $  & $4 $  & $1.4 $  \\
NGC5394 & $168.6 \pm 2.4 $  & $ 2.0 $  & $ 0.2 $  & $ 12.0 $  & $ -1.0 $  & $71 $  & $1 $  & $1.7 $  \\
NGC5406 & $62.6 \pm 3.6 $  & $ 0.9 $  & $ 0.6 $  & $ 7.2 $  & $ -1.7 $  & $46 $  & $-37 $  & $2.6 $  \\
NGC5480 & $136.6 \pm 4.2 $  & $ 1.0 $  & \nodata  & $ 10.9 $  & $ -1.0 $  & $41 $  & $2 $  & $0.9 $  \\
NGC5520 & $65.7 \pm 2.9 $  & $ 0.8 $  & \nodata  & $ 2.1 $  & $ -1.1 $  & $64 $  & $-26 $  & $0.9 $  \\
NGC5614 & $228.0 \pm 4.0 $  & $ 1.7 $  & $ 0.2 $  & $ 24.6 $  & $ -2.0 $  & $36 $  & $-20 $  & $1.9 $  \\
NGC5633 & $134.3 \pm 3.6 $  & $ 1.0 $  & \nodata  & $ 10.0 $  & $ -1.2 $  & $51 $  & $10 $  & $1.1 $  \\
NGC5657 & $28.7 \pm 2.1 $  & $ 1.3 $  & \nodata  & $ 10.1 $  & $ -1.5 $  & $71 $  & $20 $  & $1.9 $  \\
NGC5732 & $17.9 \pm 1.9 $  & $ 0.6 $  & \nodata  & \nodata  & $ -1.2 $  & $61 $  & $35 $  & $1.8 $  \\
NGC5784 & $32.4 \pm 1.9 $  & \nodata  & $ 0.2 $  & $ 8.6 $  & \nodata  & $28 $  & $-34 $  & $2.7 $  \\
NGC5908 & $392.6 \pm 4.5 $  & $ 2.1 $  & $ 0.3 $  & $ 60.9 $  & $ -1.9 $  & $65 $  & $-28 $  & $1.6 $  \\
NGC5930 & $136.4 \pm 2.8 $  & $ 1.7 $  & \nodata  & $ 10.1 $  & $ -1.2 $  & $71 $  & $-44 $  & $1.3 $  \\
NGC5934 & $90.7 \pm 2.5 $  & $ 2.1 $  & $ 0.2 $  & $ 32.3 $  & $ -1.7 $  & $69 $  & $3 $  & $2.8 $  \\
NGC5947 & $11.0 \pm 1.3 $  & $ 0.5 $  & $ 0.2 $  & $ 5.9 $  & $ -1.4 $  & $47 $  & $24 $  & $2.9 $  \\
NGC5953 & $358.0 \pm 4.4 $  & $ 1.7 $  & \nodata  & $ 12.8 $  & $ -1.0 $  & $44 $  & $-39 $  & $1.0 $  \\
NGC5980 & $137.7 \pm 3.0 $  & $ 1.3 $  & $ 0.2 $  & $ 10.0 $  & $ -1.3 $  & $76 $  & $11 $  & $2.0 $  \\
NGC6004 & $66.8 \pm 3.3 $  & $ 0.8 $  & $ 0.1 $  & $ 3.5 $  & $ -1.6 $  & $39 $  & $10 $  & $1.9 $  \\
NGC6060 & $132.7 \pm 4.0 $  & $ 1.5 $  & $ 0.4 $  & $ 9.4 $  & $ -1.5 $  & $59 $  & $11 $  & $2.1 $  \\
NGC6155 & $81.1 \pm 3.4 $  & $ 1.0 $  & \nodata  & $ 10.9 $  & $ -1.1 $  & $51 $  & $-33 $  & $1.2 $  \\
NGC6168 & $29.8 \pm 2.4 $  & $ 1.1 $  & $ 0.1 $  & $ 9.8 $  & $ -1.0 $  & $90 $  & $20 $  & $1.2 $  \\
NGC6186 & $164.3 \pm 3.9 $  & $ 1.6 $  & $ 0.2 $  & $ 11.4 $  & $ -1.4 $  & $71 $  & $-18 $  & $1.4 $  \\
NGC6301 & $59.2 \pm 2.7 $  & $ 1.3 $  & $ 0.9 $  & $ 7.8 $  & $ -1.4 $  & $57 $  & $-28 $  & $4.1 $  \\
NGC6310 & $8.6 \pm 1.3 $  & $ 0.9 $  & \nodata  & \nodata  & $ -2.1 $  & $90 $  & $-20 $  & $1.6 $  \\
NGC6314 & $33.2 \pm 1.7 $  & $ 2.0 $  & $ 0.2 $  & $ 8.5 $  & $ -2.3 $  & $66 $  & $-2 $  & $3.2 $  \\
NGC6361 & $376.9 \pm 3.8 $  & $ 2.6 $  & $ 0.3 $  & $ 48.3 $  & $ -1.3 $  & $84 $  & $-36 $  & $1.9 $  \\
NGC6394 & $38.1 \pm 2.2 $  & $ 1.5 $  & $ 0.4 $  & $ 7.9 $  & $ -1.6 $  & $73 $  & $30 $  & $4.2 $  \\
NGC6478 & $144.1 \pm 3.3 $  & $ 1.7 $  & $ 0.4 $  & $ 10.8 $  & $ -1.5 $  & $77 $  & $34 $  & $3.3 $  \\
NGC7738 & $98.9 \pm 1.5 $  & $ 3.0 $  & $ 0.1 $  & $ 29.1 $  & $ -1.2 $  & $64 $  & $42 $  & $3.3 $  \\
NGC7819 & $32.1 \pm 1.8 $  & $ 0.9 $  & \nodata  & $ 8.5 $  & $ -1.1 $  & $60 $  & $14 $  & $2.4 $  \\
UGC00809 & $6.0 \pm 1.1 $  & $ 1.0 $  & $ 0.2 $  & \nodata  & $ -1.1 $  & $83 $  & $22 $  & $2.0 $  \\
UGC03253 & $14.3 \pm 1.6 $  & $ 1.1 $  & $ 0.2 $  & $ 8.6 $  & $ -1.7 $  & $61 $  & $30 $  & $2.0 $  \\
UGC03539 & $56.1 \pm 2.5 $  & $ 1.3 $  & $ 0.2 $  & $ 11.8 $  & $ -1.2 $  & $90 $  & $20 $  & $1.6 $  \\
UGC03969 & $46.4 \pm 2.3 $  & $ 1.6 $  & $ 0.4 $  & $ 13.2 $  & $ -1.5 $  & $80 $  & $44 $  & $4.0 $  \\
UGC03973 & $33.1 \pm 2.1 $  & $ 0.8 $  & $ 0.3 $  & $ 9.0 $  & $ -1.1 $  & $37 $  & $26 $  & $3.3 $  \\
UGC04029 & $56.6 \pm 2.4 $  & $ 1.4 $  & $ 0.3 $  & $ 17.7 $  & $ -1.8 $  & $90 $  & $-27 $  & $2.2 $  \\
UGC04132 & $182.8 \pm 3.6 $  & $ 1.7 $  & $ 0.3 $  & $ 32.5 $  & $ -1.2 $  & $77 $  & $27 $  & $2.6 $  \\
UGC04280 & $14.1 \pm 1.5 $  & $ 0.9 $  & \nodata  & $ 8.2 $  & $ -1.7 $  & $90 $  & $-3 $  & $1.7 $  \\
UGC04461 & $32.3 \pm 2.0 $  & $ 0.8 $  & $ 0.2 $  & $ 9.3 $  & $ -1.1 $  & $78 $  & $-42 $  & $2.4 $  \\
UGC05108 & $27.2 \pm 1.6 $  & $ 2.0 $  & $ 0.3 $  & $ 8.5 $  & $ -1.5 $  & $74 $  & $5 $  & $4.0 $  \\
UGC05111 & $90.7 \pm 2.5 $  & $ 2.2 $  & $ 0.5 $  & $ 35.2 $  & $ -1.6 $  & $90 $  & $29 $  & $3.3 $  \\
UGC05359 & $11.6 \pm 1.2 $  & $ 1.0 $  & $ 0.5 $  & $ 2.9 $  & $ -1.6 $  & $72 $  & $0 $  & $4.2 $  \\
UGC05598 & $20.1 \pm 1.8 $  & $ 1.1 $  & $ 0.2 $  & $ 10.3 $  & $ -1.3 $  & $79 $  & $36 $  & $2.7 $  \\
UGC07012 & $9.0 \pm 1.3 $  & $ 0.4 $  & \nodata  & $ 6.6 $  & $ -0.9 $  & $59 $  & $15 $  & $1.5 $  \\
UGC08107 & $80.0 \pm 2.4 $  & $ 1.5 $  & $ 0.3 $  & $ 52.1 $  & $ -1.5 $  & $84 $  & $43 $  & $4.1 $  \\
UGC08267 & $50.2 \pm 2.2 $  & $ 2.2 $  & $ 0.3 $  & $ 12.9 $  & $ -1.5 $  & $85 $  & $40 $  & $3.5 $  \\
UGC09067 & $45.3 \pm 2.4 $  & $ 1.0 $  & $ 0.3 $  & $ 7.7 $  & $ -1.3 $  & $69 $  & $14 $  & $3.9 $  \\
UGC09476 & $47.2 \pm 3.1 $  & $ 0.7 $  & $ 0.2 $  & \nodata  & $ -1.3 $  & $53 $  & $2 $  & $1.6 $  \\
UGC09537 & $38.7 \pm 2.3 $  & $ 1.4 $  & $ 0.6 $  & $ 10.1 $  & $ -1.7 $  & $90 $  & $-43 $  & $4.4 $  \\
UGC09542 & $28.9 \pm 2.0 $  & $ 1.2 $  & $ 0.3 $  & $ 11.1 $  & $ -1.4 $  & $76 $  & $31 $  & $2.7 $  \\
UGC09665 & $61.6 \pm 2.8 $  & $ 1.2 $  & $ 0.2 $  & $ 29.0 $  & $ -1.3 $  & $90 $  & $-35 $  & $1.2 $  \\
UGC09759 & $46.0 \pm 2.2 $  & $ 1.6 $  & \nodata  & $ 9.7 $  & $ -1.7 $  & $63 $  & $-43 $  & $1.7 $  \\
UGC09873 & $10.4 \pm 1.2 $  & $ 1.3 $  & $ 0.2 $  & $ 9.3 $  & $ -1.2 $  & $83 $  & $36 $  & $3.1 $  \\
UGC09892 & $18.0 \pm 1.6 $  & $ 1.0 $  & $ 0.3 $  & $ 8.7 $  & $ -1.5 $  & $90 $  & $13 $  & $2.8 $  \\
UGC09919 & $5.8 \pm 1.1 $  & $ 1.0 $  & \nodata  & \nodata  & $ -1.3 $  & $90 $  & $-10 $  & $1.6 $  \\
UGC10043 & $66.0 \pm 2.9 $  & $ 1.3 $  & \nodata  & $ 10.2 $  & $ -1.6 $  & $90 $  & $-29 $  & $1.1 $  \\
UGC10123 & $97.4 \pm 2.9 $  & $ 1.7 $  & $ 0.2 $  & $ 27.6 $  & $ -1.5 $  & $90 $  & $-37 $  & $1.8 $  \\
UGC10205 & $37.4 \pm 2.2 $  & $ 2.0 $  & $ 0.2 $  & $ 10.1 $  & $ -1.9 $  & $59 $  & $36 $  & $3.2 $  \\
UGC10380 & $17.0 \pm 1.7 $  & $ 1.7 $  & $ 0.2 $  & $ 8.9 $  & $ -1.7 $  & $90 $  & $-29 $  & $3.8 $  \\
UGC10384 & $69.1 \pm 2.2 $  & $ 1.5 $  & $ 0.2 $  & $ 13.5 $  & $ -1.0 $  & $90 $  & $23 $  & $2.4 $  \\
UGC10710 & $36.2 \pm 2.2 $  & $ 1.3 $  & $ 0.4 $  & $ 7.5 $  & $ -1.8 $  & $84 $  & $-36 $  & $4.1 $  \\

%% file: res_tab.tex
$I_{12}$ & \twco~ intensity & 1 & 
 $0.74$ & $ < 0.001$ & & $0.62$ & $ < 0.001$ & &$0.62$ & $ < 0.001$ \\

$I_{13}$  & \ttco~ intensity & 2 &
 $0.32$ & $ < 0.001$ & & $0.21$ & $0.0022$ & &$0.20$ & $0.079$ \\
 
MOM2  &  \twco~ Moment2& 1 & 
 $0.24$ & $ < 0.001$ & & $0.25$ & $ < 0.001$ & &$0.09$ & $0.43$ \\
 
$\Sigma_{*}$ & Stellar mass surface density & 3 & 
 $0.16$ & $0.0069$ & & $-0.14$ & $0.048$ & &$-0.02$ & $0.87$ \\
 
 $Z_{\rm mwt}$ & Mass weighted stellar metallicity&  3 & 
 $0.01$ & $0.88$ & & $0.21$ & $0.0024$ & &$-0.18$ & $0.12$ \\
 
Age$_{\rm mwt}$  & Mass weighted stellar age & 3 & 
 $0.11$ & $0.073$ & & $0.04$ & $0.6$ & &$0.06$ & $0.63$ \\
 
 Av & Nebular extinction from Pipe3D  & 3 & 
 $0.26$ & $ < 0.001$ & & $0.33$ & $ < 0.001$ & &$0.35$ & $0.002$ \\
 
 12 + $\log$(O/H) & Gas phase metallicity from O3N2 & 3 & 
 $-0.03$ & $0.65$ & & $-0.00$ & $0.98$ & &$0.38$ & $0.0027$ \\
 
$\Sigma_{\mathrm{SFR}}$ & Star formation rate surface density&  3 & 
 $0.24$ & $ < 0.001$ & & $-0.16$ & $0.02$ & &$0.06$ & $0.64$ \\
 
  $sSFR$ & Specific star formation rate & 3 &
 $0.32$ & $ < 0.001$ & & $0.07$ & $0.33$ & &$0.37$ & $0.0036$ \\
 
 $\Sigma_{\mathrm{mol}}/\Sigma_{*}$ & Molecular gas to stellar mass fraction & 2 &
 $0.65$ & $ < 0.001$ & & $0.52$ & $ < 0.001$ & &$0.60$ & $ < 0.001$ \\
 
 $t_{\mathrm{dep}}$ & Depletion time assuming constant $\rm X_{\rm ^{12}CO}$ & 2 & 
 $0.04$ & $0.56$ & & $0.31$ & $ < 0.001$ & &$0.07$ & $0.57$ \\